\documentclass{article}
\usepackage[utf8]{inputenc}
\usepackage[margin=1in]{geometry}
\usepackage[dvipsnames]{xcolor}
\usepackage{amsmath}
\usepackage{physics}
\usepackage{graphicx}
\usepackage{amssymb}
\usepackage{enumitem}
\usepackage{mathtools}
\usepackage{authblk}
\usepackage{hyperref}
\usepackage{caption}
\usepackage[biblabel=brackets,style=phys,eprint=true]{biblatex}

\newcommand{\bbeta}{\mathfrak b}

\begin{document}
\title{Spectral Form Factor of a Quantum Spin Glass}
\author[1]{Michael Winer}
\author[1,2]{Richard Barney}
\author[1]{Christopher L. Baldwin}
\author[1]{Victor Galitski}
\author[3]{Brian Swingle}
\affil[1]{Joint Quantum Institute, Department of Physics, University
of Maryland, College Park, Maryland 20742, USA}
\affil[2]{Condensed Matter Theory Center, Department of Physics, University
of Maryland, College Park, Maryland 20742, USA}
\affil[3]{Department of Physics, Brandeis University, Waltham, Massachusetts 02453, USA}

\maketitle

\begin{abstract}
    It is widely expected that systems which fully thermalize are chaotic in the sense of exhibiting random-matrix statistics of their energy level spacings, whereas integrable systems exhibit Poissonian statistics. In this paper, we investigate a third class: spin glasses. These systems are partially chaotic but do not achieve full thermalization due to large free energy barriers. We examine the level spacing statistics of a canonical infinite-range quantum spin glass, the quantum $p$-spherical model, using an analytic path integral approach. We find statistics consistent with a direct sum of independent random matrices, and show that the number of such matrices is equal to the number of distinct metastable configurations---the exponential of the spin glass ``complexity'' as obtained from the quantum Thouless-Anderson-Palmer equations. We also consider the statistical properties of the complexity itself and identify a set of contributions to the path integral which suggest a Poissonian distribution for the number of metastable configurations. Our results show that level spacing statistics can probe the ergodicity-breaking in quantum spin glasses and provide a way to generalize the notion of spin glass complexity beyond models with a semi-classical limit.

\end{abstract}

\tableofcontents

\section{Introduction} \label{sec:introduction}



An isolated quantum many-body system which reaches an effective thermal equilibrium state starting from an out-of-equilibrium initial state is often called ``quantum chaotic." As commonly used, quantum chaos is a loose term referring to a family of phenomena that typically co-occur, including the ability of the system to serve as its own heat bath ~\cite{Deutsch1991,Srednicki1994,Rigol2008Thermalization}, hydrodynamic behavior of conserved quantities ~\cite{glorioso2018lectures,crossley,PhysRevD.91.105031,Haehl_2018,Jensen_2018}, and random-matrix-like energy eigenvalues~\cite{PhysRevLett.52.1,Dyson1962I,mehta2004random,guhr1998random}. Given this variety, it is crucial to understand the relationships between different manifestations of quantum chaos~\cite{DAlessio2016From,Santos2010}.

These relationships are complicated and interesting in large part because the systems in question have structure, such as locality and symmetry. For example, if the Hamiltonian has spatial locality, energy conservation implies the existence of slow hydrodynamic modes and an associated long time scale, the Thouless time, such that random-matrix behavior is only present for energy levels closer than the inverse Thouless time~\cite{Chan_2018,Moudgalya_2021}. Similarly, if the Hamiltonian possesses a symmetry, then it can be organized into blocks labelled by irreducible representations of the symmetry. One finds random-matrix statistics within each individual block, but full ergodicity is broken because matrix elements between different blocks are forbidden~\cite{2019Santos,PhysRevE.102.060202,winer2020hydrodynamic,winer2021spontaneous,Roy2022zig}.

It is natural to ask whether there are other ways in which ergodicity can be lost, and if so, what the resulting spectral statistics of the Hamiltonians are. In particular, we will better understand the relations between different measures of quantum chaos by understanding how they are lost and what replaces them. 

Quantum spin glasses provide one well-established context to explore these questions, since they exhibit a rich phenomenology associated with the inability to fully thermalize~\cite{Binder1986Spin,Mezard1987,Fischer1991,Nishimori2001,Castellani2005Spin,Mezard2009,Stein2013}
In this paper, we determine the spectral statistics of an analytically tractable spin glass model, the quantum $p$-spherical model. We find that up to times polynomial in the system size, the Hamiltonian can effectively be described as approximately block-diagonal. Each block behaves as a random matrix independent of the others, and the number of blocks depends on the energy per particle. At high energies, there is only one block and the system is ergodic. Below a critical energy density, the Hamiltonian breaks into exponentially many blocks --- the average number of blocks jumps discontinuously from the high energy regime and then decreases as the energy density decreases further. We establish these results via a path integral computation of the spectral form factor (SFF), which measures correlations between pairs of energy levels~\cite{saad2019semiclassical,babyUniverse,Winer_2020,PhysRevE.72.046207,yiming}.

In the remainder of the introduction, we give some physical context by reviewing the spectral form factor and mean-field spin glasses, and then summarize our results.
In Sec.~\ref{sec:overview_PSM}, we review the $p$-spherical model in detail.
In Sec.~\ref{sec:ergodic_ramp}, we calculate the SFF of this model in the high-temperature ergodic regime, and in Sec.~\ref{sec:nonergodic_ramp}, we do so in the non-ergodic regime. Finally, in Sec.~\ref{sec:HigherMoments}, we investigate higher-moment analogues of the SFF.
We then discuss implications of these results and directions for future work in Sec.~\ref{sec:conclusion}.

\begin{figure}[t]
    \centering
    \includegraphics[width=.9\textwidth]{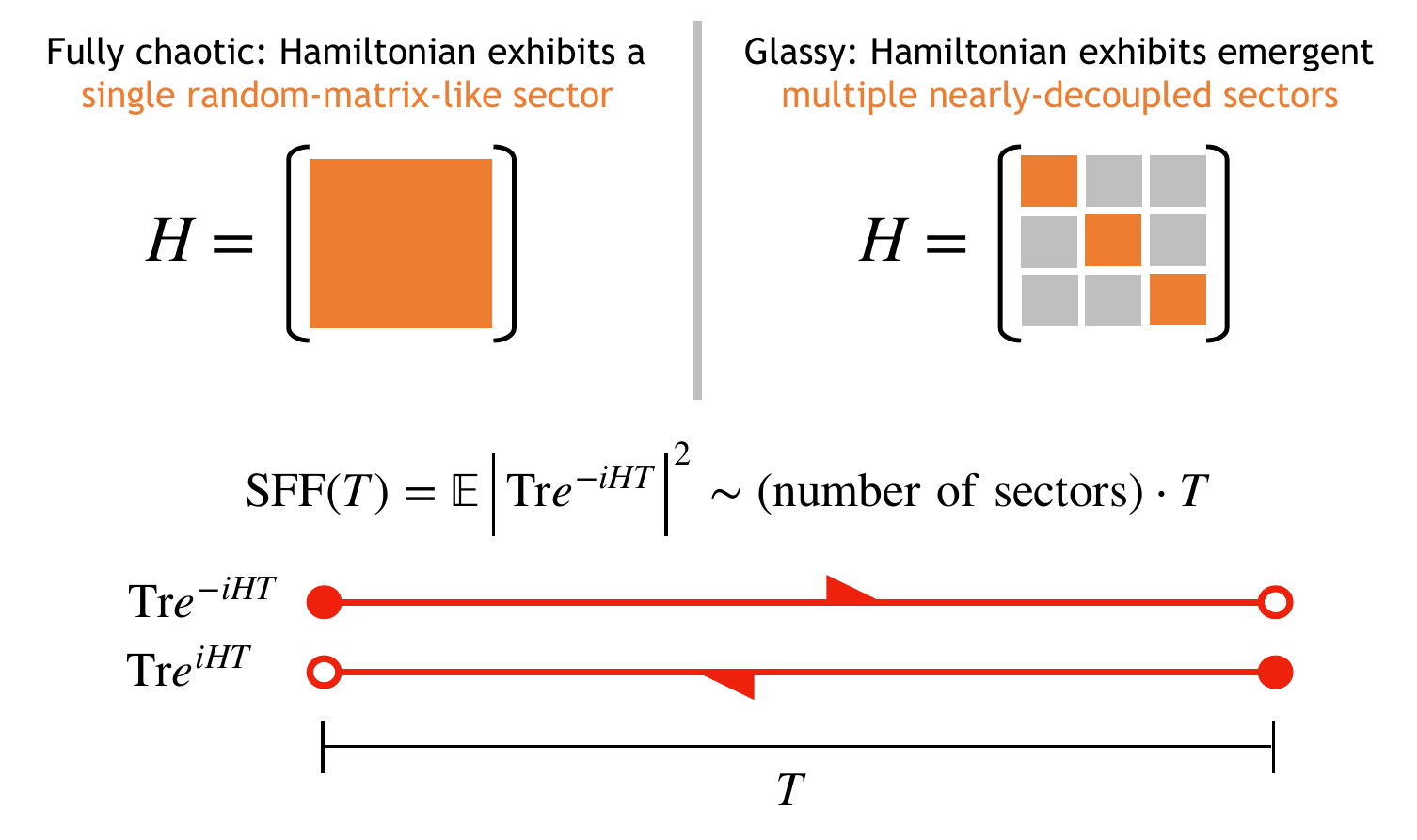}
    \caption{(Top left) Fully chaotic systems have energy levels that are statistically similar to a Gaussian random matrix, indicated by the orange block. (Top right) By contrast, quantum spin glasses in the non-ergodic phase have spectral statistics that resemble a collection of many nearly-decoupled random matrices (Bottom) Spectral statistics can be diagnosed via the spectral form factor, denoted $\textrm{SFF}(T)$, which consists of a path integral over a pair of real-time contours as indicated by the red lines. The universal part of $\textrm{SFF}(T)$, which is proportional to $T$, is enhanced by the number of effectively uncoupled sectors (other non-universal contributions are not indicated here).}
    \label{fig:my_label}
\end{figure}

\subsection{Review of the spectral form factor} \label{subsec:review_spectral_form_factor}

To study the spectral correlations of a Hamiltonian $H$, a standard tool is the spectral form factor (SFF) \cite{Cotler2017,brezin1997spectral}, defined as
\begin{equation} \label{eq:general_SFF_definition}
\textrm{SFF}(T) \equiv \big| \textrm{Tr} e^{-iHT} \big|^2.
\end{equation}
In situations where the spectrum is unbounded, or when one wishes to concentrate on a portion of the spectrum, the trace in Eq.~\eqref{eq:general_SFF_definition} is regulated by a filter function $f(H)$:
\begin{equation} \label{eq:filtered_SFF_definition}
\textrm{SFF}(T, f) \equiv \big| \textrm{Tr} f(H) e^{-iHT} \big|^2.
\end{equation}
One common choice is $f(H) = e^{-\beta H}$\cite{saad2019semiclassical,Papadodimas_2015}, and another is $f(H) = e^{-c(H - E_0)^2}$.
The latter allows one to study level statistics near a specified energy $E_0$.

For a single Hamiltonian, the SFF is an erratic function of time \cite{brezin1997spectral}.
Thus one usually considers an ensemble of Hamiltonians and defines the SFF as the average of Eq.~\eqref{eq:filtered_SFF_definition} over the ensemble.
Throughout this paper, we use the notation $\mathbb{E}[ \, \cdot \, ]$ to denote the ensemble average.

The SFF is closely related to the correlation function of the density of states.
Formally, the (filtered) density of states is given by
\begin{equation} \label{eq:general_density_of_states_definition}
\rho(E, f) \equiv \sum_n f(E_n) \delta(E - E_n) = \textrm{Tr} f(H) \delta(E - H),
\end{equation}
where $n$ labels the eigenstate of $H$ with eigenvalue $E_n$, and its correlation function is
\begin{equation} \label{eq:density_of_states_correlation_definition}
C(E, \omega, f) \equiv \mathbb{E} \left[ \rho \left( E + \frac{\omega}{2}, f \right) \rho \left( E - \frac{\omega}{2}, f \right) \right].
\end{equation}
We have that
\begin{equation} \label{eq:SFF_density_of_states_relationship}
\begin{aligned}
\textrm{SFF}(T, f) &= \mathbb{E} \Big[ \textrm{Tr} f(H) e^{-iHT} \textrm{Tr} f(H) e^{iHT} \Big] \\
&= \int dE d\omega \, e^{-i \omega T} \mathbb{E} \left[ \textrm{Tr} f(H) \delta \left( E + \frac{\omega}{2} - H \right) \textrm{Tr} f(H) \delta \left( E - \frac{\omega}{2} - H \right) \right] \\
&= \int d\omega \, e^{-i \omega T} \int dE \, C(E, \omega, f).
\end{aligned}
\end{equation}
The SFF is simply the Fourier transform of the correlation function with respect to $\omega$, integrated over $E$ (although the filter function allows one to concentrate on an arbitrary subset of the spectrum).

\begin{figure}
\centering
\includegraphics[width=0.9\textwidth]{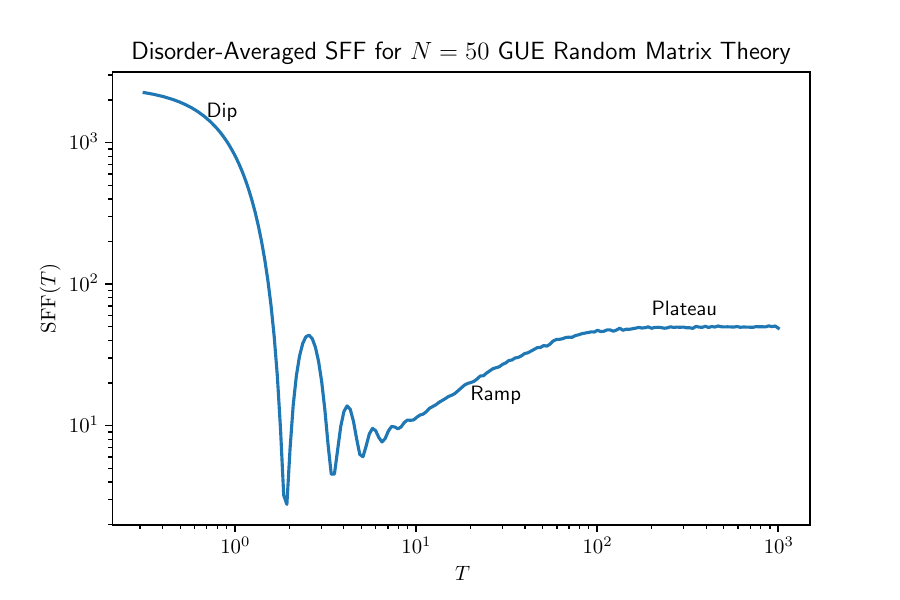}
\caption{The disorder-averaged SFF for the Gaussian unitary ensemble (GUE) of matrix dimension $N = 50$, computed numerically by averaging over ten thousand realizations. The three distinct regimes --- dip, ramp, plateau --- are indicated.}
\label{fig:SFFgraph}
\end{figure}

It is conceptually useful to split the SFF into two contributions:
\begin{equation} \label{eq:general_SFF_decomposed}
\textrm{SFF}(T, f) = \big| \mathbb{E} \textrm{Tr} f(H) e^{-iHT} \big|^2 + \bigg( \mathbb{E} \left[ \big| \textrm{Tr} f(H) e^{-iHT} \big|^2 \right] - \big| \mathbb{E} \textrm{Tr} f(H) e^{-iHT} \big|^2 \bigg).
\end{equation}
The first term, the disconnected piece of the SFF, comes solely from the average density of states.
It is the second term, the connected piece, that contains information on the correlation between energy levels.
The assertion of ``random matrix universality''~\cite{PhysRevLett.52.1,Altland1997} can be phrased as the statement that an ensemble of quantum chaotic Hamiltonians will generically have the same \textit{connected} SFF as the canonical Gaussian ensembles of random matrix theory~\cite{mehta2004random,tao2012topics}.
This conjectured universal behavior is illustrated in Fig.~\ref{fig:SFFgraph}, which plots the disorder-averaged SFF of the Gaussian unitary ensemble (one of the aforementioned canonical ensembles).
Note the three distinct regimes:
\begin{itemize}
\item The ``dip'', occurring at short times, comes from the disconnected piece of the SFF (and thus its precise shape is non-universal).
It reflects a loss of constructive interference --- the different terms of $\textrm{Tr} e^{-iHT}$ acquire different phase factors as $T$ increases.
\item The ``ramp'', occurring at intermediate times, is arguably the most interesting regime.
In the canonical matrix ensembles, it is a consequence of the result\cite{mehta2004random}
\begin{equation} \label{eq:matrix_ensembles_ramp_result}
\mathbb{E} \left[ \rho \left( E + \frac{\omega}{2} \right) \rho \left( E - \frac{\omega}{2} \right) \right] - \mathbb{E} \left[ \rho \left( E + \frac{\omega}{2} \right) \right] \mathbb{E} \left[ \rho \left( E - \frac{\omega}{2} \right) \right] \sim -\frac{1}{\bbeta \pi^2 \omega^2},
\end{equation}
where $\bbeta = 1$, $2$, $4$ in the orthogonal, unitary, and sympletic ensembles respectively \cite{mehta2004random}.
The right-hand side being negative is a reflection of the well-known level repulsion in quantum chaotic systems \cite{wigner1959group}.
Taking the Fourier transform with respect to $\omega$ gives a term proportional to $T$ for the connected SFF.
Such a linear-in-$T$ ramp is often taken as a defining signature of quantum chaos.
\item The ``plateau'', occurring at late times, results from the discreteness of the spectrum.
At times much larger than the inverse level spacing, one expects that all off-diagonal terms in the double-trace of the SFF sum to effectively zero, meaning that
\begin{equation} \label{eq:SFF_plateau_derivation}
\textrm{SFF}(T, f) = \sum_{mn} e^{-i(E_m - E_n)T} f(E_m) f(E_n) \sim \sum_n f(E_n)^2.
\end{equation}
As the plateau regime is both challenging to access analytically and not particularly informative, we shall not consider it further in this work.
\end{itemize}

The bulk of our analysis in this paper is devoted to calculation of the ramp in a well-known quantum spin glass model, the $p$-spherical model (discussed below).
The results can be understood via the elementary observation that when a Hamiltonian is block diagonal,
\begin{equation} \label{eq:block_diagonal_matrix}
H = \begin{pmatrix} H_1 & 0 & 0 \\ 0 & H_2 & 0 & \hdots \\ 0 & 0 & H_3 \\ & \vdots & & \ddots \end{pmatrix},
\end{equation}
then $\textrm{Tr} e^{-iHT} = \sum_k \textrm{Tr} e^{-iH_kT}$.
If the different blocks are independent, then the variance of $\textrm{Tr} e^{-iHT}$ is the sum of the variance of each $\textrm{Tr} e^{-iH_kT}$, i.e., \textit{the SFF is the sum of the SFF for each block.}
In particular, the coefficient of the universal linear-in-$T$ ramp is multiplied by the number of independent blocks.
Systems with only approximately block-diagonal Hamiltonians, for which there are small matrix elements between blocks, have this enhancement of the ramp up to the transition timescale between blocks.
For a more detailed analysis, see Ref.~\cite{winer2020hydrodynamic}.

\subsection{Review of mean-field spin glasses} \label{subsec:review_mean_field_spin_glasses}

\begin{figure}[t]
\centering
\includegraphics[width=0.9\textwidth]{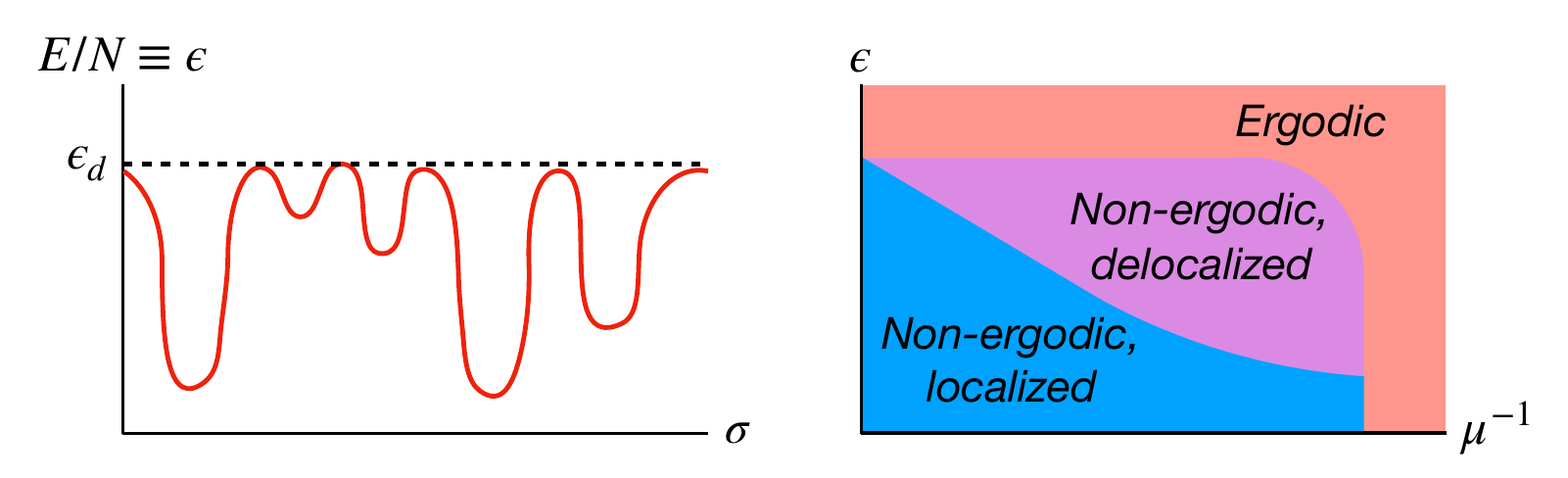}
\caption{(Left) Cartoon of the energy landscape in a 1RSB spin glass. The y-axis is energy per spin, $E/N$, where $E$ is energy and $N$ is the number of spins. Different points on the x-axis represent (very roughly, since the actual configuration space is $N$-dimensional) different spin configurations $\sigma$. The dashed line indicates the energy density $\epsilon_d$ below which the system is non-ergodic. (Right) Sketch of the dynamical phase diagram for a quantum 1RSB spin glass. The x-axis represents parameters controlling the strength of quantum fluctuations, and the y-axis is energy density. Note that many other types of phase transitions are also present, in particular equilibrium transitions, but are not indicated here. See, e.g., Refs.~\cite{Castellani2005Spin,Mezard2009,Anous:2021eqj} for more information.}
\label{fig:spin_glass_cartoons}
\end{figure}

Broadly speaking, spin glasses are systems in which the magnetic moments $\sigma_i$ are frozen but disordered at low temperatures.
However, this definition (much like that of ``quantum chaos'') encompasses a wide variety of phenomena which are in many ways quite distinct, as is made clear by the literature on the subject~\cite{Binder1986Spin,Mezard1987,Fischer1991,Nishimori2001,Castellani2005Spin,Mezard2009,Stein2013}.
In the present paper, we focus on what are known as ``one-step replica symmetry breaking'' (1RSB) spin glass phases~\cite{Mezard2009}.
We are specifically interested in quantum spin glasses, but we first review the corresponding classical case, for which configurations are labelled by a list $\sigma \equiv \{ \sigma_1, \cdots, \sigma_N \}$ and the Hamiltonian is simply a function of $\sigma$.

While the technical definition of 1RSB is somewhat involved, the qualitative physics is straightforward to understand and captured by the sketch in Fig.~\ref{fig:spin_glass_cartoons}.
The energy landscape, i.e., energy as a function of spin configuration, has many deep wells and steep barriers.
In particular, the number of wells is $e^{O(N)}$ and the heights of the energy barriers separating wells are $O(N)$, where $N$ is the number of spins. As a result, below a certain energy density $\epsilon_d$, the system is extremely non-ergodic: it remains trapped within an exponentially small fraction of the thermodynamically relevant configuration space until exponentially long timescales.
While the 1RSB phenomenon was originally studied in the context of stochastic classical dynamics~\cite{Kirkpatrick1987Dynamics,Crisanti1993Spherical,Cugliandolo1993Analytical,Barrat1996Dynamics}, it has recently been shown to imply exponentially long \textit{tunneling} timescales for isolated quantum dynamics as well~\cite{Altshuler2010Anderson,Bapst2013Quantum,Zhao2014Three,Baldwin2018Quantum,Smelyanskiy2020Nonergodic}.

TAP states (named after Thouless, Anderson, and Palmer~\cite{Thouless1977Solution}) provide a more quantitative description of such ``deep wells''.
Arguably the most general definition (see Ref.~\cite{Nishimori2001} for others) is in terms of the Legendre transform of the free energy with respect to local fields:
\begin{equation} \label{eq:TAP_Legendre_transform}
F \big( \{ m_i \} \big) = -\frac{1}{\beta} \log{\textrm{Tr} e^{-\beta H + \beta \sum_i h_i \sigma_i}} + \sum_i h_i m_i,
\end{equation}
where $H$ is the Hamiltonian of interest and the fields $\{ h_i \}$ are chosen so that $\langle \sigma_i \rangle = m_i$ (where $\langle \, \cdot \, \rangle$ indicates a thermal average).
TAP states are simply the local minima of $F(\{m_i\})$.
Physically, each corresponds to a different ``well'' of the energy landscape, including thermal fluctuations around the lowest point (thus TAP states do generically depend on temperature).
The partition function can be decomposed as a sum over TAP states:
\begin{equation} \label{eq:TAP_partition_decomposition}
Z \equiv \sum_{\sigma} e^{-\beta H(\sigma)} = \sum_{\alpha} \left[ \sum_{\sigma} \delta_{\sigma \in \alpha} e^{-\beta H(\sigma)} \right] \equiv \sum_{\alpha} Z_{\alpha},
\end{equation}
where $\alpha$ denotes a TAP state and $\delta_{\sigma \in \alpha}$ restricts the trace to only those states belonging to TAP state $\alpha$.
Note that in this discussion, $\sigma$ can refer to any set of degrees of freedom: Ising spins, vector spins, continuous coordinates, etc.
In all cases, Eqs.~\eqref{eq:TAP_Legendre_transform} and~\eqref{eq:TAP_partition_decomposition} can be interpreted accordingly.

Quantum generalizations of spin glasses are usually obtained by adding non-commuting terms to the Hamiltonian.
For example, with an Ising Hamiltonian, one often interprets $\sigma_i$ as the Pauli spin-$z$ operator $\sigma_i^z$ and includes an additional transverse field $\Gamma \sum_i \sigma_i^x$~\cite{Ishii1985Effect,Thirumalai1989Infinite,Goldschmidt1990,Buttner1990Replica}.
On the other hand, with systems having continuous degrees of freedom (including the one which we study in this paper), one can interpret $\sigma_i$ as a position coordinate and include the ``kinetic energy'' $\sum_i \pi_i^2 / 2\mu$, where $\pi_i$ is the momentum operator conjugate to $\sigma_i$~\cite{Cugliandolo1999RealTime,Cugliandolo2001}.
Generically, the resulting system has a frozen spin glass phase at low energy and small quantum fluctuations (the latter being controlled by $\Gamma$ and $\mu^{-1}$ respectively in the examples above), and has a paramagnetic phase at either high energy or large quantum fluctuations.
A sketch of the typical phase diagram is shown in Fig.~\ref{fig:spin_glass_cartoons}, with these two phases indicated by ``non-ergodic'' and ``ergodic''.

It has recently been noted that quantum 1RSB spin glasses can exhibit \textit{eigenstate} phase transitions which are distinct from the above~\cite{Laumann2014Many,Baldwin2017Clustering,Biroli2021Out}.
Qualitatively speaking, on the low energy/fluctuation side of the eigenstate phase boundary, each eigenstate of the Hamiltonian is localized on a single TAP state.
This implies that under the system's internal dynamics alone (i.e., as given by the Schrodinger equation), the system cannot tunnel between TAP states on \textit{any} timescale, even times exponential in the number of spins.
On the other side of the phase boundary, each eigenstate is delocalized over many TAP states in accordance with random matrix behavior.
As discussed in Ref.~\cite{Baldwin2018Quantum}, while this implies that the system does tunnel between TAP states, the timescale for tunneling is necessarily exponential in system size, analogous to the activation times under open-system dynamics.
Only when there exists a single TAP state can one identify the phase as genuinely thermalizing.
As a result, one finds phase diagrams like that sketched in Fig.~\ref{fig:spin_glass_cartoons}, with ``non-ergodic''/``ergodic'' indicating whether multiple TAP states exist and ``localized''/``delocalized'' referring to the eigenstate properties.


\subsection{Summary of results} \label{subsec:implications}

In this paper, we calculate the SFF for a particular ensemble of quantum spin glasses, the quantum $p$-spherical model (PSM)~\cite{Cugliandolo1998Quantum,Cugliandolo1999RealTime,Cugliandolo2001}. 
We find that in the ergodic phase, the connected part of the SFF agrees with the expectation from random matrix theory (Eq.~\eqref{eq:SFF_connected_contribution_ergodic_phase} below), while in the non-ergodic phase, it is enhanced by a factor which is precisely the number of TAP states (Eq.~\eqref{eq:SFF_final_result}).
Given the discussion in Secs.~\ref{subsec:review_spectral_form_factor} and~\ref{subsec:review_mean_field_spin_glasses}, this makes precise and validates the idea that each metastable state (i.e., TAP state) corresponds to a block of the Hamiltonian that is quantum chaotic on its own but is nearly decoupled from all others, thus making the system as a whole non-ergodic~\cite{Baldwin2017Clustering}.
This is the main result of the present work.

We also consider higher moments of the evolution operator and identify a set of saddle points (Eq.~\eqref{eq:higher_moment_special_final_expression}) which, in addition to confirming the picture that different TAP states have independent level statistics, suggest that at least at low complexity, the number of TAP states at a given energy is Poisson-distributed and independent of other energies.
Yet as we shall discuss, since our analysis does not consider the perturbative corrections around each saddle point, this does not constitute a complete calculation and serves more as motivation for future investigation.

\section{Real-time dynamics of the quantum $p$-spherical model} \label{sec:overview_PSM}

\subsection{The model} \label{subsec:model}

The classical $p$-spherical model (PSM)~\cite{Crisanti1992} is a disordered spin model with all-to-all $p$-body interactions.
It is defined by the classical Hamiltonian
\begin{equation} \label{eq:classical_Hamiltonian}
H_{\textrm{cl}} \equiv \sum_{(i_1 \cdots i_p)} J_{i_1 \cdots i_p} \sigma_{i_1} \cdots \sigma_{i_p},
\end{equation}
where the couplings $J_{i_1\dots i_p}$ are independent Gaussian random variables with mean zero and variance
\begin{equation} \label{eq:p_spin_coupling_variance}
\mathbb{E} {J_{i_1 \cdots i_p}}^2 = \frac{J^2(p-1)!}{C_{i_1 \cdots i_p} N^{p-1}}.
\end{equation}
Here and throughout, $\mathbb{E}$ indicates an average over couplings.
The notation $(i_1 \cdots i_p)$ denotes sets of $p$ indices such that $1 \leq i_1 \leq \cdots \leq i_p \leq N$. The sum in Eq.~\eqref{eq:classical_Hamiltonian} is over all such sets.
Our treatment differs from the standard convention by including a parameter $J$ for the overall strength of the disorder.
To recover the standard expressions, simply set $J^2=p/2$.
We also include the combinatorial factor $C_{i_1 \cdots i_p} = \prod_{1 \leq i \leq N} n_i!$, where $n_i$ is the number of indices set equal to $i$.
This term is almost always one, but its inclusion avoids $1/N$ corrections in the action. 

The $\sigma_i$ are real, continuous spin variables subject to the spherical constraint
\begin{equation} \label{eq:spherical_constraint}
\sum_{i=1}^N \sigma_i^2 = N,
\end{equation}
which ensures that the system has an extensive free energy.
It is apparent that this is a mean-field model without any spatial structure.
This allows for infinite free energy barriers around metastable states in the thermodynamic limit, making the model ideal for examining the impact of metastability on the spectral statistics of spin glasses.

In this work, we follow Refs.~\cite{Cugliandolo1998Quantum,Cugliandolo1999RealTime,Cugliandolo2001} in generalizing Eq.~\eqref{eq:classical_Hamiltonian} to a quantum Hamiltonian $H$.
We treat the $\sigma_i$ as commuting position operators, and define conjugate momentum operators $\pi_i$ which satisfy the commutation relations
\begin{equation} \label{eq:canonical_commutation_relations}
[\sigma_i, \pi_j] = i \delta_{ij}.
\end{equation}
The \textit{quantum} PSM simply includes a kinetic energy term in the Hamiltonian:
\begin{equation} \label{eq:quantum_Hamiltonian}
H = \sum_{i=1}^N \frac{\pi_i^2}{2\mu} + \sum_{(i_1 \cdots i_p)} J_{i_1 \cdots i_p} \sigma_{i_1} \cdots \sigma_{i_p}.
\end{equation}
The mass $\mu$ is an additional parameter controlling the strength of quantum fluctuations.
To incorporate the spherical constraint, we take the Hilbert space to be the subspace in which $\sum_i \sigma_i^2$ has eigenvalue $N$.

The quantum PSM may be interpreted as a soft-spin version of the Ising $p$-spin model in an external transverse field --- itself the subject of much study~\cite{Gardner1985,Goldschmidt1990,Nieuwenhuizen1998,Dobrosavljevic1990,DeCesare1996} --- where $\mu^{-1}$ is analogous to the transverse field.
Alternatively, if we think of $\sigma \equiv \{ \sigma_1, \cdots, \sigma_N \}$ as a position vector in $N$-dimensional space, the quantum PSM has a natural interpretation as a particle of mass $\mu$ moving on a hypersphere of radius $\sqrt{N}$.
This particle experiences the Gaussian random potential
\begin{equation} \label{eq:hypersphere_random_potential}
V(\sigma) = \sum_{(i_1 \cdots i_p)} J_{i_1 \cdots i_p} \sigma_{i_1} \cdots \sigma_{i_p},
\end{equation}
whose correlation function is
\begin{equation} \label{eq:random_potential_correlation}
\mathbb{E} V(\sigma) V(\sigma') = \frac{J^2 (p-1)!}{C_{i_1 \cdots i_p} N^{p-1}} \sum_{(i_1 \cdots i_p)} \sigma_{i_1} \sigma'_{i_1} \cdots \sigma_{i_p} \sigma'_{i_p} = \frac{J^2}{pN^{p-1}} \big( \sigma \cdot \sigma' \big)^p.
\end{equation}

Note that there is a very important difference between $p=2$ and $p>2$: the former is a Gaussian model, essentially (but for the spherical constraint) a system of linearly coupled harmonic oscillators.
It therefore has qualitatively different behavior than the $p>2$ models, which are genuinely interacting and serve as reasonable toy models for rugged energy landscapes.
In this work, we exclusively consider $p>2$.

\subsection{Schwinger-Keldysh path integral} \label{subsec:Schwinger_Keldysh_path_integral}

\begin{figure}[t]
\centering
\includegraphics[width=1.0\textwidth]{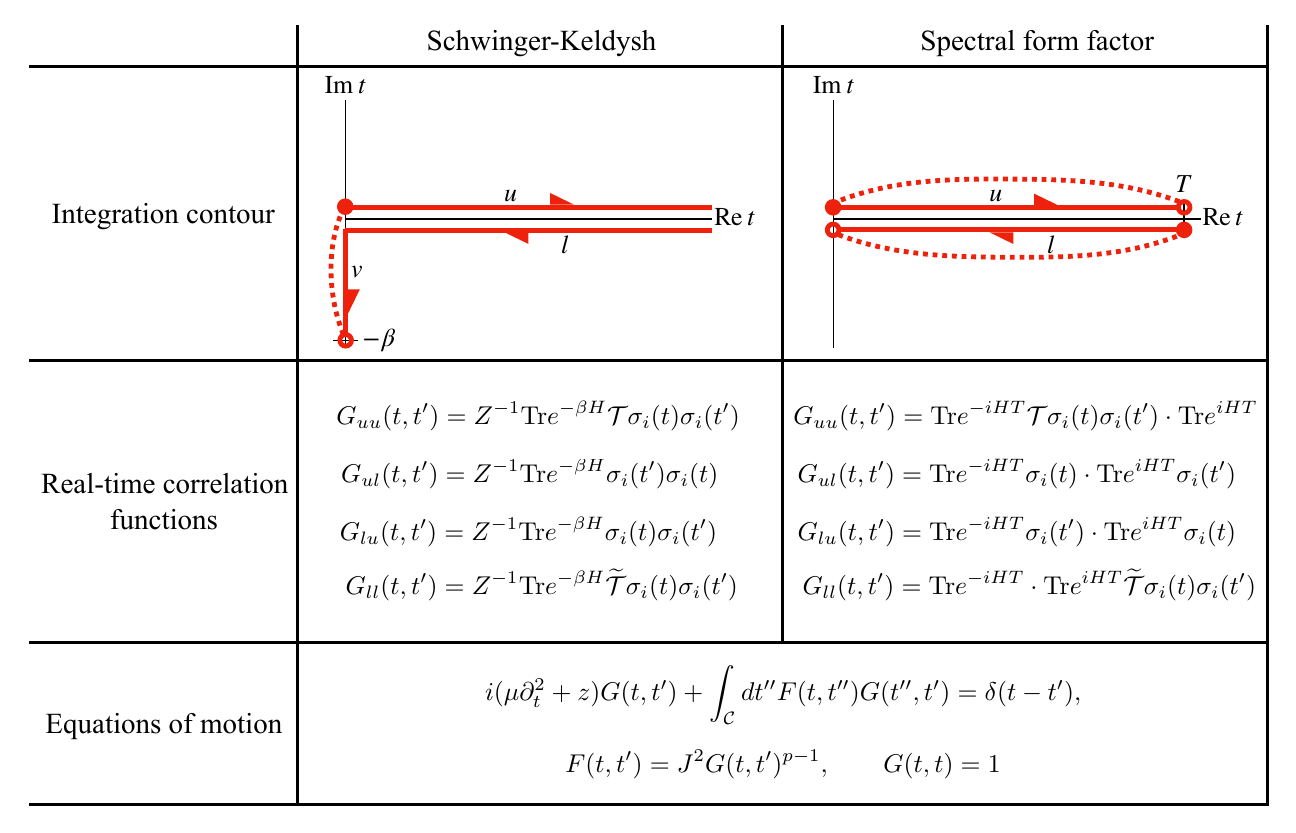}
\caption{Summary of the contours, order parameters, and (at least at high temperature) equations of motion considered in this work. The left column gives the quantities appropriate to the Schwinger-Keldysh path integral, and the right column to the spectral form factor (SFF) path integral. \\ (Top row) Contours for the respective path integrals. Each of the different branches is labelled, and directions are indicated by arrowheads. Points connected by dashed lines are identified, making the contours periodic. \\
(Middle row) Relationship between order parameters of the theory and observable quantities. $H$ and $Z$ are the $p$-spin Hamiltonian and partition function respectively. $\mathcal{T}$ and $\widetilde{\mathcal{T}}$ denote time ordering and anti-ordering. \\
(Bottom row) Equations of motion. These take the same form for both path integrals, differing only in the contour $\mathcal{C}$ being used.}
\label{fig:contour_summary}
\end{figure}

Just as other all-to-all models have a saddle-point/mean-field description at large $N$, so too does the PSM.
We start with the disorder-averaged (i.e., ``annealed'') path integral on the Schwinger-Keldysh contour at inverse temperature $\beta$, illustrated in the left column of Fig.~\ref{fig:contour_summary}.
While it is in general incorrect (often grossly) to disorder-average the path integral itself, it is known that the annealed approximation is accurate in the PSM as long as $\beta$ is less than a critical value $\beta_s$~\cite{Castellani2005Spin,Thomson_2020}.
We shall assume that this is true throughout.
The annealed path integral is
\begin{equation} \label{eq:Keldysh_path_integral_start_1}
\begin{aligned}
\mathbb{E} Z_{\textrm{SK}} &= \int \mathcal{D} \sigma^N \exp \left[ \int_{\mathcal{C}} dt \sum_i \left( \frac{i \mu}{2} \big( \partial_t \sigma_i(t) \big)^2 - \frac{i z(t)}{2} \big( \sigma_i(t)^2 - 1 \big) \right) \right] \\
&\qquad \qquad \quad \cdot \int dP(J) \exp \left[ -i \int_{\mathcal{C}} dt \sum_{(i_1 \cdots i_p)} J_{i_1 \cdots i_p} \sigma_{i_1}(t) \cdots \sigma_{i_p}(t) \right],
\end{aligned}
\end{equation}
where
\begin{equation} \label{eq:PSM_coupling_distribution}
dP(J) \propto \prod_{(i_1 \cdots i_p)} dJ_{i_1 \cdots i_p} \exp \left[ -\frac{N^{p-1} C_{i_1 \cdots i_p} J_{i_1 \cdots i_p}^2}{2(p-1)!J^2} \right].
\end{equation}
For brevity, we use $\mathcal{C}$ to denote the entire contour.
Thus $\int_{\mathcal{C}} dt$ indicates a contour integral within the complex-$t$ plane.
The Lagrange multiplier $z(t)$ is included to enforce the spherical constraint.
It can be interpreted as a time-dependent harmonic potential whose value is chosen such that $\sum_i \sigma_i(t)^2 = N$ at all times.
Thus the measure $\mathcal{D}\sigma^N$ is simply the product measure over each $\sigma_i$ independently.
From here, the same manipulations used to get Schwinger-Dyson equations for the SYK model will give us equations of motion for the PSM.

One can immediately perform the Gaussian integrals over the couplings to obtain
\begin{equation} \label{eq:disorder_averaged_partition_function}
\mathbb{E} Z_{\textrm{SK}} = \int \mathcal{D} \sigma^N e^{-NS'},
\end{equation}
where
\begin{equation} \label{eq:annealed_effective_action}
\begin{aligned}
NS' &\equiv \int_{\mathcal{C}} dt \sum_i \left( -\frac{i\mu}{2} \big( \partial_t \sigma_i(t) \big)^2 + \frac{iz(t)}{2} \big( \sigma_i(t)^2 - 1 \big) \right) \\
& \qquad \qquad + \frac{J^2 (p-1)!}{2 C_{i_1 \cdots i_p} N^{p-1}} \sum_{(i_1 \cdots i_p)} \int_{\mathcal{C}} dt dt' \sigma_{i_1}(t) \sigma_{i_1}(t') \cdots \sigma_{i_p}(t) \sigma_{i_p}(t') \\
&= \int_{\mathcal{C}} dt \sum_i \left( -\frac{i\mu}{2} \big( \partial_t \sigma_i(t) \big)^2 + \frac{iz(t)}{2} \big( \sigma_i(t)^2 - 1 \big) \right) + \frac{NJ^2}{2p} \int_{\mathcal{C}} dt dt' \left( \frac{1}{N} \sum_i \sigma_i(t) \sigma_i(t') \right)^p.
\end{aligned}
\end{equation}
Next introduce a ``fat unity'',
\begin{equation} \label{eq:partition_function_fat_unity}
\begin{aligned}
1 =& \int \mathcal{D}\mathcal{G} \prod_{tt'} \delta \Big( N\mathcal{G}(t, t') - \sum_i \sigma_i(t) \sigma_i(t') \Big) \\
=& \int \mathcal{D}\mathcal{G} \mathcal{D}\mathcal{F} \exp\left[ \frac{N}{2} \int_{\mathcal{C}} dt dt' \mathcal{F}(t, t') \left( \mathcal{G}(t, t') - \frac{1}{N} \sum_i \sigma_i(t) \sigma_i(t') \right) \right].
\end{aligned}
\end{equation}
The integral over the self-energy $\mathcal{F}(t, t')$ runs along the imaginary axis, making the second line simply the identity $\int dp e^{ipx} = 2\pi \delta(x)$ (we absorb factors of $2\pi$ into the measure $\mathcal{D}\mathcal{F}$).
However, when we ultimately evaluate the path integral by saddle point, we shall find that the saddle point value of $\mathcal{F}(t, t')$ is real.
Inserting Eq.~\eqref{eq:partition_function_fat_unity} into the path integral gives
\begin{equation} \label{eq:disorder_averaged_expanded_partition_function}
\mathbb{E} Z_{\textrm{SK}} = \int \mathcal{D}\mathcal{G} \mathcal{D}\mathcal{F} \int \mathcal{D} \sigma^N e^{-NS''},
\end{equation}
where
\begin{equation} \label{eq:annealed_expanded_effective_action}
\begin{aligned}
NS'' &\equiv -\frac{iN}{2} \int_{\mathcal{C}} dt z(t) + \frac{N}{2} \int_{\mathcal{C}} dt dt' \left( \frac{J^2}{p} \mathcal{G}(t, t')^p - \mathcal{F}(t, t') \mathcal{G}(t, t') \right) \\
&\qquad \qquad + \frac{1}{2} \sum_i \left[ \int_{\mathcal{C}} dt \left( -i \mu \big( \partial_t \sigma_i(t) \big)^2 + iz(t) \sigma_i(t)^2 \right) + \int_{\mathcal{C}} dt dt' \sigma_i(t) \mathcal{F}(t, t') \sigma_i(t') \right].
\end{aligned}
\end{equation}
We can now perform the integral over $\sigma_i$, resulting in
\begin{equation} \label{eq:disorder_averaged_final_partition_function}
\mathbb{E} Z_{\textrm{SK}} = \int \mathcal{D}\mathcal{G} \mathcal{D}\mathcal{F} e^{-NS_{\textrm{eff}}},
\end{equation}
where
\begin{equation} \label{eq:annealed_final_action}
S_{\textrm{eff}} \equiv -\frac{i}{2} \int_{\mathcal{C}} dt z(t) + \frac{1}{2} \int_{\mathcal{C}} dt dt' \left( \frac{J^2}{p} \mathcal{G}(t, t')^p - \mathcal{F}(t, t') \mathcal{G}(t, t') \right) + \frac{1}{2} \log{\textrm{Det}} \Big[ i(\mu \partial_t^2 + z) + \mathcal{F} \Big].
\end{equation}

At large $N$, the remaining path integral can be evaluated within the saddle point approximation.
The locations of the saddle points are determined by setting to zero the functional derivatives of Eq.~\eqref{eq:annealed_final_action}:
\begin{equation} \label{eq:Keldysh_EOM}
\begin{gathered}
i \big( \mu \partial_t^2 + z(t) \big) \mathcal{G}(t, t') + \int_{\mathcal{C}} dt'' \mathcal{F}(t, t'') \mathcal{G}(t'', t') = \delta(t - t'), \\
\mathcal{F}(t, t') = J^2 \mathcal{G}(t, t')^{p-1}, \qquad \mathcal{G}(t, t) = 1.
\end{gathered}
\end{equation}
Keep in mind that the time arguments in Eq.~\eqref{eq:Keldysh_EOM} are complex and range over the entire Schwinger-Keldysh contour.
In particular, although it is hidden in this compact notation, the infinitesimals $dt$ acquire different phases depending on the branch of the contour: $dt$ is a positive real infinitesimal on the upper (``forward'') real-time branch, a negative real infinitesimal on the lower (``backward'') real-time branch, and a negative imaginary infinitesimal on the thermal branch.

$\mathcal{G}(t, t')$ is the order parameter of this theory.
As is clear from the manner by which it was introduced (top line of Eq.~\eqref{eq:partition_function_fat_unity}), expectation values of $\mathcal{G}(t, t')$ within the path integral are equivalent to expectation values of $N^{-1} \sum_i \sigma_i(t) \sigma_i(t')$.
The latter are simply time-ordered correlation functions.
We shall focus on the real-time correlation functions, for which it is more transparent to explicitly indicate the branches by $\alpha \in \{u, l\}$ and have $t$ be simply a real variable.
Formally, we have that
\begin{equation} \label{eq:Keldysh_formal_expectation_values}
\begin{aligned}
\big< \mathcal{G}_{uu}(t, t') \big> &= \mathbb{E} \Big[ Z_{\textrm{SK}}^{-1} \textrm{Tr} e^{-\beta H} \mathcal{T} \sigma_i(t) \sigma_i(t') \Big], & \qquad \big< \mathcal{G}_{ul}(t, t') \big> &= \mathbb{E} \Big[ Z_{\textrm{SK}}^{-1} \textrm{Tr} e^{-\beta H} \sigma_i(t') \sigma_i(t) \Big], \\
\big< \mathcal{G}_{lu}(t, t') \big> &= \mathbb{E} \Big[ Z_{\textrm{SK}}^{-1} \textrm{Tr} e^{-\beta H} \sigma_i(t) \sigma_i(t') \Big], & \qquad \big< \mathcal{G}_{ll}(t, t') \big> &= \mathbb{E} \Big[ Z_{\textrm{SK}}^{-1} \textrm{Tr} e^{-\beta H} \widetilde{\mathcal{T}} \sigma_i(t) \sigma_i(t') \Big],
\end{aligned}
\end{equation}
where $\mathcal{T}$ denotes time ordering and $\widetilde{\mathcal{T}}$ denotes time anti-ordering.
Note that we can omit the sum over $i$ because the different spins (upon disorder-averaging) have equivalent behavior.

A number of formal properties of $\mathcal{G}_{\alpha \alpha'}(t, t')$ are evident from Eq.~\eqref{eq:Keldysh_formal_expectation_values}.
For one thing, $\mathcal{G}_{\alpha \alpha'}(t, t')$ clearly depends only on the time difference $t - t'$, and we shall often write $\mathcal{G}_{\alpha \alpha'}(t)$ with $t'=0$.
Since the four components differ only in time ordering, we see that for \textit{any} function $f(x)$,
\begin{equation} \label{eq:Keldysh_component_symmetry_identity}
f \Big( \mathcal{G}_{uu}(t) \Big) + f \Big( \mathcal{G}_{ll}(t) \Big) = f \Big( \mathcal{G}_{ul}(t) \Big) + f \Big( \mathcal{G}_{lu}(t) \Big).
\end{equation}
We can further express all four components in terms of a single complex-valued function (equivalently two real-valued functions).
For example, write $\mathcal{G}_{lu}(t)$ in terms of its real and imaginary parts as $\mathcal{G}^R(t) + i \mathcal{G}^I(t)$.
Since $\mathcal{G}_{lu}(t)^* = \mathcal{G}_{lu}(-t)$, $\mathcal{G}^R(t)$ is even and $\mathcal{G}^I(t)$ is odd.
One can easily confirm that
\begin{equation} \label{eq:Keldysh_correlation_relationships}
\begin{aligned}
\mathcal{G}_{uu}(t) &= \mathcal{G}^R(t) + i \textrm{sgn}[t] \mathcal{G}^I(t), & \qquad \mathcal{G}_{ul}(t) &= \mathcal{G}^R(t) - i \mathcal{G}^I(t), \\
\mathcal{G}_{lu}(t) &= \mathcal{G}^R(t) + i \mathcal{G}^I(t), & \qquad \mathcal{G}_{ll}(t) &= \mathcal{G}^R(t) - i \textrm{sgn}[t] \mathcal{G}^I(t).
\end{aligned}
\end{equation}

One of the most important features of $\mathcal{G}_{\alpha \alpha'}(t, t')$ is the limiting behavior at large $|t - t'|$, as a function of the inverse temperature $\beta$.
Numerical solution of Eq.~\eqref{eq:Keldysh_EOM} demonstrates that there is a critical value $\beta_d$ (which is less than $\beta_s$):
\begin{itemize}
\item For $\beta < \beta_d$, $\lim_{|t-t'| \rightarrow \infty} \mathcal{G}_{\alpha \alpha'}(t, t') = 0$.
We call this the ``ergodic'' phase ($\mathbb{E} \langle \sigma_i(t) \rangle = 0$ by symmetry regardless of temperature, and so in this phase $\mathbb{E} \langle \sigma_i(t) \sigma_i(t') \rangle \rightarrow \mathbb{E} \langle \sigma_i(t) \rangle \mathbb{E} \langle \sigma_i(t') \rangle$).
\item For $\beta_d < \beta < \beta_s$, $\lim_{|t-t'| \rightarrow \infty} \mathcal{G}_{\alpha \alpha'}(t, t') = q_{\textrm{EA}} > 0$.
We call this the ``non-ergodic'' phase.
The quantity $q_{\textrm{EA}}$ is referred to as the ``Edwards-Anderson'' order parameter.
\item For $\beta_s < \beta$, our initial annealed approximation is no longer valid.
The replica trick is required to obtain accurate results~\cite{Mezard1987,Fischer1991}, but (at least for finite-time dynamical properties) the behavior is qualitatively similar to that of the non-ergodic phase.
\end{itemize}

\subsection{TAP equations on the Schwinger-Keldysh contour}
\label{subsec:QTAP}

The dynamical calculation described above only hints at the complexity of the non-ergodic phase.
A more complete picture emerges from a generalization in the spirit of the TAP equations.
Our treatment follows that of Ref.~\cite{Biroli_2001}, which derived TAP equations on the thermal circle for the quantum PSM.
While the extension to real-time dynamics is straightforward, we are not aware of any explicit calculation in the literature.
Thus we present a detailed derivation of the following equations in App.~\ref{sec:TAP_derivation}.

As discussed in Sec.~\ref{subsec:review_mean_field_spin_glasses}, the TAP free energy (or Gibbs potential) is the Legendre transform of the free energy with respect to local fields.
It is therefore a function of the magnetization $m_i$ of each spin.
For the free energy of quantum systems, the magnetization should also have an imaginary time index $m_i(\tau)$.
The imaginary-time correlation function $\mathcal{G}(\tau, \tau')$ becomes an additional order parameter.

We define the TAP \textit{action} on the Schwinger-Keldysh contour analogously.
It is a function of the magnetizations $m_i(t)$ and the correlation function $\mathcal{G}(t, t')$, with $t$ again being complex-valued and ranging over the entire contour.
Specifically,
\begin{equation} \label{eq:Keldysh_TAP_action}
\begin{aligned}
iNS_{\textrm{TAP}}[m, \mathcal{G}] &\equiv \log{\int \mathcal{D}\sigma^N \exp \left[ i \sum_i S_i^0 - i \int_{\mathcal{C}} dt \sum_{(i_1 \cdots i_p)} J_{i_1 \cdots i_p} \sigma_{i_1}(t) \cdots \sigma_{i_p}(t) \right]} \\
&\qquad \qquad + \frac{iN}{2} \int_{\mathcal{C}} dt z(t) - i \int_{\mathcal{C}} dt \sum_i h_i(t) m_i(t) + \frac{iN}{2} \int_{\mathcal{C}} dt dt' \Lambda(t, t') \mathcal{G}(t, t'),
\end{aligned}
\end{equation}
where $\mathcal{C}$ denotes the Schwinger-Keldysh contour and
\begin{equation} \label{eq:Keldysh_TAP_noninteracting_action}
S_i^0 \equiv \int_{\mathcal{C}} dt \left( \frac{\mu}{2} \big( \partial_t \sigma_i(t) \big)^2 - \frac{z(t)}{2} \sigma_i(t)^2 + h_i(t) \sigma_i(t) \right) - \frac{1}{2} \int_{\mathcal{C}} dt dt' \Lambda(t, t') \sigma_i(t) \sigma_i(t').
\end{equation}
The fields $h_i(t)$ and $\Lambda(t, t')$ are \textit{not} independent parameters.
They are instead chosen so that $\langle \sigma_i(t) \rangle = m_i(t)$ and $N^{-1} \sum_i \langle \sigma_i(t) \sigma_i(t') \rangle = \mathcal{G}(t, t')$, just as $z(t)$ is again chosen to enforce $N^{-1} \sum_i \langle \sigma_i(t)^2 \rangle = 1$, where the expectation value is with respect to the action in Eq.~\eqref{eq:Keldysh_TAP_action}.

Due to the Legendre-transform structure of $S_{\textrm{TAP}}$, we have that
\begin{equation} \label{eq:quantum_TAP_Legendre_relations}
N \frac{\partial S_{\textrm{TAP}}}{\partial m_i(t)} = -h_i(t), \qquad \frac{\partial S_{\textrm{TAP}}}{\partial \mathcal{G}(t, t')} = \frac{1}{2} \Lambda(t, t').
\end{equation}
The TAP equations are those for $m_i(t)$ and $\mathcal{G}(t, t')$ which one gets by setting the right-hand sides of Eq.~\eqref{eq:quantum_TAP_Legendre_relations} to zero.
The solutions are therefore the values of magnetization and correlation function which the system can consistently possess ``on its own,'' without any external fields.
In this sense, each solution corresponds to a distinct metastable state.
There is no reason why there cannot be many self-consistent solutions, and indeed, spin glass models such as the PSM do have many at sufficiently low temperature.

We calculate the TAP equations in App.~\ref{sec:TAP_derivation}.
They are simplified by the fact that we can take $m_i(t) = m$ and $z(t) = z$.
We also define $q_{\textrm{EA}} \equiv N^{-1} \sum_i m_i^2$.
The equations come out to be (together with $\mathcal{G}(t, t) = 1$)
\begin{equation} \label{eq:Keldysh_TAP_EOM}
i \big( \mu \partial_t^2 + z \big) \Big( \mathcal{G}(t, t') - q_{\textrm{EA}} \Big) + J^2 \int_{\mathcal{C}} dt'' \Big( \mathcal{G}(t, t'')^{p-1} - q_{\textrm{EA}}^{p-1} \Big) \Big( \mathcal{G}(t'', t') - q_{\textrm{EA}} \Big) = \delta(t - t'),
\end{equation}
\begin{equation} \label{eq:Keldysh_TAP_magnetization_equation}
J^2 \int_{\mathcal{C}} dt' \Big( \mathcal{G}(t, t')^{p-1} - (p-1) q_{\textrm{EA}}^{p-2} \mathcal{G}(t, t') + (p-2) q_{\textrm{EA}}^{p-1} \Big) m_i = -izm_i - i \sum_{(i_1 \cdots i_p)} J_{i_1 \cdots i_p} \frac{\partial (m_{i_1} \cdots m_{i_p})}{\partial m_i}.
\end{equation}
Note that Eq.~\eqref{eq:Keldysh_TAP_magnetization_equation} is $N$ equations, one for each spin $i$, and that it holds equally for any value of $t$ due to time translation invariance.
Defining $\mathcal{F}(t, t') \equiv J^2 \mathcal{G}(t, t')^{p-1}$, Eq.~\eqref{eq:Keldysh_TAP_EOM} is quite similar to Eq.~\eqref{eq:Keldysh_EOM}.
The only difference is that Eq.~\eqref{eq:Keldysh_TAP_EOM} uses $\Delta \mathcal{G}(t, t') \equiv \mathcal{G}(t, t') - q_{\textrm{EA}}$ and $\Delta \mathcal{F}(t, t') \equiv \mathcal{F}(t, t') - J^2 q_{\textrm{EA}}^{p-1}$, which decay to zero at large $|t - t'|$, rather than $\mathcal{G}(t, t')$ and $\mathcal{F}(t, t')$ themselves.

Despite the more involved derivation, $\mathcal{G}(t, t')$ remains a contour-ordered expectation value.
Thus, returning to the notation in which $\alpha \in \{u, l\}$ labels branches and $t$ is real, $\mathcal{G}_{\alpha \alpha'}(t - t')$ possesses the same formal properties as discussed in the previous subsection (Eqs.~\eqref{eq:Keldysh_component_symmetry_identity} and~\eqref{eq:Keldysh_correlation_relationships}).
Of particular importance will be the Fourier transform of $\Delta \mathcal{G}_{\alpha \alpha'}(t)$ at zero frequency, denoted $\Delta \widetilde{\mathcal{G}}_{\alpha \alpha'}(0)$, as well as its (matrix) inverse, $\Delta \widetilde{\mathcal{G}}_{\alpha \alpha'}^{-1}(0)$.
Also define $L \equiv \int_{-\infty}^{\infty} dt \Delta \mathcal{G}^R(t)$ and $\Lambda \equiv \int_0^{\infty} dt \Delta \mathcal{G}^I(t)$.
Then from Eq.~\eqref{eq:Keldysh_correlation_relationships}, we see that
\begin{equation} \label{eq:Keldysh_zero_frequency_matrix}
\begin{pmatrix} \Delta \widetilde{\mathcal{G}}_{uu}(0) & \Delta \widetilde{\mathcal{G}}_{ul}(0) \\ \Delta \widetilde{\mathcal{G}}_{lu}(0) & \Delta \widetilde{\mathcal{G}}_{ll}(0) \end{pmatrix} = \begin{pmatrix} L + 2i \Lambda & L \\ L & L - 2i \Lambda \end{pmatrix},
\end{equation}
\begin{equation} \label{eq:Keldysh_zero_frequency_matrix_inverse}
\begin{pmatrix} \Delta \widetilde{\mathcal{G}}_{uu}(0) & \Delta \widetilde{\mathcal{G}}_{ul}(0) \\ \Delta \widetilde{\mathcal{G}}_{lu}(0) & \Delta \widetilde{\mathcal{G}}_{ll}(0) \end{pmatrix}^{-1} = \frac{1}{4 \Lambda^2} \begin{pmatrix} L - 2i \Lambda & -L \\ -L & L + 2i \Lambda \end{pmatrix}.
\end{equation}

The multiplicity of solutions to the TAP equations comes from Eq.~\eqref{eq:Keldysh_TAP_magnetization_equation}.
By use of Eqs.~\eqref{eq:Keldysh_TAP_EOM},~\eqref{eq:Keldysh_zero_frequency_matrix}, and~\eqref{eq:Keldysh_zero_frequency_matrix_inverse}, it can be written (associating $u$ with 0 and $l$ with 1)
\begin{equation} \label{eq:Keldysh_TAP_magnetization_alternate}
\begin{aligned}
\left[ (-1)^{\alpha} \sum_{\alpha'} \Delta \widetilde{\mathcal{G}}_{\alpha \alpha'}^{-1}(0) - (p-1)J^2 q_{\textrm{EA}}^{p-2} \sum_{\alpha'} (-1)^{\alpha'} \Delta \widetilde{\mathcal{G}}_{\alpha \alpha'}(0) \right] &m_i \\
= \left[ \frac{1}{2i \Lambda} - (p-1)J^2 q_{\textrm{EA}}^{p-2} 2i \Lambda \right] &m_i = -i \sum_{(i_1 \cdots i_p)} J_{i_1 \cdots i_p} \frac{\partial (m_{i_1} \cdots m_{i_p})}{\partial m_i}.
\end{aligned}
\end{equation}
Eq.~\eqref{eq:Keldysh_TAP_magnetization_alternate} is identical to that which appears and has been well-studied for the \textit{classical} PSM \cite{Crisanti1992,Crisanti1993Spherical,Crisanti1995ThoulessAndersonPalmerAT,Castellani2005Spin}.
Thus we simply quote the following results.
In addition to the inverse temperature $\beta$, solutions to Eq.~\eqref{eq:Keldysh_TAP_magnetization_alternate} are parametrized by the quantity
\begin{equation} \label{eq:normalized_energy_definition}
\mathcal{E} \equiv \frac{1}{NJq_{\textrm{EA}}^{p/2}} \sum_{(i_1 \cdots i_p)} J_{i_1 \cdots i_p} m_{i_1} \cdots m_{i_p},
\end{equation}
which can be interpreted as a ``normalized'' potential energy density: each magnetization has a value which is (very roughly) comparable to $q_{\textrm{EA}}^{1/2}$, and thus the natural scale for the interaction energy is $Jq_{\textrm{EA}}^{p/2}$.
The value of $q_{\textrm{EA}}$ for a given $\mathcal{E} < 0$ is given by the largest solution to
\begin{equation} \label{eq:Keldysh_TAP_overlap_closed_equation}
-2Jq_{\textrm{EA}}^{p/2 - 1} \Lambda = \frac{p}{2(p-1)} \left( -\mathcal{E} - \sqrt{\mathcal{E}^2 - \mathcal{E}_{\textrm{th}}^2} \right), \qquad \mathcal{E}_{\textrm{th}} \equiv -\frac{2\sqrt{p-1}}{p},
\end{equation}
where $\Lambda$ depends on $q_{\textrm{EA}}$ through Eq.~\eqref{eq:Keldysh_TAP_EOM}.
One can show that solutions to Eq.~\eqref{eq:Keldysh_TAP_overlap_closed_equation} exist only for $\beta > \beta_d$, with $\beta_d$ the same as defined in Sec.~\ref{subsec:Schwinger_Keldysh_path_integral}.
Furthermore, Eq.~\eqref{eq:Keldysh_TAP_overlap_closed_equation} only makes sense if $\mathcal{E} \leq \mathcal{E}_{\textrm{th}}$.
In that case, the number of solutions $\mathcal{N}(\beta, \mathcal{E})$ to Eq.~\eqref{eq:Keldysh_TAP_magnetization_alternate} --- in addition to the trivial solution $m_i = 0$ --- is exponential in system size: $N^{-1} \log{\mathcal{N}(\beta, \mathcal{E})} \sim \Sigma(\mathcal{E})$, with\footnote{
As written, Eq.~\eqref{eq:TAP_complexity} is a bit sloppy.
$\mathcal{N}(\mathcal{E})$ is given by Eq.~\eqref{eq:TAP_complexity} when the latter is non-negative and $\beta$ is such that solutions to Eq.~\eqref{eq:Keldysh_TAP_overlap_closed_equation} exist.
In all other cases, $\mathcal{N}(\mathcal{E}) = 0$.
}
\begin{equation} \label{eq:TAP_complexity}
\Sigma(\mathcal{E}) = \frac{1}{2} \left( 1 + 2\log{\frac{p}{2}} \right) - \frac{p \mathcal{E}^2}{2} + \frac{p^2}{8(p-1)} \Big( \mathcal{E} + \sqrt{\mathcal{E}^2 - \mathcal{E}_{\textrm{th}}^2} \Big)^2 + \log{\Big( -\mathcal{E} + \sqrt{\mathcal{E}^2 - \mathcal{E}_{\textrm{th}}^2} \Big)}.
\end{equation}
The exponent $\Sigma(\mathcal{E})$ is referred to as the ``complexity'' in the spin glass literature.

The connection between this TAP approach and the conventional Schwinger-Keldysh path integral lies in the fact that: i) the inverse temperature $\beta_d$ at which TAP states with non-zero magnetization appear is identical to that at which the autocorrelation function acquires a non-zero late-time limit; ii) the overlap determined by Eq.~\eqref{eq:Keldysh_TAP_overlap_closed_equation} is identical to the late-time value of the autocorrelation function.
This strongly suggests the following picture:
\begin{itemize}
\item For $\beta < \beta_d$ (the ``ergodic'' phase), there exists a single equilibrium state with zero magnetization, and the correlation function decays to zero on a finite timescale.
\item For $\beta_d < \beta$ (the ``non-ergodic'' phase), there exist exponentially many metastable states having non-zero magnetization.
The number of states is given by the exponential of the complexity $\Sigma(\mathcal{E})$.
Dynamically, in the $N \rightarrow \infty$ limit, a system prepared in one metastable state will remain in that state for all time.
At finite $N$, it is only on a timescale exponential in $N$ that the system can transition between states.
\end{itemize}
Much more can be said about these phases (in particular how the replica-symmetry-breaking transition at $\beta_s$ appears within the TAP approach), and a large body of literature is devoted to this topic.
We refer in particular to Ref.~\cite{Castellani2005Spin} as an excellent starting point.

In the present work, we determine the spectral statistics of the PSM in both the ergodic and non-ergodic phase.
Those of the former can be computed very much along the lines of Ref.~\cite{saad2019semiclassical}, which we do in Sec.~\ref{sec:ergodic_ramp}.
Those of the latter, however, require novel calculations which we present in Sec.~\ref{sec:nonergodic_ramp}.
Unsurprisingly, the properties of TAP states shall play an essential role.

\section{The semiclassical ramp in the ergodic phase}
\label{sec:ergodic_ramp}

To reiterate, we are evaluating
\begin{equation} \label{eq:generic_SFF_definition}
\textrm{SFF}(T, f) \equiv \mathbb{E} \big| \textrm{Tr} f(H) e^{-iHT} \big|^2 = \mathbb{E} \Big[ \textrm{Tr} f(H) e^{-iHT} \textrm{Tr} f(H) e^{iHT} \Big],
\end{equation}
where $H$ is the PSM Hamiltonian (Eq.~\eqref{eq:quantum_Hamiltonian}) and $f$ is a filter function as discussed in Sec.~\ref{subsec:review_spectral_form_factor}.
Here we consider the ergodic phase, for which the results are analogous to those of SYK~\cite{saad2019semiclassical}. We then consider the non-ergodic phase in Sec.~\ref{sec:nonergodic_ramp}.

\subsection{Effective action} \label{subsec:ergodic_effective_action}

The calculation begins by retracing the steps described in Sec.~\ref{subsec:Schwinger_Keldysh_path_integral}, only on a modified contour.
We still have upper and lower branches indicated by $\alpha \in \{u, l\}$ (with $u = 0$ and $l = 1$), but now each is separately periodic.
Furthermore, we no longer have a thermal branch.
See the right column of Fig.~\ref{fig:contour_summary}, as compared to the left column.
While some care is required to account for the filter functions (as discussed in Appendix~\ref{sec:filter_functions}), we ultimately arrive at an expression analogous to Eq.~\eqref{eq:annealed_final_action}:
\begin{equation} \label{eq:SFF_formal_path_integral}
\textrm{SFF}(T, f) = \int \mathcal{D}G \mathcal{D}F \, f \big( \epsilon_u[G] \big) f \big( \epsilon_l[G] \big) e^{-NS_{\textrm{eff}}[G, F]},
\end{equation}
\begin{equation} \label{eq:SFF_formal_action}
\begin{aligned}
S_{\textrm{eff}}[G, F] &= -\frac{i}{2} \int_0^T dt \sum_{\alpha} (-1)^{\alpha} z_{\alpha}(t) + \frac{1}{2} \int_0^T dt dt' \sum_{\alpha \alpha'} (-1)^{\alpha + \alpha'} \left( \frac{J^2}{p} G_{\alpha \alpha'}(t, t')^p - F_{\alpha \alpha'}(t, t') G_{\alpha \alpha'}(t, t') \right) \\
&\qquad \qquad + \frac{1}{2} \log{\textrm{Det}} \Big[ i (-1)^{\alpha} \delta_{\alpha \alpha'} \big( \mu \partial_t^2 + z_{\alpha} \big) + (-1)^{\alpha + \alpha'} F_{\alpha \alpha'} \Big],
\end{aligned}
\end{equation}
where the ``energy density'' $\epsilon_{\alpha}[G]$ is defined as ($0^+$ denotes a positive infinitesimal)
\begin{equation} \label{eq:energy_density_definition}
\epsilon_{\alpha}[G] \equiv -\frac{\mu}{2} \partial_t^2 G_{\alpha \alpha}(0^+, 0) - \frac{iJ^2}{p} \int_0^T dt \sum_{\alpha'} (-1)^{\alpha'} G_{\alpha \alpha'}(t, 0)^p.
\end{equation}
See App.~\ref{sec:filter_functions} for details.
The saddle point of $S_{\textrm{eff}}$ is given by the equations (compare to Eq.~\eqref{eq:Keldysh_EOM})
\begin{equation} \label{eq:SFF_general_saddle_point_equations}
\begin{gathered}
i \big( \mu \partial_t^2 + z_{\alpha}(t) \big) G_{\alpha \alpha'}(t, t') + \int_0^T dt'' \sum_{\alpha''} (-1)^{\alpha''} F_{\alpha \alpha''}(t, t'') G_{\alpha'' \alpha'}(t'', t') = (-1)^{\alpha} \delta_{\alpha \alpha'} \delta(t - t'), \\
F_{\alpha \alpha'}(t, t') = J^2 G_{\alpha \alpha'}(t, t')^{p-1}, \qquad G_{\alpha \alpha}(t, t) = 1.
\end{gathered}
\end{equation}

Denoting averages with respect to the path integral of Eq.~\eqref{eq:SFF_formal_path_integral} by $\langle \, \cdot \, \rangle$, the expectation value of $G$ is related to the original degrees of freedom as follows (we omit the filter functions here for brevity):
\begin{equation} \label{eq:SFF_order_parameter_spin_relationship}
\begin{aligned}
\big< G_{uu}(t, t') \big> &= \mathbb{E} \Big[ \textrm{Tr} e^{-iHT} \mathcal{T} \sigma_i(t) \sigma_i(t') \textrm{Tr} e^{iHT} \Big], &\qquad \big< G_{ul}(t, t') \big> &= \mathbb{E} \Big[ \textrm{Tr} e^{-iHT} \sigma_i(t) \textrm{Tr} e^{iHT} \sigma_i(t') \Big], \\
\big< G_{lu}(t, t') \big> &= \mathbb{E} \Big[ \textrm{Tr} e^{-iHT} \sigma_i(t') \textrm{Tr} e^{iHT} \sigma_i(t) \Big], &\qquad \big< G_{ll}(t, t') \big> &= \mathbb{E} \Big[ \textrm{Tr} e^{-iHT} \textrm{Tr} e^{iHT} \widetilde{\mathcal{T}} \sigma_i(t) \sigma_i(t') \Big],
\end{aligned}
\end{equation}
where $\mathcal{T}$ denotes time ordering and $\widetilde{\mathcal{T}}$ denotes time anti-ordering.
One immediately sees from Eq.~\eqref{eq:SFF_order_parameter_spin_relationship} that:
\begin{enumerate}[label=\roman*)]
\item all components of $\langle G_{\alpha \alpha'}(t, t') \rangle$ are time-translation invariant and have period $T$;
\item $\langle G_{uu}(t, t') \rangle$ and $\langle G_{ll}(t, t') \rangle$ are even functions of $t - t'$;
\item $\langle G_{ul}(t, t') \rangle$ and $\langle G_{lu}(t, t') \rangle$ are in fact independent of both time arguments;
\item $\langle G_{uu}(t, t') \rangle^* = \langle G_{ll}(t, t') \rangle$;
\item $\langle G_{ul}(t, t') \rangle^* = \langle G_{lu}(t, t') \rangle$.
\end{enumerate}
Solutions to Eq.~\eqref{eq:SFF_general_saddle_point_equations} do \textit{not} necessarily share all these properties, since some of the symmetries may be spontaneously broken.

However, one simple solution that obeys all of the above is to take $G_{ul}(t, t') = G_{lu}(t, t') = 0$.
The resulting action is precisely what one would get from averaging each factor of $\textrm{Tr} e^{-iHT}$ separately, i.e., this solution gives the disconnected contribution to the SFF:
\begin{equation} \label{eq:SFF_disconnected_contribution}
\mathbb{E} \Big[ \textrm{Tr} f(H) e^{-iHT} \textrm{Tr} f(H) e^{iHT} \Big] = \mathbb{E} \Big[ \textrm{Tr} f(H) e^{-iHT} \Big] \mathbb{E} \Big[ \textrm{Tr} f(H) e^{iHT} \Big] + \cdots,
\end{equation}
where $\cdots$ denotes the contribution to the path integral from non-zero $G_{ul}$ and/or $G_{lu}$.
Eq.~\eqref{eq:SFF_disconnected_contribution} holds equally well in the non-ergodic phase, and thus the remainder of this paper will be concerned with determining those additional contributions.

\subsection{Connected solutions} \label{subsec:ergodic_connected_solutions}

Following Ref.~\cite{saad2019semiclassical}, we construct approximate solutions to Eq.~\eqref{eq:SFF_general_saddle_point_equations} which become accurate at large $T$.
We first present the solutions and justify them afterwards.
Take $\mathcal{G}_{\alpha \alpha'}(t, t')$ to be the Schwinger-Keldysh correlation function at inverse temperature $\beta_{\textrm{aux}}$, exactly as given in Sec.~\ref{subsec:Schwinger_Keldysh_path_integral} (Eq.~\eqref{eq:Keldysh_formal_expectation_values} in particular).
Again define $\mathcal{F}_{\alpha \alpha'}(t, t') \equiv J^2 \mathcal{G}_{\alpha \alpha'}(t, t')^{p-1}$.
A solution to the SFF saddle point equations (up to terms which vanish at large $T$) is
\begin{equation} \label{eq:SFF_connected_solution_G}
G_{\alpha \alpha'}(t, t') = \sum_{n = -\infty}^{\infty} \mathcal{G}_{\alpha \alpha'}(t - t' + \delta_{\alpha \neq \alpha'} \Delta + nT),
\end{equation}
\begin{equation} \label{eq:SFF_connected_solution_Sigma}
F_{\alpha \alpha'}(t, t') = \sum_{n = -\infty}^{\infty} \mathcal{F}_{\alpha \alpha'}(t - t' + \delta_{\alpha \neq \alpha'} \Delta + nT).
\end{equation}
Here $\Delta$ can be any real number between 0 and $T$.
Thus Eqs.~\eqref{eq:SFF_connected_solution_G} and~\eqref{eq:SFF_connected_solution_Sigma} constitute a two-parameter family of solutions, the parameters being $\beta_{\textrm{aux}}$ and $\Delta$.
Every such solution contributes to the SFF.

As for the Lagrange multipliers $z_{\alpha}(t)$, they are independent of $t$ due to time translation invariance.
We further have that $z_u = z_l \equiv z$: both equal the value of the chemical potential needed to satisfy the \textit{equilibrium} spherical constraint, i.e., $N^{-1} \sum_i \textrm{Tr} Z_{\textrm{SK}}^{-1} e^{-\beta_{\textrm{aux}} H} \sigma_i^2 = 1$ (time translation invariance then implies that $\mathcal{G}_{\alpha \alpha}(t, t) = 1$ for all times and both branches).

To justify Eqs.~\eqref{eq:SFF_connected_solution_G} and~\eqref{eq:SFF_connected_solution_Sigma}, it is essential that $\mathcal{G}_{\alpha \alpha'}(t - t')$ decay exponentially to zero as $|t - t'| \rightarrow \infty$.
Thus these solutions only apply in the ergodic phase. 
With this in mind, the following comments together establish their validity:
\begin{itemize}
\item The sum over $n$ ensures that $G_{\alpha \alpha'}(t - t')$ has period $T$, even though $\mathcal{G}_{\alpha \alpha'}(t - t')$ does not.
\item Since $\mathcal{G}_{\alpha \alpha}(t - t')$ decays exponentially, $G_{\alpha \alpha}(0) \sim 1$ up to terms which are exponentially small in $T$.
\item The equation $F_{\alpha \alpha'}(t, t') = J^2 G_{\alpha \alpha'}(t, t')^{p-1}$ is satisfied up to exponentially small terms because, when raising Eq.~\eqref{eq:SFF_connected_solution_G} to the $p-1$'th power, all cross terms are exponentially small (as is the sum over them).
In other words,
\begin{equation} \label{eq:cross_terms_neglecting}
\left( \sum_n \mathcal{G}_{\alpha \alpha'}(t - t' + nT + \delta_{\alpha \neq \alpha'} \Delta) \right)^{p-1} \sim \sum_n \mathcal{G}_{\alpha \alpha'}(t - t' + nT + \delta_{\alpha \neq \alpha'} \Delta)^{p-1}.
\end{equation}
\item $\mathcal{G}_{\alpha \alpha'}(t, t')$ obeys Eq.~\eqref{eq:Keldysh_EOM}, written explicitly in terms of components as
\begin{equation} \label{eq:Keldysh_EOM_explicit}
\begin{aligned}
&i \big( \mu \partial_t^2 + z \big) \mathcal{G}_{\alpha \alpha'}(t - t') + \int_0^{\infty} dt'' \sum_{\alpha''} (-1)^{\alpha''} \mathcal{F}_{\alpha \alpha''}(t - t'') \mathcal{G}_{\alpha'' \alpha'}(t'' - t') \\
&\qquad \qquad \qquad \qquad \; \; \; - i \int_0^{\beta_{\textrm{aux}}} d\tau'' \mathcal{F}_{\alpha v}(t + i\tau'') \mathcal{G}_{v \alpha'}(-i\tau'' - t') = (-1)^{\alpha} \delta_{\alpha \alpha'} \delta(t - t'),
\end{aligned}
\end{equation}
where $v$ denotes the thermal branch of the contour.
For $t, t' \gg 1$ (which still allows $t - t'$ to take any value), $\mathcal{G}_{\alpha v}(t + i\tau)$ is exponentially small for all $\tau$ and the last term on the left-hand side can be neglected.
We can also take the lower limit of the $t''$ integral to $-\infty$.
Thus when checking whether Eq.~\eqref{eq:SFF_connected_solution_G} satisfies Eq.~\eqref{eq:SFF_general_saddle_point_equations}, we have that
\begin{equation} \label{eq:SFF_connected_solution_demonstration}
\begin{aligned}
&i \big( \mu \partial_t^2 + z \big) G_{\alpha \alpha'}(t - t') + \int_0^T dt'' \sum_{\alpha''} (-1)^{\alpha''} F_{\alpha \alpha''}(t - t'') G_{\alpha'' \alpha'}(t'' - t') \\
&\; \; \sim i \big( \mu \partial_t^2 + z \big) \mathcal{G}_{\alpha \alpha'}(t - t') + \int_{-\infty}^{\infty} dt'' \sum_{\alpha''} (-1)^{\alpha''} \mathcal{F}_{\alpha \alpha''}(t - t'') \mathcal{G}_{\alpha'' \alpha'}(t'' - t') \\
&\; \; \sim (-1)^{\alpha} \delta_{\alpha \alpha'} \delta(t - t'),
\end{aligned}
\end{equation}
again making use of the fact that $\mathcal{G}_{\alpha \alpha'}(t - t')$ is exponentially small when $|t - t'|$ is large.
The equation is indeed satisfied.
\item Finally, the off-diagonal components $G_{ul}(t, t')$ and $G_{lu}(t, t')$ contain the parameter $\Delta$ because they break the \textit{separate} time translation symmetries in $t$ and $t'$ (see property iii above).
Thus if any choice of $\Delta$ solves Eq.~\eqref{eq:SFF_general_saddle_point_equations}, so do all choices of $\Delta \in [0, T)$.
\end{itemize}

As noted above, we have thus identified a two-parameter family of solutions to the SFF saddle point equations.
In what follows it will be more convenient to parametrize the solutions by the equilibrium energy density $\epsilon(\beta_{\textrm{aux}})$ corresponding to inverse temperature $\beta_{\textrm{aux}}$.
We can express $\epsilon(\beta)$ in terms of $\mathcal{G}$ (and thus $G$) by inserting a factor of $H$ into the Schwinger-Keldysh contour.
Since $H$ clearly commutes with the evolution operator $e^{-\beta H} e^{iHt} e^{-iHt}$, it can be inserted at any point, in particular at a late time for which (again because $\mathcal{G}_{\alpha \alpha'}(t - t')$ decays exponentially) the thermal branch can be neglected.
By following the same steps as in Appendix~\ref{sec:filter_functions}, we find that $\epsilon(\beta)$ is given precisely by Eq.~\eqref{eq:energy_density_definition}, evaluated on either branch:
\begin{equation} \label{eq:thermal_energy_density_expression}
\begin{aligned}
\epsilon =& -\frac{\mu}{2} \partial_t^2 \mathcal{G}_{uu}(0^+) - \frac{iJ^2}{p} \int_{-\infty}^{\infty} dt \Big( \mathcal{G}_{uu}(t)^p - \mathcal{G}_{ul}(t)^p \Big) \\
=& -\frac{\mu}{2} \partial_t^2 \mathcal{G}_{ll}(0^+) + \frac{iJ^2}{p} \int_{-\infty}^{\infty} dt \Big( \mathcal{G}_{ll}(t)^p - \mathcal{G}_{lu}(t)^p \Big).
\end{aligned}
\end{equation}

\subsection{Contribution of connected solutions} \label{subsec:ergodic_connected_action}

Having demonstrated that Eqs.~\eqref{eq:SFF_connected_solution_G} and~\eqref{eq:SFF_connected_solution_Sigma} solve the SFF saddle point equations, it remains to calculate the action (Eq.~\eqref{eq:SFF_formal_action}) evaluated at the solutions.
First note that, since each solution obeys Eq.~\eqref{eq:SFF_general_saddle_point_equations}, we can rewrite the action as
\begin{equation} \label{eq:SFF_on_shell_action}
S_{\textrm{eff}} = -\frac{J^2 (p-1)T}{2p} \int_0^T dt \sum_{\alpha \alpha'} (-1)^{\alpha + \alpha'} G_{\alpha \alpha'}(t)^p - \frac{1}{2} \sum_{\omega} \log{\textrm{Det}} \widetilde{G}_{\alpha \alpha'}(\omega),
\end{equation}
where $\omega \in 2\pi \mathbb{Z}/T$ and $\widetilde{G}_{\alpha \alpha'}(\omega) \equiv \int_0^T dt e^{i \omega t} G_{\alpha \alpha'}(t)$.
Note that the Lagrange multiplier terms have dropped out since $z_u = z_l$.
Furthermore, since $\int dt G_{\alpha \alpha'}(t)^p \sim \int dt \mathcal{G}_{\alpha \alpha'}(t)^p$, the general relation in Eq.~\eqref{eq:Keldysh_component_symmetry_identity} implies that the first term of Eq.~\eqref{eq:SFF_on_shell_action} in fact vanishes.

For the second term, note that by Eq.~\eqref{eq:SFF_connected_solution_G},
\begin{equation} \label{eq:SFF_connected_solution_Fourier_transform}
\widetilde{G}_{\alpha \alpha'}(\omega) = e^{-i \delta_{\alpha \neq \alpha'} \omega \Delta} \widetilde{\mathcal{G}}_{\alpha \alpha'}(\omega), \qquad \widetilde{\mathcal{G}}_{\alpha \alpha'}(\omega) \equiv \int_{-\infty}^{\infty} dt e^{i \omega t} \mathcal{G}_{\alpha \alpha'}(t).
\end{equation}
The exponential decay of $\mathcal{G}_{\alpha \alpha'}(t)$ implies that $\widetilde{\mathcal{G}}_{\alpha \alpha'}(\omega)$ (and thus $\widetilde{G}_{\alpha \alpha'}(\omega)$) is an infinitely differentiable function of $\omega$.
Strictly speaking, since the path integral is regularized by a timestep $\Delta t \rightarrow 0$, $\widetilde{\mathcal{G}}_{\alpha \alpha'}(\omega)$ is furthermore periodic with period $2\pi/\Delta t$.
The same is true of $\log{\textrm{Det}} \widetilde{G}_{\alpha \alpha'}(\omega)$.
Thus the Euler-Maclaurin formula~\cite{Knuth1994} gives
\begin{equation} \label{eq:SFF_on_shell_action_second_term}
\sum_{n=-\pi/\Delta t}^{\pi/\Delta t} \log{\textrm{Det}} \widetilde{G}_{\alpha \alpha'} \left( \frac{2\pi n}{T} \right) \sim \frac{T}{2\pi} \int_{-\pi/\Delta t}^{\pi/\Delta t} d\omega \log{\textrm{Det}} \widetilde{G}_{\alpha \alpha'}(\omega) \rightarrow \frac{T}{2\pi} \int_{-\infty}^{\infty} d\omega \log{\textrm{Det}} \widetilde{G}_{\alpha \alpha'}(\omega),
\end{equation}
up to terms which vanish faster than any polynomial in $T^{-1}$.
Thus $S_{\textrm{eff}}$ is proportional to $T$, and we only need to evaluate the proportionality constant.

Rather than calculate the integral directly, we follow Ref.~\cite{saad2019semiclassical} and evaluate the derivative $dS_{\textrm{eff}}/dT$ starting from Eq.~\eqref{eq:SFF_formal_action}.
It is convenient to rescale time as $t \rightarrow Tt$, so that $T$ becomes simply another parameter:
\begin{equation} \label{eq:SFF_formal_action_rescaled}
\begin{aligned}
S_{\textrm{eff}} &= \frac{T^2}{2} \int_0^1 dt dt' \sum_{\alpha \alpha'} (-1)^{\alpha + \alpha'} \left( \frac{J^2}{p} G_{\alpha \alpha'}(t, t')^p - F_{\alpha \alpha'}(t, t') G_{\alpha \alpha'}(t, t') \right) \\
&\qquad \qquad + \frac{1}{2} \log{\textrm{Det}} \Big[ i (-1)^{\alpha} \delta_{\alpha \alpha'} \big( \mu T^{-2} \partial_t^2 + z_{\alpha} \big) + (-1)^{\alpha + \alpha'} F_{\alpha \alpha'} \Big].
\end{aligned}
\end{equation}
Note that, since $S_{\textrm{eff}}$ is evaluated at a solution of the saddle point equations, we only need to differentiate the explicit factors of $T$:
\begin{equation} \label{eq:SFF_formal_action_total_time_derivative_v1}
\begin{aligned}
\frac{dS_{\textrm{eff}}}{dT} =& \; T \int_0^1 dt dt' \sum_{\alpha \alpha'} (-1)^{\alpha + \alpha'} \left( \frac{J^2}{p} G_{\alpha \alpha'}(t, t')^p - F_{\alpha \alpha'}(t, t') G_{\alpha \alpha'}(t, t') \right) \\
&\qquad \qquad - \frac{i\mu}{T^3} \int_0^1 dt \sum_{\alpha} (-1)^{\alpha} \partial_t^2 \Big[ i (-1)^{\alpha} \delta_{\alpha \alpha'} \big( \mu T^{-2} \partial_t^2 + z_{\alpha}(t) \big) + (-1)^{\alpha + \alpha'} F_{\alpha \alpha'} \Big]^{-1} \bigg|_{\alpha = \alpha', t = t'^+}.
\end{aligned}
\end{equation}
Returning to unscaled time and using Eq.~\eqref{eq:SFF_general_saddle_point_equations}, we have that
\begin{equation} \label{eq:SFF_formal_action_total_time_derivative_v2}
\frac{dS_{\textrm{eff}}}{dT} = -\frac{(p-1)J^2}{p} \int_0^T dt \sum_{\alpha \alpha'} (-1)^{\alpha + \alpha'} G_{\alpha \alpha'}(t)^p - \frac{i\mu}{T} \sum_{\alpha} (-1)^{\alpha} \partial_t^2 G_{\alpha \alpha}(0^+) = 0,
\end{equation}
again using Eqs.~\eqref{eq:Keldysh_component_symmetry_identity} and~\eqref{eq:thermal_energy_density_expression}.
Thus the proportionality constant is in fact zero, i.e., $S_{\textrm{eff}} = 0$.

\subsection{Evaluation of the SFF} \label{subsec:ergodic_SFF_evaluation}

To finally compute the SFF, we simply need to sum over all connected solutions, i.e., integrate over $\epsilon_{\textrm{aux}}$ and $\Delta$.
However, there are additional discrete symmetries which give further solutions: i) we can time-reverse the off-diagonal components, i.e., take $G_{ul}(t) = \mathcal{G}_{ul}(-t)$ and $G_{lu}(t) = \mathcal{G}_{lu}(-t)$; ii) if $p$ is even, we can take $G_{ul}(t) = -\mathcal{G}_{ul}(t)$ and $G_{lu}(t) = -\mathcal{G}_{lu}(t)$.
These must be summed over as well, giving an additional factor of $2(1 + \delta_{p \textrm{ even}})$, where $\delta_{p \textrm{ even}}$ is the indicator function on $p$ being even (1 if true, 0 if false).
Thus our final expression is
\begin{equation} \label{eq:SFF_connected_contribution_ergodic_phase}
\begin{aligned}
\mathbb{E} \Big[ \textrm{Tr} f(H) e^{-iHT} \textrm{Tr} f(H) e^{iHT} \Big] &\sim \left| \mathbb{E} \textrm{Tr} f(H) e^{-iHT} \right|^2 + \int \frac{d\epsilon_{\textrm{aux}}}{2\pi} f(\epsilon_{\textrm{aux}})^2 \int_0^T d\Delta \, 2 \big( 1 + \delta_{p \textrm{ even}} \big) e^0 \\
&= \left| \mathbb{E} \textrm{Tr} f(H) e^{-iHT} \right|^2 + 2 \big( 1 + \delta_{p \textrm{ even}} \big) T \int \frac{d\epsilon_{\textrm{aux}}}{2\pi} f(\epsilon_{\textrm{aux}})^2.
\end{aligned}
\end{equation}
The measure $1/2\pi$ can be derived using hydrodynamic methods~\cite{winer2020hydrodynamic,saad2019semiclassical}, but its precise value is not essential for our purposes.
The key feature is simply that the linear-in-$T$ ramp has emerged.

However, keep in mind that Eq.~\eqref{eq:SFF_connected_contribution_ergodic_phase} is only valid if the filter function is such that all contributing values of $\epsilon_{\textrm{aux}}$ lie in the ergodic phase.
In the following section we modify this analysis to hold in the non-ergodic phase as well.
We shall see that it is necessary to incorporate the structure of multiple TAP states.

\section{The semiclassical ramp in the non-ergodic phase}
\label{sec:nonergodic_ramp}

As we have stressed repeatedly, the results of Sec.~\ref{sec:ergodic_ramp} rely heavily on having an equilibrium correlation function which decays to zero at late times.
Thus a new approach is needed to calculate the SFF in the non-ergodic phase, where $\mathcal{G}_{\alpha \alpha'}(t - t') \rightarrow q_{\textrm{EA}} \neq 0$ as $|t - t'| \rightarrow \infty$.
More specifically, we can no longer neglect the integral over the thermal branch in Eq.~\eqref{eq:Keldysh_EOM}, and $\mathcal{G}_{\alpha \alpha'}(t - t')$ no longer solves the SFF equations of motion (Eq.~\eqref{eq:SFF_general_saddle_point_equations}).

However, in the \textit{TAP} equations of motion, Eq.~\eqref{eq:Keldysh_TAP_EOM}, we \textit{can} neglect the thermal branch since $\mathcal G(t)-q_{\textrm{EA}}$ does decay to zero exponentially quickly.
This suggests that a viable strategy is to construct solutions for the SFF using the TAP correlation function.
Since TAP states are parametrized by the quantity $\mathcal{E}$ in Eq.~\eqref{eq:normalized_energy_definition}, it will be necessary to first modify the SFF path integral so as to involve $\mathcal{E}$.
We associate the magnetizations and overlap from the TAP approach with time-averaged functions of the spin configuration, namely
\begin{equation} \label{eq:spin_TAP_quantity_definitions}
m_i[\sigma] \equiv \frac{1}{T} \int_0^T dt \sigma_{iu}(t), \qquad q[\sigma] \equiv \frac{1}{T^2} \int_0^T dt dt' \frac{1}{N} \sum_i \sigma_{iu}(t) \sigma_{iu}(t').
\end{equation}
The choice to use only the upper contour in defining $m_i[\sigma]$ and $q[\sigma]$ will become convenient in Sec.~\ref{sec:HigherMoments}, but for now one could equally well use any other combination of branches, say the average of $\sigma_i(t)$ over the lower branch or over both branches symmetrically.
With these definitions, we introduce $\mathcal{E}$ via Eq.~\eqref{eq:normalized_energy_definition}.

\subsection{Effective action} \label{eq:nonergodic_effective_action}

To begin, insert an additional fat unity into the path integral:
\begin{equation} \label{eq:TAP_resolving_fat_unity}
\begin{aligned}
1 &= \int d\mathcal{E}_{\textrm{aux}} \delta \left[ N \mathcal{E}_{\textrm{aux}} - \frac{1}{J q[\sigma]^{p/2}} \sum_{(i_1 \cdots i_p)} J_{i_1 \cdots i_p} \left( \frac{1}{T} \int_0^T dt \sigma_{i_1 u}(t) \right) \cdots \left( \frac{1}{T} \int_0^T dt \sigma_{i_p u}(t) \right) \right].
\end{aligned}
\end{equation}
With this addition, the full path integral is
\begin{equation} \label{eq:TAP_resolved_original_path_integral}
\begin{aligned}
\textrm{SFF} &= \int \mathcal{D}P(J)\mathcal D\sigma^N d\mathcal{E}_{\textrm{aux}} d\lambda \exp \left[ \frac{i}{2} \sum_i \int_0^T dt \sum_{\alpha} (-1)^{\alpha} \Big[ \mu \big( \partial_t \sigma_{i \alpha}(t) \big)^2 - z_{\alpha}(t) \big( \sigma_{i \alpha}(t)^2 - 1 \big) \Big] \right] \\
&\qquad \cdot \exp \left[ -i \sum_{(i_1 \cdots i_p)} J_{i_1 \cdots i_p} \int_0^T dt \sum_{\alpha} (-1)^{\alpha} \sigma_{i_1 \alpha}(t) \cdots \sigma_{i_p \alpha}(t) \right] \\
&\qquad \qquad \cdot \exp \left[ iN\lambda \mathcal{E}_{\textrm{aux}} - \frac{i \lambda}{J q[\sigma]^{p/2}} \sum_{(i_1 \cdots i_p)} J_{i_1 \cdots i_p} \left( \frac{1}{T} \int_0^T dt \sigma_{i_1 u}(t) \right) \cdots \left( \frac{1}{T} \int_0^T dt \sigma_{i_p u}(t) \right) \right].
\end{aligned}
\end{equation}
Proceeding as usual --- averaging over disorder, introducing $G_{\alpha \alpha'}(t, t')$ and $F_{\alpha \alpha'}(t, t')$ as before, integrating out spins --- we arrive at
\begin{equation} \label{eq:TAP_resolved_SFF_path_integral}
\textrm{SFF}(T, f) = \int d\mathcal{E}_{\textrm{aux}} d\lambda \mathcal{D}G \mathcal{D}F \, f \big( \epsilon_u[\lambda, G] \big) f \big( \epsilon_l[\lambda, G] \big) e^{-NS_{\textrm{eff}}[\mathcal{E}_{\textrm{aux}}, \lambda, G, F]},
\end{equation}
\begin{equation} \label{eq:TAP_resolved_SFF_action}
\begin{aligned}
S_{\textrm{eff}}[\mathcal{E}_{\textrm{aux}}, \lambda, G, F] &= -i \lambda \mathcal{E}_{\textrm{aux}} - \frac{i}{2} \int_0^T dt \sum_{\alpha} (-1)^{\alpha} z_{\alpha}(t) \\
&\qquad + \frac{\lambda^2}{2p} + \frac{J \lambda}{p q[G]^{p/2}} \int_0^T dt \sum_{\alpha} (-1)^{\alpha} \left( \frac{1}{T} \int_0^T dt' G_{\alpha u}(t, t') \right)^p \\
&\qquad \qquad + \frac{1}{2} \int_0^T dt dt' \sum_{\alpha \alpha'} (-1)^{\alpha + \alpha'} \left( \frac{J^2}{p} G_{\alpha \alpha'}(t, t')^p - F_{\alpha \alpha'}(t, t') G_{\alpha \alpha'}(t, t') \right) \\
&\qquad \qquad \qquad + \frac{1}{2} \log{\textrm{Det}} \Big[ i (-1)^{\alpha} \delta_{\alpha \alpha'} \big( \mu \partial_t^2 + z_{\alpha} \big) + (-1)^{\alpha + \alpha'} F_{\alpha \alpha'} \Big],
\end{aligned}
\end{equation}
where we are denoting $q[G] \equiv T^{-2} \int_0^T dt dt' G_{uu}(t, t')$.
The argument of the filter function is modified as well; it is now
\begin{equation} \label{eq:TAP_resolved_energy_density_definition}
\epsilon_{\alpha}[\lambda, G] \equiv -\frac{\mu}{2} \partial_t^2 G_{\alpha \alpha}(0^+, 0) - \frac{iJ^2}{p} \int_0^T dt \sum_{\alpha'} (-1)^{\alpha'} G_{\alpha \alpha'}(t, 0)^p - \frac{iJ \lambda}{p q[G]^{p/2}} \left( \frac{1}{T} \int_0^T dt G_{\alpha u}(t, 0) \right)^p.
\end{equation}

Note that $\mathcal{E}_{\textrm{aux}}$ enters linearly into the action.
Thus if we were to integrate over $\mathcal{E}_{\textrm{aux}}$ at this point, we would obtain a $\delta$-function forcing $\lambda = 0$.
The action would then reduce to the ergodic-phase expression, Eq.~\eqref{eq:SFF_formal_action}.
While reassuring, this would not have accomplished anything, so we instead treat $\mathcal{E}_{\textrm{aux}}$ as a fixed parameter for now.
We obtain saddle point equations by differentiating Eq.~\eqref{eq:TAP_resolved_SFF_action} only with respect to $\lambda$, $z$, $G$, and $F$.

The saddle point equations, assuming time translation invariance from the outset, are
\begin{equation} \label{eq:TAP_resolved_EOM}
i \big( \mu \partial_t^2 + z_{\alpha} \big) G_{\alpha \alpha'}(t - t') + \int_0^T dt'' \sum_{\alpha''} (-1)^{\alpha''} F_{\alpha \alpha''}(t - t'') G_{\alpha'' \alpha'}(t'' - t') = (-1)^{\alpha} \delta_{\alpha \alpha'} \delta(t - t'),
\end{equation}
\begin{equation} \label{eq:TAP_resolved_self_energy_equation}
\begin{aligned}
F_{\alpha \alpha'}(t) = J^2 G_{\alpha \alpha'}(t)^{p-1} &+ \frac{J \lambda}{T q[G]^{p/2}} \big( \delta_{\alpha u} + \delta_{\alpha' u} \big) \left( \frac{\widetilde{G}_{\alpha \alpha'}(0)}{T} \right)^{p-1} \\
&- \frac{J \lambda}{T q[G]^{p/2 + 1}} \delta_{\alpha u} \delta_{\alpha' u} \sum_{\alpha''} (-1)^{\alpha''} \left( \frac{\widetilde{G}_{\alpha'' u}(0)}{T} \right)^p, 
\end{aligned}
\end{equation}
\begin{equation} \label{eq:TAP_resolved_potential_equation}
\mathcal{E}_{\textrm{aux}} = -\frac{iJT}{p q[G]^{p/2}} \sum_{\alpha} (-1)^{\alpha} \left( \frac{\widetilde{G}_{\alpha u}(0)}{T} \right)^p - \frac{i \lambda}{p},
\end{equation}
as well as the usual requirement $G_{\alpha \alpha}(0) = 1$.
Here $\widetilde{G}_{\alpha \alpha'}(\omega)$ is the Fourier transform of $G_{\alpha \alpha'}(t)$.

\subsection{Connected solutions} \label{subsec:TAP_resolved_connected_solutions}

With $\mathcal{E}_{\textrm{aux}}$ fixed, let $\mathcal{G}_{\alpha \alpha'}(t)$ be the solution to the TAP equation of motion (Eq.~\eqref{eq:Keldysh_TAP_EOM}) corresponding to inverse temperature $\beta_{\textrm{aux}}$.
Denote the Edwards-Anderson order parameter at $\mathcal{E}_{\textrm{aux}}$ and $\beta_{\textrm{aux}}$ by $q_{\textrm{EA}}$.
Also recall the various auxiliary quantities we defined in Sec.~\ref{subsec:QTAP}: the self-energy $\mathcal{F}_{\alpha \alpha'}(t) \equiv J^2 \mathcal{G}_{\alpha \alpha'}(t)^{p-1}$, the deviations $\Delta \mathcal{G}_{\alpha \alpha'}(t) \equiv \mathcal{G}_{\alpha \alpha'}(t) - q_{\textrm{EA}}$ and $\Delta \mathcal{F}_{\alpha \alpha'}(t) \equiv \mathcal{F}_{\alpha \alpha'}(t) - J^2 q_{\textrm{EA}}^{p-1}$, and the quantity $\Lambda \equiv \int_0^{\infty} dt \Delta \mathcal{G}^I(t)$.
We have that $\mathcal{G}_{\alpha \alpha'}(t)$ and $\mathcal{F}_{\alpha \alpha'}(t)$ obey Eq.~\eqref{eq:Keldysh_TAP_EOM}, which by taking $t$ and $t'$ to be far from the thermal branch can be written
\begin{equation} \label{eq:Keldysh_TAP_EOM_simplified}
i \big( \mu \partial_t^2 + z \big) \Delta \mathcal{G}_{\alpha \alpha'}(t - t') + \int_{-\infty}^{\infty} dt'' \sum_{\alpha''} (-1)^{\alpha''} \Delta \mathcal{F}_{\alpha \alpha''}(t - t'') \Delta \mathcal{G}_{\alpha'' \alpha'}(t'' - t') = (-1)^{\alpha} \delta_{\alpha \alpha'} \delta(t - t').
\end{equation}
We also have, as a result of Eq.~\eqref{eq:Keldysh_TAP_magnetization_alternate}, the relationship
\begin{equation} \label{eq:potential_GF_relationship}
\mathcal{E}_{\textrm{aux}} = \frac{2(p-1)J q_{\textrm{EA}}^{p/2-1} \Lambda}{p} + \frac{1}{2pJ q_{\textrm{EA}}^{p/2-1} \Lambda}.
\end{equation}
Finally, recall the expression for the complexity $\Sigma(\mathcal{E})$, the logarithm of the number of solutions to the TAP magnetization equations at $\mathcal{E}$:
\begin{equation} \label{eq:TAP_complexity_repeat}
\Sigma(\mathcal{E}) = \frac{1}{2} \left( 1 + 2\log{\frac{p}{2}} \right) - \frac{p \mathcal{E}^2}{2} + \frac{p^2}{8(p-1)} \left( \mathcal{E} + \sqrt{\mathcal{E}^2 - \mathcal{E}_{\textrm{th}}^2} \right)^2 + \log{\left( -\mathcal{E} + \sqrt{\mathcal{E}^2 - \mathcal{E}_{\textrm{th}}^2} \right)},
\end{equation}
where $\mathcal{E}_{\textrm{th}}^2 = 4(p-1)/p^2$.
Using Eq.~\eqref{eq:potential_GF_relationship}, we can express $\Sigma(\mathcal{E}_{\textrm{aux}})$ in terms of $\Lambda$ and $q_{\textrm{EA}}$:
\begin{equation} \label{eq:TAP_complexity_rewrite}
\Sigma(\mathcal{E}_{\textrm{aux}}) = -\frac{p-2}{2p} - \frac{1}{8pJ^2 q_{\textrm{EA}}^{p-2} \Lambda^2} + \frac{2(p-1)J^2 q_{\textrm{EA}}^{p-2} \Lambda^2}{p} - \frac{1}{2} \log{4J^2 q_{\textrm{EA}}^{p-2} \Lambda^2}.
\end{equation}

Our solution to the SFF saddle point equations, Eqs.~\eqref{eq:TAP_resolved_EOM} through~\eqref{eq:TAP_resolved_potential_equation}, is best written in the frequency domain (tildes denote Fourier transforms):
\begin{equation} \label{eq:TAP_resolved_G_solution_frequency}
\widetilde{G}_{\alpha \alpha'}(\omega) = Tq_{\textrm{EA}} \delta_{\omega 0} + \Delta \widetilde{\mathcal{G}}_{\alpha \alpha'}(\omega) + \frac{\widetilde{g}_{\alpha \alpha'}(\omega)}{T},
\end{equation}
\begin{equation} \label{eq:TAP_resolved_F_solution_frequency}
\widetilde{F}_{\alpha \alpha'}(\omega) = \left( TJ^2 q_{\textrm{EA}}^{p-1} + ipJ q_{\textrm{EA}}^{p/2-1} \big( \mathcal{E}_{\textrm{aux}} - 2J q_{\textrm{EA}}^{p/2-1} \Lambda \big) \big( \delta_{\alpha u} + \delta_{\alpha' u} \big) \right) \delta_{\omega 0} + \Delta \widetilde{\mathcal{F}}_{\alpha \alpha'}(\omega) + \frac{\widetilde{f}_{\alpha \alpha'}(\omega)}{T},
\end{equation}
\begin{equation} \label{eq:TAP_resolved_lambda_solution}
\lambda = ip \big( \mathcal{E}_{\textrm{aux}} - 2J q_{\textrm{EA}}^{p/2-1} \Lambda \big) + \frac{\delta}{T}.
\end{equation}
We again take $z_{\alpha}$ to be the equilibrium value corresponding to $\beta_{\textrm{aux}}$.
The precise form of the correction terms $\widetilde{g}_{\alpha \alpha'}(\omega)$, $\widetilde{f}_{\alpha \alpha'}(\omega)$, and $\delta$ is largely unimportant --- the essential feature is simply that they are $O(1)$ and the corrections are thus $O(T^{-1})$.
Note that in the time domain, this solution amounts to
\begin{equation} \label{eq:TAP_resolved_G_solution_time}
G_{\alpha \alpha'}(t) = q_{\textrm{EA}} + \sum_{n=-\infty}^{\infty} \Delta \mathcal{G}_{\alpha \alpha'}(t + nT) + \frac{g_{\alpha \alpha'}(t)}{T},
\end{equation}
\begin{equation} \label{eq:TAP_resolved_F_solution_time}
F_{\alpha \alpha'}(t) = J^2 q_{\textrm{EA}}^{p-1} + \sum_{n=-\infty}^{\infty} \Delta \mathcal{F}_{\alpha \alpha'}(t + nT) + \frac{ipJ q_{\textrm{EA}}^{p/2-1} \big( \mathcal{E}_{\textrm{aux}} - 2J q_{\textrm{EA}}^{p/2-1} \Lambda \big)}{T} \big( \delta_{\alpha u} + \delta_{\alpha' u} \big) + \frac{f_{\alpha \alpha'}(t)}{T}.
\end{equation}
The sums are convergent because $\Delta \mathcal{G}_{\alpha \alpha'}(t)$ and $\Delta \mathcal{F}_{\alpha \alpha'}(t)$ decay rapidly to zero as $|t| \rightarrow \infty$.

Although we have omitted it for notational simplicity, we can add a term $\delta_{\alpha \neq \alpha'} \Delta$ to the time arguments of $\Delta \mathcal{G}_{\alpha \alpha'}(t + nT)$ and $\Delta \mathcal{F}_{\alpha \alpha'}(t + nT)$ for any $\Delta \in [0, T)$, exactly as in Sec.~\ref{sec:ergodic_ramp}.
Due to the separate time translation symmetry on each branch of the SFF contour, all such solutions are equally valid and contribute the same action.
Thus we shall demonstrate the validity of Eqs.~\eqref{eq:TAP_resolved_G_solution_frequency} through~\eqref{eq:TAP_resolved_lambda_solution} and evaluate the action only for $\Delta = 0$, but then integrate over all $\Delta \in [0, T)$ in the final expression for the SFF.

Let us first confirm that our solution satisfies the saddle point equation for $\lambda$, Eq.~\eqref{eq:TAP_resolved_potential_equation}.
Referring to Eq.~\eqref{eq:Keldysh_zero_frequency_matrix}, we have that $\Delta \widetilde{\mathcal{G}}_{\alpha \alpha'}(0) = L + (-1)^{\alpha} 2i \Lambda \delta_{\alpha \alpha'}$.
Thus
\begin{equation} \label{eq:spin_TAP_overlap_expansion}
q[G] = q_{\textrm{EA}} + \frac{L + 2i \Lambda}{T} + O(T^{-2}),
\end{equation}
and Eq.~\eqref{eq:TAP_resolved_potential_equation} becomes
\begin{equation} \label{eq:TAP_resolved_potential_demonstration}
\mathcal{E}_{\textrm{aux}} = 2J q_{\textrm{EA}}^{p/2-1} \Lambda - \frac{i \lambda}{p} + O(T^{-1}).
\end{equation}
Solving for $\lambda$ indeed gives Eq.~\eqref{eq:TAP_resolved_lambda_solution}.
The $O(T^{-1})$ terms determine $\delta$ as a function of the other quantities.

Now turn to Eq.~\eqref{eq:TAP_resolved_self_energy_equation}.
In the frequency domain, the right-hand side evaluates to\footnote{Since $\widetilde{g}_{\alpha \alpha'}(\omega) = O(1)$ with respect to $T$, $g_{\alpha \alpha'}(t)$ decays to zero as $|t| \rightarrow \infty$ (at least to leading order).}
\begin{equation} \label{eq:TAP_resolved_self_energy_demonstration}
\begin{aligned}
&\int_0^T dt e^{i \omega t} J^2 \left( q_{\textrm{EA}} + \sum_{n=-\infty}^{\infty} \Delta \mathcal{G}_{\alpha \alpha'}(t + nT) \right)^{p-1} + ipJ q_{\textrm{EA}}^{p/2-1} \big( \mathcal{E}_{\textrm{aux}} - 2J q_{\textrm{EA}}^{p/2-1} \Lambda \big) \big( \delta_{\alpha u} + \delta_{\alpha' u} \big) \delta_{\omega 0} + O(T^{-1}) \\
&\qquad \qquad \stackrel{\textrm{set}}{=} \left( TJ^2 q_{\textrm{EA}}^{p-1} + ipJ q_{\textrm{EA}}^{p/2-1} \big( \mathcal{E}_{\textrm{aux}} - 2J q_{\textrm{EA}}^{p/2-1} \Lambda \big) \big( \delta_{\alpha u} + \delta_{\alpha' u} \big) \right) \delta_{\omega 0} + \Delta \widetilde{\mathcal{F}}_{\alpha \alpha'}(\omega) + \frac{\widetilde{f}_{\alpha \alpha'}(\omega)}{T}.
\end{aligned}
\end{equation}
Along the lines of Eq.~\eqref{eq:cross_terms_neglecting}, we have that
\begin{equation} \label{eq:non_ergodic_cross_terms_neglecting}
\begin{aligned}
J^2 \left( q_{\textrm{EA}} + \sum_{n=-\infty}^{\infty} \Delta \mathcal{G}_{\alpha \alpha'}(t + nT) \right)^{p-1} &= J^2 q_{\textrm{EA}}^{p-1} + J^2 \sum_{r=1}^{p-1} \binom{p-1}{r} q_{\textrm{EA}}^{p-1-r} \left( \sum_{n=-\infty}^{\infty} \Delta \mathcal{G}_{\alpha \alpha'}(t + nT) \right)^r \\
&\sim J^2 q_{\textrm{EA}}^{p-1} + J^2 \sum_{r=1}^{p-1} \binom{p-1}{r} q_{\textrm{EA}}^{p-1-r} \sum_{n=-\infty}^{\infty} \Delta \mathcal{G}_{\alpha \alpha'}(t + nT)^r \\
&= J^2 q_{\textrm{EA}}^{p-1} + \sum_{n=-\infty}^{\infty} \Delta \mathcal{F}_{\alpha \alpha'}(t + nT). 
\end{aligned}
\end{equation}
Thus, up to $O(1)$, both sides of Eq.~\eqref{eq:TAP_resolved_self_energy_demonstration} agree.

Finally, we confirm that Eq.~\eqref{eq:TAP_resolved_EOM} is satisfied.
At non-zero frequencies, we have
\begin{equation} \label{eq:TAP_resolved_EOM_non_zero_demonstration}
i \big( -\mu \omega^2 + z \big) \Delta \widetilde{\mathcal{G}}_{\alpha \alpha'}(\omega) + \sum_{\alpha''} (-1)^{\alpha''} \Delta \widetilde{\mathcal{F}}_{\alpha \alpha''}(\omega) \Delta \widetilde{\mathcal{G}}_{\alpha'' \alpha'}(\omega) + O(T^{-1}) \stackrel{\textrm{set}}{=} (-1)^{\alpha} \delta_{\alpha \alpha'},
\end{equation}
which agrees at $O(1)$ due to Eq.~\eqref{eq:Keldysh_TAP_EOM_simplified}.
At zero frequency, we instead have
\begin{equation} \label{eq:TAP_resolved_EOM_zero_demonstration_1}
\begin{aligned}
&iz \left( Tq_{\textrm{EA}} + \Delta \widetilde{\mathcal{G}}_{\alpha \alpha'}(0) + \frac{\widetilde{g}_{\alpha \alpha'}(0)}{T} \right) \\
&\quad + \sum_{\alpha''} (-1)^{\alpha''} \left( TJ^2 q_{\textrm{EA}}^{p-1} + ipJ q_{\textrm{EA}}^{p/2-1} \big( \mathcal{E}_{\textrm{aux}} - 2J q_{\textrm{EA}}^{p/2-1} \Lambda \big) \big( \delta_{\alpha u} + \delta_{\alpha'' u} \big) + \Delta \widetilde{\mathcal{F}}_{\alpha \alpha''}(0) + \frac{\widetilde{f}_{\alpha \alpha''}(0)}{T} \right) \\
&\qquad \qquad \qquad \qquad \qquad \qquad \qquad \qquad \qquad \qquad \qquad \cdot \left( Tq_{\textrm{EA}} + \Delta \widetilde{\mathcal{G}}_{\alpha'' \alpha'}(0) + \frac{\widetilde{g}_{\alpha'' \alpha'}(0)}{T} \right) \stackrel{\textrm{set}}{=} (-1)^{\alpha} \delta_{\alpha \alpha'}.
\end{aligned}
\end{equation}
The $O(T)$ terms come out to be
\begin{equation} \label{eq:TAP_resolved_EOM_zero_demonstration_2}
T \left( izq_{\textrm{EA}} + \sum_{\alpha''} (-1)^{\alpha''} \Big( J^2 q_{\textrm{EA}}^{p-1} \Delta \widetilde{\mathcal{G}}_{\alpha'' \alpha'}(0) + q_{\textrm{EA}} \Delta \widetilde{\mathcal{F}}_{\alpha \alpha''}(0) \Big) + ipJ q_{\textrm{EA}}^{p/2} \big( \mathcal{E}_{\textrm{aux}} - 2J q_{\textrm{EA}}^{p/2-1} \Lambda \big) \right) \stackrel{\textrm{set}}{=} 0.
\end{equation}
Yet from the TAP magnetization equations, Eq.~\eqref{eq:Keldysh_TAP_magnetization_equation}, it follows that
\begin{equation} \label{eq:Keldysh_TAP_magnetization_double_alternate}
\sum_{\alpha''} (-1)^{\alpha''} \Big( q_{\textrm{EA}} \Delta \widetilde{\mathcal{F}}_{\alpha \alpha''}(0) - (p-1)J^2 q_{\textrm{EA}}^{p-1} \Delta \widetilde{\mathcal{G}}_{\alpha \alpha''}(0) \Big) = -izq_{\textrm{EA}} - ipJ q_{\textrm{EA}}^{p/2} \mathcal{E}_{\textrm{aux}}.
\end{equation}
Thus Eq.~\eqref{eq:TAP_resolved_EOM_zero_demonstration_2} evaluates to
\begin{equation} \label{eq:TAP_resolved_EOM_zero_leading_term}
-ipJ q_{\textrm{EA}}^{p/2} \mathcal{E}_{\textrm{aux}} + pJ^2 q_{\textrm{EA}}^{p-1} \sum_{\alpha''} (-1)^{\alpha''} \Delta \widetilde{\mathcal{G}}_{\alpha \alpha''}(0) + ipJ q_{\textrm{EA}}^{p/2} \mathcal{E}_{\textrm{aux}} - 2ipJ^2 q_{\textrm{EA}}^{p-1} \Lambda = 0,
\end{equation}
using Eq.~\eqref{eq:Keldysh_zero_frequency_matrix}.
The $O(1)$ terms of Eq.~\eqref{eq:TAP_resolved_EOM_zero_demonstration_1} determine $\widetilde{g}_{\alpha \alpha'}(0)$ and $\widetilde{f}_{\alpha \alpha'}(0)$. We have therefore confirmed that all saddle point equations are solved by Eqs.~\eqref{eq:TAP_resolved_G_solution_frequency} through~\eqref{eq:TAP_resolved_lambda_solution}.

\subsection{Contribution of connected solutions} \label{subsec:nonergodic_contribution_connected_solutions}

It remains only to evaluate the action, Eq.~\eqref{eq:TAP_resolved_SFF_action}, at the above solution.
The action can be written as
\begin{equation} \label{eq:TAP_resolved_action_starting_point}
\begin{aligned}
S_{\textrm{eff}} &= -i \lambda \mathcal{E}_{\textrm{aux}} + \frac{\lambda^2}{2p} + \frac{JT \lambda}{p q[G]^{p/2}} \sum_{\alpha} (-1)^{\alpha} \left( \frac{\widetilde{G}_{\alpha u}(0)}{T} \right)^p \\
&\qquad + \frac{T}{2} \sum_{\alpha \alpha'} (-1)^{\alpha + \alpha'} \int_0^T dt \left( \frac{J^2}{p} G_{\alpha \alpha'}(t)^p - F_{\alpha \alpha'}(t) G_{\alpha \alpha'}(t) \right) \\
&\qquad \qquad + \frac{1}{2} \sum_{\omega} \log{\textrm{Det}} \Big[ i(-1)^{\alpha} \delta_{\alpha \alpha'} \big( -\mu \omega^2 + z \big) + (-1)^{\alpha + \alpha'} \widetilde{F}_{\alpha \alpha'}(\omega) \Big].
\end{aligned}
\end{equation}
Interestingly, we can determine $S_{\textrm{eff}}$ up to a single additive constant simply by noting that $dS_{\textrm{eff}}/d\mathcal{E}_{\textrm{aux}} = -i \lambda$ (recall that $S_{\textrm{eff}}$ is stationary with respect to variations in all quantities other than $\mathcal{E}_{\textrm{aux}}$).
With $\lambda$ given by Eq.~\eqref{eq:TAP_resolved_lambda_solution} and $q_{\textrm{EA}}^{p/2-1} \Lambda$ given by Eq.~\eqref{eq:Keldysh_TAP_overlap_closed_equation}, we can carry out the integral to obtain that
\begin{equation} \label{eq:nonergodic_action_almost_evaluation}
S_{\textrm{eff}}[\mathcal{E}_{\textrm{aux}}] = \frac{p \mathcal{E}_{\textrm{aux}}^2}{2} - \frac{p^2}{8(p-1)} \Big( \mathcal{E}_{\textrm{aux}} + \sqrt{\mathcal{E}_{\textrm{aux}}^2 - \mathcal{E}_{\textrm{th}}^2} \Big)^2 - \log{\Big( -\mathcal{E}_{\textrm{aux}} + \sqrt{\mathcal{E}_{\textrm{aux}}^2 - \mathcal{E}_{\textrm{th}}^2} \Big)} + C,
\end{equation}
for some unknown constant $C$.
Comparing to Eq.~\eqref{eq:TAP_complexity}, this is highly suggestive that $S_{\textrm{eff}} = -\Sigma(\mathcal{E}_{\textrm{aux}})$.
Of course, we do need to determine the remaining constant, and so we now turn to a more elaborate calculation.

Rather than substitute Eqs.~\eqref{eq:TAP_resolved_G_solution_frequency} and~\eqref{eq:TAP_resolved_F_solution_frequency} into Eq.~\eqref{eq:TAP_resolved_action_starting_point}, we instead use the simpler functions
\begin{equation} \label{eq:simplified_G_solution_frequency}
\widetilde{G}'_{\alpha \alpha'}(\omega) = Tq_{\textrm{EA}} \delta_{\omega 0} + \Delta \widetilde{\mathcal{G}}_{\alpha \alpha'}(\omega),
\end{equation}
\begin{equation} \label{eq:simplified_F_solution_frequency}
\widetilde{F}'_{\alpha \alpha'}(\omega) = \left( TJ^2 q_{\textrm{EA}}^{p-1} + ipJ q_{\textrm{EA}}^{p/2-1} \big( \mathcal{E}_{\textrm{aux}} - 2J q_{\textrm{EA}}^{p/2-1} \Lambda \big) \big( \delta_{\alpha u} + \delta_{\alpha' u} \big) \right) \delta_{\omega 0} + \Delta \widetilde{\mathcal{F}}_{\alpha \alpha'}(\omega) + \frac{\widetilde{f}_{\alpha \alpha'}(0)}{T} \delta_{\omega 0},
\end{equation}
and show that the error incurred in doing so vanishes at large $T$.

Let us demonstrate that the error is negligible first.
At any non-zero frequency, we have that
\begin{equation} \label{eq:simplified_solutions_nonzero_frequency}
\widetilde{G}_{\alpha \alpha'}(\omega) = \widetilde{G}'_{\alpha \alpha'}(\omega) + \frac{\widetilde{g}_{\alpha \alpha'}(\omega)}{T}, \qquad \widetilde{F}_{\alpha \alpha'}(\omega) = \widetilde{F}'_{\alpha \alpha'}(\omega) + \frac{\widetilde{f}_{\alpha \alpha'}(\omega)}{T}.
\end{equation}
The partial derivatives of $S_{\textrm{eff}}$ at nonzero $\omega$ are
\begin{equation} \label{eq:action_G_derivative_nonzero_frequency}
\frac{\partial S_{\textrm{eff}}}{\partial \widetilde{G}_{\alpha \alpha'}(\omega)} = \frac{1}{2} (-1)^{\alpha + \alpha'} \int_0^T dt e^{-i \omega t} \Big( J^2 G_{\alpha \alpha'}(t)^{p-1} - F_{\alpha \alpha'}(t) \Big),
\end{equation}
\begin{equation} \label{eq:action_F_derivative_nonzero_frequency}
\frac{\partial S_{\textrm{eff}}}{\partial \widetilde{F}_{\alpha \alpha'}(\omega)} = \frac{1}{2} (-1)^{\alpha + \alpha'} \bigg( \Big[ i(-1)^{\alpha} \delta_{\alpha \alpha'} \big( -\mu \omega^2 + z \big) + (-1)^{\alpha + \alpha'} \widetilde{F}_{\alpha' \alpha}(\omega) \Big]_{\alpha \alpha'}^{-1} - \int_0^T dt e^{-i \omega t} G_{\alpha \alpha'}(t) \bigg),
\end{equation}
which vanish when evaluated at $\widetilde{G}'_{\alpha \alpha'}(\omega) = \Delta \widetilde{\mathcal{G}}_{\alpha \alpha'}(\omega)$ and $\widetilde{F}'_{\alpha \alpha'}(\omega) = \Delta \widetilde{\mathcal{F}}_{\alpha \alpha'}(\omega)$.
Thus the $O(T^{-1})$ difference between $\widetilde{G}_{\alpha \alpha'}(\omega)$ and $\widetilde{G}'_{\alpha \alpha'}(\omega)$, as with $\widetilde{F}_{\alpha \alpha'}(\omega)$ and $\widetilde{F}'_{\alpha \alpha'}(\omega)$, translates only to an $O(T^{-2})$ difference in the action.
Even after summing over all $\omega \neq 0$, the total error\footnote{
Since the $G(t)^p$ term is not diagonal in the frequency domain, this argument requires a bit more care.
One can easily show that $\partial^2 S_{\textrm{eff}} / \partial \widetilde{G}(\omega) \partial \widetilde{G}(\omega')$ is $O(T^{-1})$ for $\omega \neq \pm \omega'$ and $O(1)$ for $\omega = \pm \omega'$.
Summing over all frequencies, the former case gives a total contribution $O(T^{-3}) O(T^2) = O(T^{-1})$ and the latter gives $O(T^{-2}) O(T) = O(T^{-1})$.
The total error is thus $O(T^{-1})$ as claimed.
} is only $O(T^{-1})$.

Neglecting non-zero frequencies, $\widetilde{F}'_{\alpha \alpha'}(\omega)$ is identical to $\widetilde{F}_{\alpha \alpha'}(\omega)$ and $\widetilde{G}'_{\alpha \alpha'}(\omega)$ differs only by $\widetilde{g}_{\alpha \alpha'}(0) \delta_{\omega 0}/T$.
In the time domain, the latter corresponds to
\begin{equation} \label{eq:semi_simplified_G_solution_time}
G_{\alpha \alpha'}(t) = G'_{\alpha \alpha'}(t) + \frac{\widetilde{g}_{\alpha \alpha'}(0)}{T^2} = q_{\textrm{EA}} + \sum_{n=-\infty}^{\infty} \Delta \mathcal{G}_{\alpha \alpha'}(t + nT) + \frac{\widetilde{g}_{\alpha \alpha'}(0)}{T^2}.
\end{equation}
Yet
\begin{equation} \label{eq:action_G_derivative_time}
\frac{\partial S_{\textrm{eff}}}{\partial G_{\alpha \alpha'}(t)} = \frac{T}{2} (-1)^{\alpha + \alpha'} \Big( J^2 G_{\alpha \alpha'}(t)^{p-1} - F_{\alpha \alpha'}(t) \Big) + O(1).
\end{equation}
When evaluated at $G'_{\alpha \alpha'}(t)$ and $F'_{\alpha \alpha'}(t)$, the $O(T)$ contribution vanishes (see Eq.~\eqref{eq:non_ergodic_cross_terms_neglecting}).
Thus $\partial S_{\textrm{eff}} / \partial G_{\alpha \alpha'}(t)$ is $O(1)$, and an $O(T^{-2})$ change to $G_{\alpha \alpha'}(t)$ leads only to an $O(T^{-1})$ change in the action even after integrating over $t$.

Since all errors are $O(T^{-1})$, we can safely evaluate $S_{\textrm{eff}}$ at Eqs.~\eqref{eq:simplified_G_solution_frequency} and~\eqref{eq:simplified_F_solution_frequency} rather than the full solution (we still use Eq.~\eqref{eq:TAP_resolved_lambda_solution} for $\lambda$).
The first line of Eq.~\eqref{eq:TAP_resolved_action_starting_point} can be computed straightforwardly.
It comes out to be
\begin{equation} \label{eq:nonergodic_action_evaluation_line_1}
\frac{p \big( \mathcal{E}_{\textrm{aux}} - 2J q_{\textrm{EA}}^{p/2-1} \Lambda \big)^2}{2} = \frac{1}{8pJ^2 q_{\textrm{EA}}^{p-2} \Lambda^2} - \frac{1}{p} + \frac{2J^2 q_{\textrm{EA}}^{p-2} \Lambda^2}{p},
\end{equation}
where we used Eq.~\eqref{eq:potential_GF_relationship} to obtain the right-hand side.

Next consider the bottom line.
Since $\widetilde{F}'_{\alpha \alpha'}(\omega) = \Delta \widetilde{\mathcal{F}}_{\alpha \alpha'}(\omega)$ for $\omega \neq 0$, while $\widetilde{F}'_{\alpha \alpha'}(0) = \widetilde{F}_{\alpha \alpha'}(0)$, we can write the determinant term as
\begin{equation} \label{eq:nonergodic_action_evaluation_determinant_1}
\begin{aligned}
&\frac{1}{2} \sum_{\omega} \log{\textrm{Det}} \Big[ i(-1)^{\alpha} \delta_{\alpha \alpha'} \big( -\mu \omega^2 + z \big) + (-1)^{\alpha + \alpha'} \Delta \widetilde{\mathcal{F}}_{\alpha \alpha'}(\omega) \Big] \\
&\qquad \qquad + \frac{1}{2} \log{\textrm{Det}} \Big[ iz(-1)^{\alpha} \delta_{\alpha \alpha'} + (-1)^{\alpha + \alpha'} \widetilde{F}_{\alpha \alpha'}(0) \Big] \\
&\qquad \qquad \qquad \qquad - \frac{1}{2} \log{\textrm{Det}} \Big[ iz(-1)^{\alpha} \delta_{\alpha \alpha'} + (-1)^{\alpha + \alpha'} \Delta \widetilde{\mathcal{F}}_{\alpha \alpha'}(0) \Big].
\end{aligned}
\end{equation}
The top line vanishes by exactly the same reasoning as in Sec.~\ref{subsec:ergodic_connected_action}: it is proportional to $T$ by the Euler-Maclaurin formula, and then must be zero since the derivative with respect to $T$ vanishes.
Given Eq.~\eqref{eq:Keldysh_zero_frequency_matrix}, the bottom line is simply
\begin{equation} \label{eq:nonergodic_action_evaluation_determinant_2}
\frac{1}{2} \log{\textrm{Det}} \Delta \widetilde{\mathcal{G}}_{\alpha \alpha'}(0) = \frac{1}{2} \log{4 \Lambda^2}.
\end{equation}
For the middle line we take an indirect approach.
We have that $iz(-1)^{\alpha} \delta_{\alpha \alpha'} + (-1)^{\alpha + \alpha'} \widetilde{F}_{\alpha \alpha'}(0)$ is the matrix inverse to $\widetilde{G}_{\alpha \alpha'}(0)$ (using the full solution for the latter, Eq.~\eqref{eq:TAP_resolved_G_solution_frequency}).
Written out,
\begin{equation} \label{eq:SFF_functions_matrix_inverse_z_basis}
\begin{pmatrix} iz + \widetilde{F}_{uu}(0) & -\widetilde{F}_{ul}(0) \\ -\widetilde{F}_{lu}(0) & -iz + \widetilde{F}_{ll}(0) \end{pmatrix} = \begin{pmatrix} \widetilde{G}_{uu}(0) & \widetilde{G}_{ul}(0) \\ \widetilde{G}_{lu}(0) & \widetilde{G}_{ll}(0) \end{pmatrix}^{-1} = \frac{1}{\textrm{Det} \widetilde{G}(0)} \begin{pmatrix} \widetilde{G}_{ll}(0) & -\widetilde{G}_{ul}(0) \\ -\widetilde{G}_{lu}(0) & \widetilde{G}_{uu}(0) \end{pmatrix}.
\end{equation}
Rather than this $(u, l)$ basis, express Eq.~\eqref{eq:SFF_functions_matrix_inverse_z_basis} in the $(u + l, u - l)$ basis (called ``classical''/``quantum'' in the Keldysh literature), denoted $(+, -)$:
\begin{equation} \label{eq:SFF_functions_matrix_inverse_x_basis}
\begin{pmatrix} \widetilde{F}_{--}(0) & iz + \widetilde{F}_{-+}(0) \\ iz + \widetilde{F}_{+-}(0) & \widetilde{F}_{++}(0) \end{pmatrix} = \frac{1}{\textrm{Det} \widetilde{G}(0)} \begin{pmatrix} \widetilde{G}_{--}(0) & -\widetilde{G}_{+-}(0) \\ -\widetilde{G}_{-+}(0) & \widetilde{G}_{++}(0) \end{pmatrix}.
\end{equation}
We can read off that $\textrm{Det} \widetilde{G}(0)^{-1} = \widetilde{F}_{++}(0) / \widetilde{G}_{++}(0)$.
Note that we only need $\widetilde{G}_{++}(0)$ and $\widetilde{F}_{++}(0)$ to $O(T)$ in order to calculate the determinant to $O(1)$.
Thus the middle line of Eq.~\eqref{eq:nonergodic_action_evaluation_determinant_1} evaluates to $(\log{J^2 q_{\textrm{EA}}^{p-2}})/2$, and the total contribution of the determinant term is
\begin{equation} \label{eq:nonergodic_action_evaluation_determinant_3}
\frac{1}{2} \log{4J^2 q_{\textrm{EA}}^{p-2} \Lambda^2}.
\end{equation}

Lastly consider the middle line of Eq.~\eqref{eq:TAP_resolved_action_starting_point}.
Since $G'_{\alpha \alpha'}(t) = \mathcal{G}_{\alpha \alpha'}(t)$ (up to exponentially small corrections), $\sum_{\alpha \alpha'} (-1)^{\alpha + \alpha'} G'_{\alpha \alpha'}(t)^p = 0$ by virtue of Eq.~\eqref{eq:Keldysh_component_symmetry_identity}.
We are left with
\begin{equation} \label{eq:nonergodic_action_evaluation_middle_line_1}
\begin{aligned}
&-\frac{T}{2} \sum_{\alpha \alpha'} (-1)^{\alpha + \alpha'} \int_0^T dt F'_{\alpha \alpha'}(t) G'_{\alpha \alpha'}(t) \\
&\qquad \sim -\frac{T}{2} \sum_{\alpha \alpha'} (-1)^{\alpha + \alpha'} \int_0^T dt \mathcal{F}_{\alpha \alpha'}(t) \mathcal{G}_{\alpha \alpha'}(t) \\
&\qquad \qquad - \frac{1}{2} \sum_{\alpha \alpha'} (-1)^{\alpha + \alpha'} \left( ipJ q_{\textrm{EA}}^{p/2-1} \big( \mathcal{E}_{\textrm{aux}} - 2J q_{\textrm{EA}}^{p/2-1} \Lambda \big) \big( \delta_{\alpha u} + \delta_{\alpha' u} \big) + \frac{\widetilde{f}_{\alpha \alpha'}(0)}{T} \right) \widetilde{G}'_{\alpha \alpha'}(0).
\end{aligned}
\end{equation}
The first term is again proportional to $\sum_{\alpha \alpha'} (-1)^{\alpha + \alpha'} G'_{\alpha \alpha'}(t)^p = 0$.
The second term would appear to be more problematic, since $\widetilde{f}_{\alpha \alpha'}(0)$ (for which we have not given an explicit expression) contributes at $O(1)$ due to $\widetilde{G}'_{\alpha \alpha'}(0)$ being $O(T)$.
However, we only need the component $\widetilde{f}_{--}(0)/T = \widetilde{F}_{--}(0)$, and from Eq.~\eqref{eq:SFF_functions_matrix_inverse_x_basis} we see that
\begin{equation} \label{eq:F_minus_component_evaluation}
\widetilde{F}_{--}(0) = \frac{1}{\textrm{Det} \widetilde{G}(0)} \widetilde{G}_{--}(0) = \frac{1}{\textrm{Det} \widetilde{G}(0)} \frac{\textrm{Det} \widetilde{G}(0) + \widetilde{G}_{+-}(0) \widetilde{G}_{-+}(0)}{\widetilde{G}_{++}(0)} = \frac{1 - 4J^2 q_{\textrm{EA}}^{p-2} \Lambda^2}{2Tq_{\textrm{EA}}} + O \left( \frac{1}{T^2} \right).
\end{equation}
Eq.~\eqref{eq:nonergodic_action_evaluation_middle_line_1} evaluates to
\begin{equation} \label{eq:nonergodic_action_evaluation_middle_line_2}
2pJ q_{\textrm{EA}}^{p/2-1} \Lambda \big( \mathcal{E}_{\textrm{aux}} - 2J q_{\textrm{EA}}^{p/2-1} \Lambda \big) - \frac{1}{2} + 2J^2 q_{\textrm{EA}}^{p-2} \Lambda^2 = \frac{1}{2} - 2J^2 q_{\textrm{EA}}^{p-2} \Lambda^2,
\end{equation}
again using Eq.~\eqref{eq:potential_GF_relationship}.

We finally have the large-$T$ limit of the action, given by the sum of Eqs.~\eqref{eq:nonergodic_action_evaluation_line_1},~\eqref{eq:nonergodic_action_evaluation_determinant_3}, and~\eqref{eq:nonergodic_action_evaluation_middle_line_2}:
\begin{equation} \label{eq:nonergodic_action_result}
S_{\textrm{eff}}[\mathcal{E}_{\textrm{aux}}] = \frac{p-2}{2p} + \frac{1}{8pJ^2 q_{\textrm{EA}}^{p-2} \Lambda^2} - \frac{2(p-1)J^2 q_{\textrm{EA}}^{p-2} \Lambda^2}{p} + \frac{1}{2} \log{4J^2 q_{\textrm{EA}}^{p-2} \Lambda^2}.
\end{equation}
Comparing to the complexity $\Sigma(\mathcal{E})$ given in Eq.~\eqref{eq:TAP_complexity_rewrite}, we see that $S_{\textrm{eff}}$ is precisely $-\Sigma(\mathcal{E}_{\textrm{aux}})$.

\subsection{Evaluation of the SFF} \label{subsec:final_SFF_evaluation}

We have shown that, at a given $\mathcal{E}_{\textrm{aux}}$ and for each value of inverse temperature $\beta_{\textrm{aux}}$, there is a solution to the SFF saddle point equations with $S_{\textrm{eff}} = -\Sigma(\mathcal{E}_{\textrm{aux}})$.
The full (connected) SFF is obtained by integrating over all $\mathcal{E}_{\textrm{aux}}$ and $\beta_{\textrm{aux}}$, as well as the symmetry-broken order parameter $\Delta$ (which contributes an overall factor of $T$) and an additional factor $2(1 + \delta_{p \textrm{ even}})$ from the discrete symmetries.
As in Sec.~\ref{sec:ergodic_ramp}, it is more convenient to integrate over the energy density $\epsilon(\mathcal{E}_{\textrm{aux}}, \beta_{\textrm{aux}})$.
We show in App.~\ref{sec:TAP_energy} that $\epsilon(\mathcal{E}, \beta)$ comes out to be precisely the argument of the filter function, Eq.~\eqref{eq:TAP_resolved_energy_density_definition}, when evaluated at the saddle point solution.

Our final result is that\footnote{The factor $\sqrt{pN/2\pi}$ comes from the integral over fluctuations in $\lambda$ --- the variance is $p/N$ (see Eq.~\eqref{eq:TAP_resolved_action_starting_point}), and the original fat unity introducing $\mathcal{E}_{\textrm{aux}}$ comes with a prefactor $N/2\pi$.}
\begin{equation} \label{eq:SFF_final_result}
\textrm{SFF}(T, f) = \big| \mathbb{E} \textrm{Tr} f(H) e^{-iHT} \big|^2 + 2 \big( 1 + \delta_{p \textrm{ even}} \big) T \sqrt{\frac{pN}{2\pi}} \int d\mathcal{E}_{\textrm{aux}} e^{N \Sigma(\mathcal{E}_{\textrm{aux}})} \int_{\epsilon_-(\mathcal{E}_{\textrm{aux}})}^{\epsilon_+(\mathcal{E}_{\textrm{aux}})} \frac{d\epsilon_{\textrm{aux}}}{2\pi} f(\epsilon_{\textrm{aux}})^2,
\end{equation}
where the inner integral runs only over the range $[\epsilon_-(\mathcal{E}_{\textrm{aux}}), \epsilon_+(\mathcal{E}_{\textrm{aux}})]$ in which solutions to the TAP equations exist.
Furthermore, one can easily generalize Eq.~\eqref{eq:SFF_final_result} by making the filter function $\mathcal{E}$-dependent, i.e., $f(\mathcal{E}, \epsilon_{\textrm{aux}})$.
The resulting quantity is the SFF for the projection of the system into certain TAP states.

Compare Eq.~\eqref{eq:SFF_final_result} for the non-ergodic phase to Eq.~\eqref{eq:SFF_connected_contribution_ergodic_phase} for the ergodic phase, and recall the discussion of block-diagonal Hamiltonians in Sec.~\ref{subsec:review_spectral_form_factor}.
Our result demonstrates that each metastable (i.e., TAP) state can be thought of its own quantum chaotic subspace, one which is independent of any others.
This is the central result of our paper.
While the qualitative idea has been proposed in previous work~\cite{Baldwin2017Clustering}, the present analysis both makes it precise and proves it.

\section{Higher moments of the evolution operator}
\label{sec:HigherMoments}

In this final section we consider higher moments of $\textrm{Tr} e^{-iHT}$, i.e., the quantities
\begin{equation} \label{eq:higher_SFF_definition}
\textrm{SFF}^{(n)}(T, f) \equiv \mathbb{E} \Big[ \Big( \textrm{Tr} f(H) e^{-iHT} \Big)^n \Big( \textrm{Tr} f(H) e^{iHT} \Big)^n \Big].
\end{equation}
The saddle points of these higher moments exhibit an interesting structure that will shed further light on the distribution of TAP states, although care must be taken in interpreting the results.
We first present the calculation and discuss afterwards.

\subsection{Effective action} \label{subsec:higher_moment_effective_action}

The effective action governing the $n$'th moment is derived in exactly the same manner as in Sec.~\ref{sec:nonergodic_ramp}.
The only major difference is that now spins have a ``replica'' index $a \in \{1, \cdots, n\}$ in addition to a contour index $\alpha \in \{u, l\}$.
We also include a separate fat unity defining $\mathcal{E}_{\textrm{aux}, a}$ for each replica.
The result is (compare to Eqs.~\eqref{eq:TAP_resolved_SFF_path_integral} and~\eqref{eq:TAP_resolved_SFF_action})
\begin{equation} \label{eq:higher_moment_SFF_path_integral}
\textrm{SFF}^{(n)}(T, f) = \int d\mathcal{E}_{\textrm{aux}} d\lambda \mathcal{D}G \mathcal{D}F \prod_{a=1}^n f \big( \epsilon_{au}[\lambda, G] \big) f \big( \epsilon_{al}[\lambda, G] \big) e^{-NS_{\textrm{eff}}[\mathcal{E}_{\textrm{aux}}, \lambda, G, F]},
\end{equation}
\begin{equation} \label{eq:higher_moment_SFF_action}
\begin{aligned}
S_{\textrm{eff}}[\mathcal{E}_{\textrm{aux}}, \lambda, G, F] &= -i \sum_a \lambda_a \mathcal{E}_{\textrm{aux}, a} - \frac{i}{2} \int_0^T dt \sum_{a \alpha} (-1)^{\alpha} z_{a \alpha}(t) \\
&\quad + \sum_{aa'} \frac{\lambda_a \lambda_{a'}}{2p q[G_{aa}]^{p/2} q[G_{a'a'}]^{p/2}} \left( \frac{1}{T^2} \int_0^T dt dt' G_{au, a'u}(t, t') \right)^p \\
&\quad \quad + \sum_{a'} \frac{J \lambda_{a'}}{pq[G_{a'a'}]^{p/2}} \int_0^T dt \sum_{a \alpha} (-1)^{\alpha} \left( \frac{1}{T} \int_0^T dt' G_{a \alpha, a' u}(t, t') \right)^p \\
&\quad \quad \quad + \frac{1}{2} \int_0^T dt dt' \sum_{aa'} \sum_{\alpha \alpha'} (-1)^{\alpha + \alpha'} \left( \frac{J^2}{p} G_{a \alpha, a' \alpha'}(t, t')^p - F_{a \alpha, a' \alpha'}(t, t') G_{a \alpha, a' \alpha'}(t, t') \right) \\
&\quad \quad \quad \quad + \frac{1}{2} \log{\textrm{Det}} \Big[ i (-1)^{\alpha} \delta_{aa'} \delta_{\alpha \alpha'} \big( \mu \partial_t^2 + z_{a \alpha} \big) + (-1)^{\alpha + \alpha'} F_{a \alpha, a' \alpha'} \Big],
\end{aligned}
\end{equation}
with energy densities
\begin{equation} \label{eq:higher_moment_energy_density_definition}
\begin{aligned}
\epsilon_{a \alpha}[\lambda, G] &= -\frac{\mu}{2} \partial_t^2 G_{a \alpha, a \alpha}(0^+, 0) \\
&\qquad \qquad - \frac{iJ^2}{p} \int_0^T dt \sum_{a' \alpha'} (-1)^{\alpha'} G_{a \alpha, a' \alpha'}(t, 0)^p - i \sum_{a'} \frac{J \lambda_{a'}}{pq[G_{a'a'}]^{p/2}} \left( \frac{1}{T} \int_0^T dt G_{a \alpha, a' u}(t, 0) \right)^p.
\end{aligned}
\end{equation}
The saddle point equations are therefore
\begin{equation} \label{eq:higher_moment_EOM}
\begin{aligned}
i \big( \mu \partial_t^2 + z_a \big) G_{a \alpha, a' \alpha'}(t - t') + \int_0^T dt'' \sum_{a'' \alpha''} (-1)^{\alpha''} F_{a \alpha, a'' \alpha''}(t - t'') &G_{a'' \alpha'', a' \alpha'}(t'' - t') \\
&= (-1)^{\alpha} \delta_{aa'} \delta_{\alpha \alpha'} \delta(t - t'),
\end{aligned}
\end{equation}
\begin{equation} \label{eq:higher_moment_self_energy_equation}
F_{a \alpha, a' \alpha'}(t) = J^2 G_{a \alpha, a' \alpha'}(t)^{p-1} + \frac{J}{T} \left( \frac{\lambda_a}{q[G_{aa}]^{p/2}} \delta_{\alpha u} + \frac{\lambda_{a'}}{q[G_{a'a'}]^{p/2}} \delta_{\alpha' u} \right) \left( \frac{\widetilde{G}_{a \alpha, a' \alpha'}(0)}{T} \right)^{p-1} + O(T^{-2}),
\end{equation}
\begin{equation} \label{eq:higher_moment_potential_equation}
\mathcal{E}_{\textrm{aux}, a} = -\frac{iJT}{pq[G_{aa}]^{p/2}} \sum_{a' \alpha'} (-1)^{\alpha'} \left( \frac{\widetilde{G}_{au, a' \alpha'}(0)}{T} \right)^p - \frac{i}{pq[G_{aa}]^{p/2}} \sum_{a'} \frac{\lambda_{a'}}{q[G_{a'a'}]^{p/2}} \left( \frac{\widetilde{G}_{au, a'u}(0)}{T} \right)^p.
\end{equation}

Note that Eqs.~\eqref{eq:higher_moment_EOM} through~\eqref{eq:higher_moment_potential_equation} have the following permutation symmetry with respect to replica indices.
Suppose that $G$, $F$, and $\lambda$ constitute a valid solution.
For any permutation $\pi$ of the set $\{1, \cdots, n\}$, define $\pi_{\alpha}(a)$ to be the permuted element $\pi(a)$ if $\alpha = l$ but simply the original element $a$ if $\alpha = u$.
Then the quantities $\overline{G}$, $\overline{F}$, and $\overline{\lambda}$ defined by
\begin{equation} \label{eq:higher_moment_permutation_symmetry}
\overline{G}_{a \alpha, a' \alpha'}(t, t') \equiv G_{\pi_{\alpha}(a) \alpha, \pi_{\alpha'}(a') \alpha'}(t, t'), \qquad \overline{F}_{a \alpha, a' \alpha'}(t, t') \equiv F_{\pi_{\alpha}(a) \alpha, \pi_{\alpha'}(a') \alpha'}(t, t'), \qquad \overline{\lambda}_a = \lambda_a,
\end{equation}
constitute an equally valid solution.
This symmetry has a nice graphical interpretation in terms of pairings between upper and lower contours, illustrated in Fig.~\ref{fig:multi_contour}: however contour $au$ is correlated with $a'l$ in a given solution, there is an alternate solution in which $au$ has the same correlation with $\pi(a')l$.

One trivial solution to the saddle point equations is to use the solution from Sec.~\ref{sec:nonergodic_ramp} for $a = a'$ while setting all cross-replica elements to zero.
The action then decomposes into a sum of single-replica actions, which we evaluated in Sec.~\ref{sec:nonergodic_ramp}.
In other words, this contribution to the $n$'th moment is simply $\textrm{SFF}(T, f)^n$.
However, by the permutation symmetry described above, we actually have $n!$ such contributions:
\begin{equation} \label{eq:higher_moment_disconnected_part}
\textrm{SFF}^{(n)}(T, f) = n! \cdot \textrm{SFF}(T, f)^n + \cdots,
\end{equation}
where the ellipses denote additional solutions.

\subsection{Connected solutions} \label{subsec:higher_moment_connected_solutions}

\begin{figure}
    \centering
    \includegraphics[width=.9\textwidth]{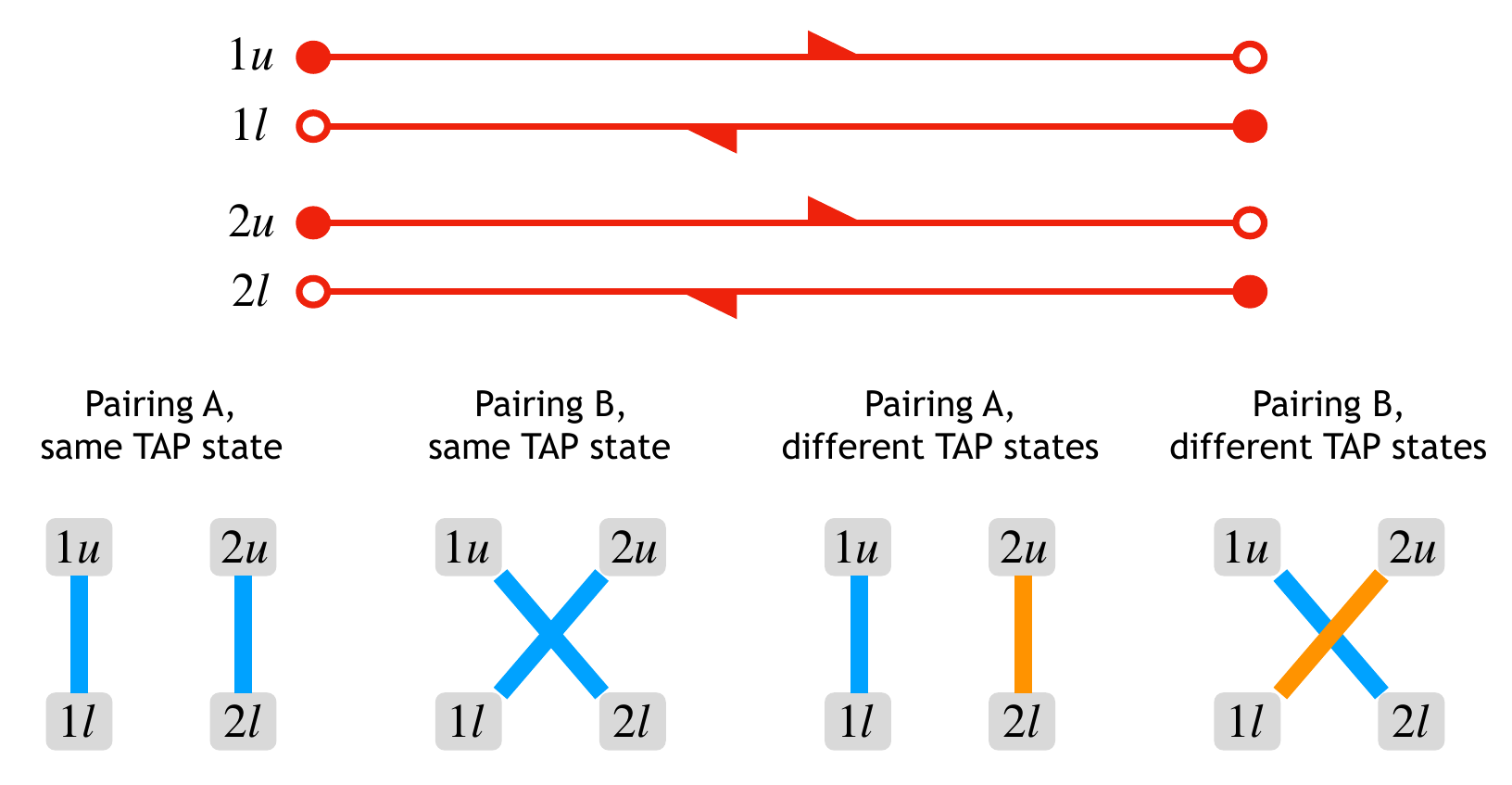}
    \caption{Graphical representation of the various saddle point solutions for the $n=2$ moment. The four contours --- $1u$, $1l$, $2u$, $2l$ --- are shown at the top. Below are the four varieties of solutions: each upper contour must be paired with a lower contour, but one is free to choose which replicas are paired, and there is further freedom in which TAP state each pair lies within (blue and orange lines indicate two different TAP states).}
    \label{fig:multi_contour}
\end{figure}

In general, for arbitrary values of $\mathcal{E}_{\textrm{aux}, a}$, we have been unable to find any further saddle points.
However, when some replicas have equal values of $\mathcal{E}_{\textrm{aux}}$, we can construct additional solutions.
Pick any set of inverse temperatures $\beta_a$ (not necessarily equal), and suppose that the replicas $\{1, \cdots, n\}$ partition into groups $A \equiv \{a_1, \cdots, a_{|A|}\}$, such that $\mathcal{E}_{\textrm{aux}, a}$ equals a common value $\mathcal{E}_{\textrm{aux}, A}$ for all $a \in A$.
We again take $G_{a \alpha, a \alpha'}(t - t')$ to be the solution from Sec.~\ref{sec:nonergodic_ramp}.
For $a$ and $a'$ in different groups, we still set $G_{a \alpha, a' \alpha'} = 0$.
For $a$ and $a'$ in the same group $A$, however, we now set
\begin{equation} \label{eq:higher_moment_cross_factor_G_ansatz}
G_{a \alpha, a' \alpha'}(t - t') = \big( q_{\textrm{EA}, a} q_{\textrm{EA}, a'} \big)^{1/2},
\end{equation}
where $q_{\textrm{EA}, a}$ is the Edwards-Anderson order parameter corresponding to $\mathcal{E}_{\textrm{aux}, A}$ and $\beta_a$.
This corresponds to the replicas lying within the same TAP state (see Fig.~\ref{fig:multi_contour}).
We can write this compactly as
\begin{equation} \label{eq:higher_moment_total_G_ansatz}
G_{a \alpha, a' \alpha'}(t) = \big( q_{\textrm{EA}, a} q_{\textrm{EA}, a'} \big)^{1/2} + \delta_{aa'} \Big( \Delta \mathcal{G}_{a, \alpha \alpha'}(t) + O(T^{-1}) \Big).
\end{equation}

Inserting into Eq.~\eqref{eq:higher_moment_potential_equation}, we have that $\lambda_a$ must obey
\begin{equation} \label{eq:higher_moment_lambda_determination_v1}
\sum_{a' \in A} \lambda_{a'} = ip \big( \mathcal{E}_{\textrm{aux}, A} - 2Jq_{\textrm{EA}, a}^{p/2-1} \Lambda_a \big) + O(T^{-1}).
\end{equation}
Note that, by virtue of Eq.~\eqref{eq:Keldysh_TAP_overlap_closed_equation}, $J q_{\textrm{EA}, a}^{p/2-1} \Lambda_a$ is a function solely of $\mathcal{E}_{\textrm{aux}, A}$.
Thus Eq.~\eqref{eq:higher_moment_lambda_determination_v1} is consistent among all $a \in A$.
The self-energy is then given by
\begin{equation} \label{eq:higher_moment_F_determination}
F_{a \alpha, a' \alpha'}(t) = J^2 \big( q_{\textrm{EA}, a} q_{\textrm{EA}, a'} \big)^{\frac{p-1}{2}} + \delta_{aa'} \Delta \mathcal{F}_{a, \alpha \alpha'}(t) + \frac{J}{T} \left( \sqrt{\frac{q_{\textrm{EA}, a'}^{p-1}}{q_{\textrm{EA}, a}}} \lambda_a \delta_{\alpha u} + \sqrt{\frac{q_{\textrm{EA}, a}^{p-1}}{q_{\textrm{EA}, a'}}} \lambda_{a'} \delta_{\alpha' u} \right) + O(T^{-1}).
\end{equation}
It remains only to check that Eq.~\eqref{eq:higher_moment_EOM} can be satisfied.
It is automatically solved at non-zero frequencies, since then $\widetilde{G}(\omega)$ and $\widetilde{F}(\omega)$ reduce to $\delta_{aa'} \Delta \widetilde{\mathcal{G}}(\omega)$ and $\delta_{aa'} \Delta \widetilde{\mathcal{F}}(\omega)$ respectively.
At zero frequency we confirm that the equation is solved to $O(T)$ (the $O(1)$ terms only determine subleading corrections).
Following the same steps as in Sec.~\ref{subsec:TAP_resolved_connected_solutions}, the left-hand side of Eq.~\eqref{eq:higher_moment_EOM} simplifies to
\begin{equation} \label{eq:higher_moment_G_determination}
JT \sqrt{q_{\textrm{EA}, a}^{p-1} q_{\textrm{EA}, a'}} \left( \sum_{a'' \in A} \lambda_{a''} + 2i(p-1) Jq_{\textrm{EA}, a}^{p/2-1} \Lambda_a + 2iJq_{\textrm{EA}, a'}^{p/2-1} \Lambda_{a'} - ip \mathcal{E}_{\textrm{aux}, A} \right) = 0,
\end{equation}
as desired.

Note that in this solution, only the sum $\sum_a \lambda_a$ is determined --- all orthogonal components of the vector $\lambda$ are free to take any values.
This does not imply that there are multiple such solutions, however.
Returning to the effective action in Eq.~\eqref{eq:higher_moment_SFF_action}, the fact that the saddle point equations determine only $G$, $F$, and $\sum_a \lambda_a$ means that, if we first integrate over them, the resulting $\lambda$-dependent action is of the form
\begin{equation} \label{eq:higher_moment_order_integrated_action}
S_{\textrm{eff}}[\mathcal{E}_{\textrm{aux}}, \lambda] = S \left[ \mathcal{E}_{\textrm{aux}}, \sum_a \lambda_a \right] - i \sum_{aa'} \sum_{b=2}^{|A|} \lambda_a u_{ab} u_{a'b} \mathcal{E}_{\textrm{aux}, a'},
\end{equation}
for some function $S$ of the single quantity $\sum_a \lambda_a$ (as well as all $\mathcal{E}_{\textrm{aux}}$) and for any choice of orthonormal basis vectors $u_{ab}$ orthogonal to the all-1 vector.
When we integrate over $\sum_a \lambda_a u_{ab}$, we thus get a $\delta$-function forcing $\sum_{a'} u_{a'b} \mathcal{E}_{\textrm{aux}, a'} = 0$.
Together, the $\delta$-functions force all $\mathcal{E}_{\textrm{aux}, a}$ to equal a common value $\mathcal{E}_{\textrm{aux}, A}$.
Not only is this consistent with our original assumption, it shows that our construction cannot work for any other values of $\mathcal{E}_{\textrm{aux}, a}$.

\subsection{Contribution of connected solutions} \label{subsec:higher_moment_connected_contribution}

To evaluate the action, note first of all that since the numbers $\beta_a$ define a continuous family of solutions, and since the action is by definition stationary at these solutions, all choices of $\beta_a$ must give the same value of the action.
We thus take all $\beta_a$ to equal a common value $\beta$ for simplicity.
The action evaluated at this solution still decomposes into a sum over groups, but now the contribution of a single group $A$ is
\begin{equation} \label{eq:higher_moment_action_evaluation_v1}
\begin{aligned}
S_{\textrm{eff}}[\mathcal{E}_{\textrm{aux}}] &= -i \mathcal{E}_{\textrm{aux}} \sum_{a \in A} \lambda_a + \frac{1}{2p} \left( \sum_{a \in A} \lambda_a \right)^2 + 2iJ q_{\textrm{EA}}^{p/2-1} \Lambda \sum_{a \in A} \lambda_a \\
&\qquad + \frac{T}{2} \sum_{aa'} \int_0^T dt \sum_{\alpha \alpha'} (-1)^{\alpha + \alpha'} \left( \frac{J^2}{p} G_{a \alpha, a \alpha'}(t)^p - F_{a \alpha, a \alpha'}(t) G_{a \alpha, a \alpha'}(t) \right) \\
&\qquad \qquad + \frac{1}{2} \log{\textrm{Det}} \Big[ i (-1)^{\alpha} \delta_{aa'} \delta_{\alpha \alpha'} \big( \mu \partial_t^2 + z \big) + (-1)^{\alpha + \alpha'} F_{a \alpha, a' \alpha'} \Big].
\end{aligned}
\end{equation}
Note that now $\mathcal{E}_{\textrm{aux}}$, $q_{\textrm{EA}}$, and $\Lambda$ are all independent of the replica $a$ (within a given group $A$).
We are also free to set all $\lambda_a = \lambda$, meaning that our saddle point solution simplifies to (in frequency space)
\begin{equation} \label{eq:higher_moment_simple_G_expression}
\widetilde{G}_{a \alpha, a' \alpha'}(\omega) = Tq_{\textrm{EA}} \delta_{\omega 0} + \delta_{aa'} \Big( \Delta \widetilde{\mathcal{G}}_{\alpha \alpha'}(\omega) + O(T^{-1}) \Big),
\end{equation}
\begin{equation} \label{eq:higher_moment_simple_F_expression}
\widetilde{F}_{a \alpha, a' \alpha'}(\omega) = \left( TJ^2 q_{\textrm{EA}}^{p-1} + \frac{ipJq_{\textrm{EA}}^{p/2-1} \big( \mathcal{E}_{\textrm{aux}} - 2Jq_{\textrm{EA}}^{p/2-1} \Lambda \big)}{|A|} \big( \delta_{\alpha u} + \delta_{\alpha' u} \big) \right) \delta_{\omega 0} + \delta_{aa'} \Delta \widetilde{\mathcal{F}}_{\alpha \alpha'}(\omega) + O(T^{-1}),
\end{equation}
\begin{equation} \label{eq:higher_moment_simple_lambda_expression}
\lambda = \frac{ip}{|A|} \big( \mathcal{E}_{\textrm{aux}} - 2Jq_{\textrm{EA}}^{p/2-1} \Lambda \big) + O(T^{-1}).
\end{equation}

Eq.~\eqref{eq:higher_moment_action_evaluation_v1} can be evaluated following the same procedure as in Sec.~\ref{subsec:nonergodic_contribution_connected_solutions}.
Directly substituting Eqs.~\eqref{eq:higher_moment_simple_G_expression} through~\eqref{eq:higher_moment_simple_lambda_expression} gives
\begin{equation} \label{eq:higher_moment_action_evaluation_v2}
S_{\textrm{eff}}[\mathcal{E}_{\textrm{aux}}] = \frac{p \mathcal{E}_{\textrm{aux}}^2}{2} - 2pJ^2 q_{\textrm{EA}}^{p-2} \Lambda^2 - \frac{Tq_{\textrm{EA}}}{2} \sum_{aa'} \sum_{\alpha \alpha'} (-1)^{\alpha + \alpha'} \widetilde{F}_{a \alpha, a' \alpha'}(0) - \frac{1}{2} \sum_{\omega} \log{\textrm{Det}} \widetilde{G}_{a \alpha, a' \alpha'}(\omega),
\end{equation}
and we again must determine certain components of $\widetilde{F}(0)$ and $\textrm{Det} \widetilde{G}(0)$.
As before, it is expedient to use the $(+,-)$ basis with respect to contour indices.
We also switch to the Fourier basis with respect to factor indices: from Eq.~\eqref{eq:higher_moment_simple_G_expression},
\begin{equation} \label{eq:higher_moment_G_Fourier_basis}
\begin{aligned}
\widetilde{G}_{b \alpha, b' \alpha'}(\omega) &\equiv \frac{1}{|A|} \sum_{aa'=1}^{|A|} e^{2\pi i(ab - a'b')/|A|} \widetilde{G}_{a \alpha, a' \alpha'}(\omega) \\
&= T|A|q_{\textrm{EA}} \delta_{b0} \delta_{b'0} \delta_{\omega 0} + \delta_{bb'} \Big( \Delta \widetilde{\mathcal{G}}_{\alpha \alpha'}(\omega) + O(T^{-1}) \Big).
\end{aligned}
\end{equation}
Thus $\textrm{Det} \widetilde{G}(\omega)$ factors with respect to $b$, and furthermore, $\sum_{\omega} \log{\textrm{Det} \widetilde{G}_b(\omega)} \sim 0$ for all $b \neq 0$ as in Secs.~\ref{subsec:ergodic_connected_action} and~\ref{subsec:nonergodic_contribution_connected_solutions}.
For $b = 0$, the determinant is calculated by comparing to the $b = 0$ block of $iz(-1)^{\alpha} + (-1)^{\alpha + \alpha'} \widetilde{F}(0)$, written in the $(+,-)$ basis (compare to Eq.~\eqref{eq:SFF_functions_matrix_inverse_x_basis}):
\begin{equation} \label{eq:higher_moment_matrix_inverse_x_basis}
\begin{pmatrix} \widetilde{F}_{0-, 0-}(0) & iz + \widetilde{F}_{0-, 0+}(0) \\ iz + \widetilde{F}_{0+, 0-}(0) & \widetilde{F}_{0+, 0+}(0) \end{pmatrix} = \frac{1}{\textrm{Det} \widetilde{G}_{0}(0)} \begin{pmatrix} \widetilde{G}_{0-, 0-}(0) & -\widetilde{G}_{0+, 0-}(0) \\ -\widetilde{G}_{0-, 0+}(0) & \widetilde{G}_{0+, 0+}(0) \end{pmatrix}.
\end{equation}
We see that $\textrm{Det} \widetilde{G}_0(0) = \widetilde{G}_{0+,0+}(0)/\widetilde{F}_{0+,0+}(0) \sim 1/J^2 q_{\textrm{EA}}^{p-2}$, and $\widetilde{F}_{0-,0-}(0)$ (which is in fact the only element of $\widetilde{F}(0)$ needed in Eq.~\eqref{eq:higher_moment_action_evaluation_v2}) is given by $\widetilde{G}_{0-,0-}(0)/\textrm{Det} \widetilde{G}_0(0) \sim (1 - 4J^2 q_{\textrm{EA}}^{p-2} \Lambda^2)/2T|A|q_{\textrm{EA}}$.
The action evaluates to
\begin{equation} \label{eq:higher_moment_action_evaluation_v3}
S_{\textrm{eff}}[\mathcal{E}_{\textrm{aux}}] = \frac{p-2}{2p} + \frac{1}{8p J^2 q_{\textrm{EA}}^{p-2} \Lambda^2} - \frac{2(p-1) J^2 q_{\textrm{EA}}^{p-2} \Lambda^2}{p} + \frac{1}{2} \log{4J^2 q_{\textrm{EA}}^{p-2} \Lambda^2},
\end{equation}
which is again precisely $-\Sigma(\mathcal{E}_{\textrm{aux}})$.

\subsection{Evaluation of the Higher SFF} \label{subsec:higher_moment_evaluation_SFF}

In the above calculation, note that we get a single contribution of complexity for the \textit{entire} group $A$.
However, there is still a factor $(2T)^{|A|} (1 + \delta_{p \textrm{ even}})^{2|A|-1}$ due to the separate time translation, time reversal, and reflection symmetries of each replica\footnote{The contribution $(1 + \delta_{p \textrm{ even}})^{2|A|-1}$, rather than $(1 + \delta_{p \textrm{ even}})^{2|A|}$, is because reflecting all spin configurations does not change the values of any overlaps.}.
Finally, the sum over all connected solutions amounts to a sum over the possible ways of partitioning $n$ elements, in addition to the $n!$ ways of pairing upper and lower contours.
Using $P \equiv \{A_1, \cdots, A_{|P|}\}$ to denote a partition, we have that
\begin{equation} \label{eq:higher_moment_final_expression}
\begin{aligned}
\textrm{SFF}^{(n)}(T, f) = n! \sum_P &\prod_{A \in P} 2^{|A|} \big( 1 + \delta_{p \textrm{ even}} \big)^{2|A| - 1} T^{|A|} \sqrt{\frac{pN}{2\pi}} \int d\mathcal{E}_{\textrm{aux}, A} e^{N \Sigma(\mathcal{E}_{\textrm{aux}, A})} \\
&\qquad \qquad \cdot \prod_{a \in A} \int_{\epsilon_-(\mathcal{E}_{\textrm{aux}, A})}^{\epsilon_+(\mathcal{E}_{\textrm{aux}, A})} \frac{\textrm{d}\epsilon_{\textrm{aux}, a}}{2\pi} f(\epsilon_{\textrm{aux}, a})^2.
\end{aligned}
\end{equation}
In particular, suppose the filter function is chosen so as to have a small width $\Delta \mathcal{E} \ll 1/N$ around a certain value $\mathcal{E}$ (as in Sec.~\ref{subsec:final_SFF_evaluation}, the above calculation can easily be modified to allow for $\mathcal{E}$-dependent filter functions).
Then the $n$'th moment simplifies to
\begin{equation} \label{eq:higher_moment_special_final_expression}
\textrm{SFF}^{(n)}(T, f) = n! \sum_P \left( \big( 1 + \delta_{p \textrm{ even}} \big)^{-1} \sqrt{\frac{pN}{2\pi}} e^{N \Sigma(\mathcal{E})} \Delta \mathcal{E} \right)^{|P|} \left( 2 \big( 1 + \delta_{p \textrm{ even}} \big)^2 T \int_{\epsilon_-(\mathcal{E})}^{\epsilon_+(\mathcal{E})} \frac{\textrm{d}\epsilon_{\textrm{aux}}}{2\pi} f(\epsilon_{\textrm{aux}})^2 \right)^n.
\end{equation}

Eq.~\eqref{eq:higher_moment_special_final_expression} has a nice interpretation as the $n$'th moment of a sum over a Poisson-distributed number of Gaussians.
To be precise, suppose we have an infinite sequence of i.i.d.\ complex Gaussians, $\{Z_i\}_{i=1}^{\infty}$, each with $\mathbb{E} Z_i = 0$ and $\mathbb{E} Z_i Z_i^* = \sigma^2$.
Consider the sum $S \equiv \sum_{i=1}^M Z_i$, where $M$ is itself a Poisson-distributed random variable with mean $\mu$.
The $n$'th moment of $SS^*$, averaging over both Gaussians and $M$, can be written
\begin{equation} \label{eq:Poisson_distributed_sum_moment_v1}
\mathbb{E} \big[ S^n S^{*n} \big] = \sum_{m=0}^{\infty} p_{\mu}(m) \mathbb{E} \left[ \left( \sum_{i=1}^m Z_i \right)^n \left( \sum_{i=1}^m Z_i^* \right)^n \right] = n! \sum_{m=0}^{\infty} p_{\mu}(m) \big( m \sigma^2 \big)^n
\end{equation}
where $p_{\mu}(m)$ denotes the Poisson distribution of mean $\mu$, and Wick's theorem is used for the latter equality.
It is known that the $n$'th moment of a Poisson distribution is $\sum_P \mu^{|P|}$, where the sum is again over all partitions of $n$ elements.
Thus
\begin{equation} \label{eq:Poisson_distributed_sum_moment_v2}
\mathbb{E} \big[ S^n S^{*n} \big] = n! \sum_P \mu^{|P|} \sigma^{2n}.
\end{equation}
If we associate $\sigma^2$ with the SFF of a single TAP state at $\mathcal{E}$,
\begin{equation} \label{eq:single_TAP_state_SFF}
\sigma^2 = 2 \big( 1 + \delta_{p \textrm{ even}} \big)^2 T \int_{\epsilon_-(\mathcal{E})}^{\epsilon_+(\mathcal{E})} \frac{\textrm{d}\epsilon_{\textrm{aux}}}{2\pi} f(\epsilon_{\textrm{aux}})^2,
\end{equation}
and associate $\mu$ with the number of TAP states,
\begin{equation} \label{eq:number_TAP_states}
\mu = \big( 1 + \delta_{p \textrm{ even}} \big)^{-1} \sqrt{\frac{pN}{2\pi}} e^{N \Sigma(\mathcal{E})} \Delta \mathcal{E},
\end{equation}
then Eqs.~\eqref{eq:higher_moment_special_final_expression} and~\eqref{eq:Poisson_distributed_sum_moment_v2} are identical.

It is quite tempting to interpret this as saying that the number of TAP states at $\mathcal{E}$ is Poisson-distributed with mean given by Eq.~\eqref{eq:number_TAP_states}, and that each TAP state has a Gaussian-distributed value of $\textrm{Tr} e^{-iHT}$ with variance (i.e., SFF) given by Eq.~\eqref{eq:single_TAP_state_SFF}.
We are not aware of any results in the literature which would contradict such a claim.
However, keep in mind that $\textrm{SFF}^{(n)}$ has perturbative corrections around each saddle point which are suppressed by powers of $N$, whereas every connected partition in Eq.~\eqref{eq:higher_moment_special_final_expression} is suppressed \textit{exponentially} relative to the fully disconnected one, whose contribution is given by Eq.~\eqref{eq:higher_moment_disconnected_part}\footnote{
At sufficiently low energies, where the function $\Sigma(\mathcal{E})$ is negative, the situation is reversed and the fully connected partition dominates.
The issue remains, however, that we do not calculate perturbative corrections around the dominant saddle point.
Furthermore, the relevance of these moment calculations to \textit{individual} realizations of the PSM is much more suspect when $\Sigma(\mathcal{E}) < 0$.
}.
Thus we cannot claim to have rigorously computed the $n$'th moment to any level of accuracy beyond the disconnected piece.
Nonetheless, the structure of saddle points which we have identified is highly suggestive and warrants further investigation.

\section{Concluding Remarks} \label{sec:conclusion}

The focus of this paper was the derivation of Eq.~\eqref{eq:SFF_final_result}, which gives the connected SFF of the quantum PSM in the non-ergodic phase.
Our result demonstrates that each metastable state (i.e., TAP state) can be considered as an independent chaotic phase with an independent RMT-like Hamiltonian, at least as far as level statistics are concerned.

It is interesting to compare our result for the PSM with Ref.~\cite{saad2019semiclassical}, which performs a similar calculation for the SYK model (the latter is structurally quite similar but with fermionic degrees of freedom).
In the ergodic phase, we find an essentially identical result using extremely similar methods.
Below the dynamical transition, however, the PSM displays an enhanced ramp quite different from that of the SYK model, which has no analogous phase.

Since we only calculate the SFF up to times polynomial in system size, our results are consistent with but do not test the distinction between localized and delocalized phases shown in Fig.~\ref{fig:spin_glass_cartoons}, which is only relevant beyond the exponentially long timescale corresponding to tunneling between TAP states.
We leave it for future work to incorporate such instanton effects into the path integral, expecting that they will reduce the SFF to the random matrix result precisely in the non-ergodic delocalized phase (and even then only beyond the exponential tunneling timescale).
At the same time, consideration of exponential scales can allow one to identify a plethora of additional dynamical phases (see Ref.~\cite{Zhao2014Three} for an example), and so the structure of instantons in these spin glass models may be quite rich.

In addition to the SFF, we have also considered higher moments of the evolution operator.
We identified an important family of saddle points and evaluated its contribution to these higher SFFs (Eq.~\eqref{eq:higher_moment_special_final_expression}).
The results suggest that: i) the number of TAP states at a given energy is Poisson-distributed, and ii) the numbers of TAP states at different energies are independent.
However, since we have not evaluated the perturbative corrections around each saddle point, which at finite complexity would generically dominate over any subleading saddle points, we cannot claim to have an accurate calculation.
It is another direction for future work to study the distribution of TAP states more systematically.

Our results can be further understood by comparing to Refs.~\cite{PhysRevE.102.060202} and~\cite{winer2020hydrodynamic} on one hand and Refs.~\cite{2001} and~\cite{Facoetti_2019} on the other.
The first set of papers argues that for a system which separates into weakly coupled sectors, the SFF enhancement is the sum of return probabilities over all configurations.
If the time evolution can be considered as an effective Markov process with transfer rates between sectors given by some matrix $M$, then the SFF enhancement factor is $\textrm{Tr} e^{MT}$.
The second set of papers argues that for a \textit{classical} spin glass undergoing Markovian stochastic dynamics with generator $M$, the number of TAP states can be calculated --- and perhaps even defined --- as $\textrm{Tr} e^{MT}$.
In this sense, the present paper can be considered as a ``missing link'' that extends the results of Refs.~\cite{2001} and~\cite{Facoetti_2019} to quantum systems.

The fact that SFF enhancement is related to return probabilities suggests that the spectral statistics of spin glasses may contain information on aging dynamics as well.
Another open question is whether the \textit{equilibrium} replica-symmetry-breaking transition has any consequences for spectral statistics.
These, as well as those already mentioned, are all promising directions for future work.

Finally, let us briefly comment on the case $p=2$, which --- being a Gaussian model but for the global spherical constraint (and still integrable in any case~\cite{Cugliandolo_2018}) --- exhibits very different behavior than the $p > 2$ models considered here.
The spectral statistics of the analogous SYK model with two-point interactions among fermions have been studied in the mean-field limit \cite{Winer_2020,Liao_2020}, and found to have ramps growing exponentially in time.
One explanation for these ramps is the spontaneous breaking of a hidden $SU(2)^k$ symmetry in the saddle point equations, since for $p=2$ the matrix $G$ has a separate $SU(2)$ conjugation symmetry at each frequency.
Inspections show that the $p=2$ spherical model has the same symmetry at tree level, although it is broken by higher loop effects.
A full analysis of this system would be yet another excellent topic for further research.

\section*{Acknowledgements}

This work was supported by the following: the U.S. Department of Energy, Office of Science, Basic Energy Sciences under award number DE-SC0001911 (V.G.); the Joint Quantum Institute (M.W.); the Air Force Office of Scientific Research under award numbers FA9550-17-1-0180 (M.W.) and FA9550-19-1-0360 (B.S.); the U.S. Department of Energy, Office of Science, Office of Advanced Scientific Computing Research, Accelerated Research for Quantum Computing program ``FAR-QC'' (R.B.); the DoE ASCR Quantum Testbed Pathfinder program under award number DE-SC0019040 (C.L.B.); the DoE ASCR Accelerated Research in Quantum Computing program under award number DE-SC0020312 (C.L.B.); the DoE QSA, AFOSR, AFOSR MURI, NSF PFCQC program, NSF QLCI under award number OMA-2120757 (C.L.B.); DoE award number DE-SC0019449 (C.L.B.), ARO MURI, and DARPA SAVaNT ADVENT (C.L.B.).
 This material is based upon work supported by the National Science Foundation Graduate Research Fellowship Program under Grant No. DGE 1840340 (R.B.), and by the National Science Foundation NRC postdoctoral fellowship program (C.L.B.).

\printbibliography

\appendix

\section{Derivation of Schwinger-Keldysh TAP equations} \label{sec:TAP_derivation}

Here we derive the TAP equations on the Schwinger-Keldysh contour, given by Eqs.~\eqref{eq:Keldysh_TAP_EOM} and~\eqref{eq:Keldysh_TAP_magnetization_equation} of the main text.
Our derivation is a straightforward generalization of that in Ref.~\cite{Biroli_2001} for the thermal circle, but since we are not aware of it appearing in the literature, we aim to make this section as self-contained as possible.

We begin with the expression for the TAP action, given by Eq.~\eqref{eq:Keldysh_TAP_action} and reproduced here:
\begin{equation} \label{eq:Keldysh_TAP_action_reproduction}
\begin{aligned}
iNS_{\textrm{TAP}}[m, \mathcal{G}, \eta] &\equiv \log{\int \mathcal{D}\sigma^N \exp \left[ i \sum_i S_i^0 - i \eta \int_{\mathcal{C}} dt \sum_{(i_1 \cdots i_p)} J_{i_1 \cdots i_p} \sigma_{i_1}(t) \cdots \sigma_{i_p}(t) \right]} \\
&\qquad \qquad + \frac{iN}{2} \int_{\mathcal{C}} dt z(t) - i \int_{\mathcal{C}} dt \sum_i h_i(t) m_i(t) + \frac{iN}{2} \int_{\mathcal{C}} dt dt' \Lambda(t, t') \mathcal{G}(t, t'),
\end{aligned}
\end{equation}
where $\mathcal{C}$ denotes the Schwinger-Keldysh contour and
\begin{equation} \label{eq:Keldysh_TAP_noninteracting_action_reproduction}
S_i^0 \equiv \int_{\mathcal{C}} dt \left( \frac{\mu}{2} \big( \partial_t \sigma_i(t) \big)^2 - \frac{z(t)}{2} \sigma_i(t)^2 + h_i(t) \sigma_i(t) \right) - \frac{1}{2} \int_{\mathcal{C}} dt dt' \Lambda(t, t') \sigma_i(t) \sigma_i(t').
\end{equation}
Note that we have included an additional parameter $\eta$ governing the strength of interactions.
We calculate $S_{\textrm{TAP}}$ by expanding in $\eta$, ultimately setting $\eta = 1$.
One can show that all terms beyond second order vanish in the thermodynamic limit~\cite{Plefka1982Convergence,Biroli_2001}, so we need only calculate $S_{\textrm{TAP}}$ and its first two derivatives at $\eta = 0$.
Then
\begin{equation} \label{eq:TAP_action_reconstruction}
S_{\textrm{TAP}}[m, \mathcal{G}, 1] \sim S_{\textrm{TAP}}[m, \mathcal{G}, 0] + \frac{\partial S_{\textrm{TAP}}[m, \mathcal{G}, 0]}{\partial \eta} + \frac{1}{2} \frac{\partial^2 S_{\textrm{TAP}}[m, \mathcal{G}, 0]}{\partial \eta^2}.
\end{equation}
We proceed term by term.

\subsubsection*{Zeroth order}

At zeroth order in $\eta$, we have that
\begin{equation} \label{eq:TAP_action_zeroth_order_start}
\begin{aligned}
iNS_{\textrm{TAP}}[m, \mathcal{G}, 0] &= \sum_i \log{\int \mathcal{D}\sigma_i e^{iS_i^0}} \\
&\quad + \frac{iN}{2} \int_{\mathcal{C}} dt z(t) - i \int_{\mathcal{C}} dt \sum_i h_i(t) m_i(t) + \frac{iN}{2} \int_{\mathcal{C}} dt dt' \Lambda(t, t') \mathcal{G}(t, t').
\end{aligned}
\end{equation}
The first term evaluates to
\begin{equation} \label{eq:noninteracting_partition_evaluation}
\begin{aligned}
&\int \mathcal{D}\sigma_i \exp \left[ -\frac{i}{2} \int_{\mathcal{C}} dt dt' \sigma_i(t) \left( \delta(t - t') \big( \mu \partial_t^2 + z(t) \big) + \Lambda(t, t') \right) \sigma_i(t') + i \int_{\mathcal{C}} dt h_i(t) \sigma_i(t) \right] \\
&\qquad \qquad \qquad = \textrm{Det} \Big[ i (\mu \partial_t^2 + z) + i \Lambda \Big]^{-\frac{1}{2}} \exp \left[ -\frac{1}{2} \int_{\mathcal{C}} dt dt' h_i(t) \Big( i (\mu \partial_t^2 + z) + i \Lambda \Big)^{-1}(t, t') h_i(t') \right].
\end{aligned}
\end{equation}
We further have that
\begin{equation} \label{eq:TAP_noninteracting_correlations}
\big< \sigma_i(t) \sigma_i(t') \big> - \big< \sigma_i(t) \big> \big< \sigma_i(t') \big> = \Big( i (\mu \partial_t^2 + z) + i \Lambda \Big)^{-1}(t, t'),
\end{equation}
\begin{equation} \label{eq:TAP_noninteracting_magnetizations}
\big< \sigma_i(t) \big> = i \int_{\mathcal{C}} dt' \Big( i (\mu \partial_t^2 + z) + i \Lambda \Big)^{-1}(t, t') h_i(t').
\end{equation}
Note that the left-hand sides are constrained to be $\mathcal{G}(t, t') - Q(t, t')$ and $m_i(t)$ (summing over $i$ for the former), where $Q(t, t') \equiv N^{-1} \sum_i m_i(t) m_i(t')$.
Thus we can use Eqs.~\eqref{eq:TAP_noninteracting_correlations} and~\eqref{eq:TAP_noninteracting_magnetizations} to express the action in terms of them.
The first term of Eq.~\eqref{eq:Keldysh_TAP_noninteracting_action_reproduction} becomes
\begin{equation} \label{eq:TAP_noninteracting_evaluation_1}
\frac{N}{2} \log{\textrm{Det}} \Big[ \mathcal{G} - Q \Big] + \frac{N}{2} \int_{\mathcal{C}} dt dt' \Big( \mathcal{G} - Q \Big)^{-1}(t, t') Q(t, t'),
\end{equation}
and the second line can be written
\begin{equation} \label{eq:TAP_noninteracting_evaluation_2}
-N \int_{\mathcal{C}} dt dt' \Big( \mathcal{G} - Q \Big)^{-1}(t, t') Q(t, t') - \frac{iN \mu}{2} \int_{\mathcal{C}} dt \partial_t^2 \mathcal{G}(t, t') \Big|_{t' = t} + \frac{N}{2} \int_{\mathcal{C}} dt dt' \Big( \mathcal{G} - Q \Big)^{-1}(t, t') \mathcal{G}(t, t').
\end{equation}
Combining the two, we have that (up to an unimportant constant)
\begin{equation} \label{eq:TAP_noninteracting_action_result}
iS_{\textrm{TAP}}[m, \mathcal{G}, 0] = \frac{1}{2} \log{\textrm{Det}} \Big[ \mathcal{G} - Q \Big] - \frac{i \mu}{2} \int_{\mathcal{C}} dt \partial_t^2 \mathcal{G}(t, t') \Big|_{t' = t}.
\end{equation}

\subsubsection*{First order}

Strictly speaking, since $h_i(t)$ and $\Lambda(t, t')$ are chosen so that $\langle \sigma_i(t) \rangle = m_i(t)$ and $\sum_i \langle \sigma_i(t) \sigma_i(t') \rangle = N\mathcal{G}(t, t')$ at every value of $\eta$, they themselves are functions of $\eta$ (as is $z(t)$).
However, the Legendre-transform structure of the TAP action ensures that the total derivative with respect to $\eta$ equals the partial derivative holding fields fixed (for example, the contribution to $\partial S_{\textrm{TAP}} / \partial \eta$ from $\partial h_i(t) / \partial \eta$ comes with a factor $\langle \sigma_i(t) \rangle - m_i(t) = 0$).
Thus we have simply that
\begin{equation} \label{eq:TAP_first_order_evaluation_general}
\frac{\partial S_{\textrm{TAP}}[m, \mathcal{G}, \eta]}{\partial \eta} = -\frac{1}{N} \int_{\mathcal{C}} dt \sum_{(i_1 \cdots i_p)} J_{i_1 \cdots i_p} \big< \sigma_{i_1}(t) \cdots \sigma_{i_p}(t) \big>.
\end{equation}
At $\eta = 0$, the spins are non-interacting and the expectation value factors:
\begin{equation} \label{eq:TAP_first_order_evaluation_noninteracting}
\frac{\partial S_{\textrm{TAP}}[m, \mathcal{G}, 0]}{\partial \eta} = -\frac{1}{N} \int_{\mathcal{C}} dt \sum_{(i_1 \cdots i_p)} J_{i_1 \cdots i_p} m_{i_1}(t) \cdots m_{i_p}(t).
\end{equation}

\subsubsection*{Second order}

The second derivative requires a bit more work.
From Eq.~\eqref{eq:TAP_first_order_evaluation_general}, we have that
\begin{equation} \label{eq:TAP_second_derivative_starting_point}
\begin{aligned}
\frac{\partial^2 S_{\textrm{TAP}}[m, \mathcal{G}, \eta]}{\partial \eta^2} &= \frac{i}{N} \int_{\mathcal{C}} dt dt' \sum_{II'} J_I J_{I'} \Big( \big< \sigma_I(t) \sigma_{I'}(t') \big> - \big< \sigma_I(t) \big> \big< \sigma_{I'}(t') \big> \Big) \\
&\quad - \frac{i}{N} \int_{\mathcal{C}} dt dt' \sum_{Ii'} J_I \frac{\partial h_{i'}(t')}{\partial \eta} \Big( \big< \sigma_I(t) \sigma_{i'}(t') \big> - \big< \sigma_I(t) \big> \big< \sigma_{i'}(t') \big> \Big) \\
&\quad + \frac{i}{2N} \int_{\mathcal{C}} dt dt' \sum_{Ii'} J_I \frac{\partial z(t')}{\partial \eta} \Big( \big< \sigma_I(t) \sigma_{i'}(t')^2 \big> - \big< \sigma_I(t) \big> \big< \sigma_{i'}(t')^2 \big> \Big) \\
&\quad + \frac{i}{2N} \int_{\mathcal{C}} dt dt' dt'' \sum_{Ii'} J_I \frac{\partial \Lambda(t', t'')}{\partial \eta} \Big( \big< \sigma_I(t) \sigma_{i'}(t') \sigma_{i'}(t'') \big> - \big< \sigma_I(t) \big> \big< \sigma_{i'}(t') \sigma_{i'}(t'') \big> \Big),
\end{aligned}
\end{equation}
where for brevity, we use $I$ to denote $(i_1 \cdots i_p)$ and $\sigma_I(t)$ to denote $\sigma_{i_1}(t) \cdots \sigma_{i_p}(t)$.
To evaluate the partial derivatives on the right-hand side, which we only need at $\eta = 0$, note that
\begin{equation} \label{eq:TAP_Legendre_relations}
N \frac{\partial S_{\textrm{TAP}}[m, \mathcal{G}, \eta]}{\partial m_i(t)} = -h_i(t), \qquad \frac{\partial S_{\textrm{TAP}}[m, \mathcal{G}, \eta]}{\partial \rho(t)} = \frac{1}{2} z(t), \qquad \frac{\partial S_{\textrm{TAP}}[m, \mathcal{G}, \eta]}{\partial \mathcal{G}(t, t')} = \frac{1}{2} \Lambda(t, t'),
\end{equation}
where we set $\sum_i \langle \sigma_i(t)^2 \rangle = N \rho(t)$ (even though ultimately $\rho(t) = 1$).
Thus, using Eq.~\eqref{eq:TAP_first_order_evaluation_noninteracting},
\begin{equation} \label{eq:TAP_field_derivatives_evaluation}
\begin{gathered}
\frac{\partial h_i(t)}{\partial \eta} = -N \frac{\partial^2 S_{\textrm{TAP}}[m, \mathcal{G}, 0]}{\partial m_i(t) \partial \eta} = \sum_{(i_1 \cdots i_p)} J_{i_1 \cdots i_p} \frac{\partial \big( m_{i_1}(t) \cdots m_{i_p}(t) \big)}{\partial m_i(t)}, \\
\frac{\partial z(t)}{\partial \eta} = 2 \frac{\partial^2 S_{\textrm{TAP}}[m, \mathcal{G}, 0]}{\partial \rho(t) \partial \eta} = 0, \qquad \frac{\partial \Lambda(t, t')}{\partial \eta} = 2 \frac{\partial^2 S_{\textrm{TAP}}[m, \mathcal{G}, 0]}{\partial \mathcal{G}(t, t') \partial \eta} = 0.
\end{gathered}
\end{equation}
The second derivative simplifies to
\begin{equation} \label{eq:TAP_second_derivative_simplified}
\begin{aligned}
\frac{\partial^2 S_{\textrm{TAP}}[m, \mathcal{G}, \eta]}{\partial \eta^2} &= \frac{i}{N} \int_{\mathcal{C}} dt dt' \sum_{II'} J_I J_{I'} \Big( \big< \sigma_I(t) \sigma_{I'}(t') \big> - \big< \sigma_I(t) \big> \big< \sigma_{I'}(t') \big> \Big) \\
&\quad - \frac{i}{N} \int_{\mathcal{C}} dt dt' \sum_{II'i'} J_I J_{I'} \Big( \big< \sigma_I(t) \sigma_{i'}(t') \big> - \big< \sigma_I(t) \big> \big< \sigma_{i'}(t') \big> \Big) \frac{\partial \big< \sigma_{I'}(t') \big>}{\partial m_{i'}(t')}.
\end{aligned}
\end{equation}
Just as one can show that higher-order $\eta$ derivatives vanish in the thermodynamic limit, one can also show that it is safe to replace $J_I J_{I'}$ by its average in Eq.~\eqref{eq:TAP_second_derivative_simplified}~\cite{Plefka1982Convergence,Biroli_2001}.
Again using that expectation values factor at $\eta = 0$, we therefore have that
\begin{equation} \label{eq:TAP_second_order_evaluation}
\frac{\partial^2 S_{\textrm{TAP}}[m, \mathcal{G}, 0]}{\partial \eta^2} = \frac{iJ^2}{p} \int_{\mathcal{C}} dt dt' \bigg[ \mathcal{G}(t, t')^p - Q(t, t')^p - p \Big( \mathcal{G}(t, t') - Q(t, t') \Big) Q(t, t')^{p-1} \bigg].
\end{equation}

\subsubsection*{Full TAP action}

Inserting Eqs.~\eqref{eq:TAP_noninteracting_action_result},~\eqref{eq:TAP_first_order_evaluation_noninteracting}, and~\eqref{eq:TAP_second_order_evaluation} into Eq.~\eqref{eq:TAP_action_reconstruction}, we find that
\begin{equation} \label{eq:TAP_action_full_result}
\begin{aligned}
iS_{\textrm{TAP}} &= \frac{1}{2} \log{\textrm{Det}} \Big[ \mathcal{G} - Q \Big] - \frac{i \mu}{2} \int_{\mathcal{C}} dt \partial_t^2 \mathcal{G}(t, t') \Big|_{t' = t} - \frac{i}{N} \int_{\mathcal{C}} dt \sum_{(i_1 \cdots i_p)} J_{i_1 \cdots i_p} m_{i_1}(t) \cdots m_{i_p}(t) \\
&\qquad \qquad - \frac{J^2}{2p} \int_{\mathcal{C}} dt dt' \bigg[ \mathcal{G}(t, t')^p + (p-1) Q(t, t')^p - p \mathcal{G}(t, t') Q(t, t')^{p-1} \bigg].
\end{aligned}
\end{equation}
As discussed in the main text, the TAP equations come from setting to zero the derivatives with respect to $m_i(t)$ and $\mathcal{G}(t, t')$.
However, since $\mathcal{G}(t, t)$ is not free to vary (it must equal 1 by the spherical constraint), we include a Lagrange multiplier that we again denote by $z(t)$.
The TAP equations are thus
\begin{equation} \label{eq:original_TAP_correlation_equation}
\Big( \mathcal{G} - Q \Big)^{-1}(t, t') - i\mu \partial_t^2 \delta(t - t') - J^2 \big( \mathcal{G}(t, t')^{p-1} - Q(t, t')^{p-1} \big) = iz(t) \delta(t - t'),
\end{equation}
\begin{equation} \label{eq:original_TAP_magnetization_equation}
\begin{aligned}
\int_{\mathcal{C}} dt' \Big( \mathcal{G} - Q \Big)^{-1}(t, t') m_i(t') &+ i \sum_{(i_1 \cdots i_p)} J_{i_1 \cdots i_p} \frac{\partial \big( m_{i_1}(t) \cdots m_{i_p}(t) \big)}{\partial m_i(t)} \\
&- (p-1)J^2 \int_{\mathcal{C}} dt' \Big( \mathcal{G}(t, t') - Q(t, t') \Big) Q(t, t')^{p-2} m_i(t') = 0.
\end{aligned}
\end{equation}
Time translation invariance allows us to further simplify by setting $m_i(t) = m_i$ and $z(t) = z$.
Note that then $Q(t, t') = N^{-1} \sum_i m_i^2 \equiv q_{\textrm{EA}}$.
After some rearrangement, the TAP equations can be written
\begin{equation} \label{eq:final_TAP_correlation_equation}
i \big( \mu \partial_t^2 + z \big) \Big( \mathcal{G}(t, t') - q_{\textrm{EA}} \Big) + J^2 \int_{\mathcal{C}} dt'' \Big( \mathcal{G}(t, t')^{p-1} - q_{\textrm{EA}}^{p-1} \Big) \Big( \mathcal{G}(t'', t') - q_{\textrm{EA}} \Big) = \delta(t - t'),
\end{equation}
\begin{equation} \label{eq:final_TAP_magnetization_equation}
J^2 \int_{\mathcal{C}} dt' \Big( \mathcal{G}(t, t')^{p-1} - (p-1) q_{\textrm{EA}}^{p-2} \mathcal{G}(t, t') + (p-2) q_{\textrm{EA}}^{p-1} \Big) m_i = -izm_i - i \sum_{(i_1 \cdots i_p)} J_{i_1 \cdots i_p} \frac{\partial (m_{i_1} \cdots m_{i_p})}{\partial m_i}.
\end{equation}
These are precisely Eqs.~\eqref{eq:Keldysh_TAP_EOM} and~\eqref{eq:Keldysh_TAP_magnetization_equation} from the main text.

\section{Energy of a TAP state} \label{sec:TAP_energy}

Here we derive an expression for the total energy density $\epsilon$ in terms of a solution $\mathcal{G}(t, t')$ to the TAP equations, Eqs.~\eqref{eq:final_TAP_correlation_equation} and~\eqref{eq:final_TAP_magnetization_equation}.

Note that, despite the complicated derivation, $\mathcal{G}(t, t')$ is simply the contour-ordered expectation value of $N^{-1} \sum_i \sigma_i(t) \sigma_i(t')$.
Thus, as we discuss below in App.~\ref{sec:filter_functions}, the kinetic energy per spin is given by
\begin{equation} \label{eq:TAP_kinetic_energy_expression}
\frac{\mu}{2N \Delta t^2} \sum_i \Big< \big[ \sigma_i(t + 2\Delta t) - \sigma_i(t + \Delta t) \big] \big[ \sigma_i(t + \Delta t) - \sigma_i(t) \big] \Big> = -\frac{\mu}{2} \partial_t^2 \mathcal{G}(t^+, t),
\end{equation}
where $t^+$ denotes $t + \Delta t$.

To evaluate the potential energy, we replace $J_{i_1 \cdots i_p}$ by $(1 + \zeta(t)) J_{i_1 \cdots i_p}$ in the original path integral.
On the one hand,
\begin{equation} \label{eq:TAP_potential_energy_formal_expression}
\frac{\partial}{\partial \zeta(t)} S_{\textrm{TAP}} \bigg|_{\zeta = 0} = -\frac{1}{N} \sum_{(i_1 \cdots i_p)} J_{i_1 \cdots i_p} \big< \sigma_{i_1}(t) \cdots \sigma_{i_p}(t) \big>.
\end{equation}
On the other hand, it is easy to see how the explicit calculation of $S_{\textrm{TAP}}$ is modified by $\zeta(t)$:
\begin{equation} \label{eq:TAP_action_zeta_modification}
\begin{aligned}
S_{\textrm{TAP}} = S_{\textrm{TAP}} \bigg|_{\zeta = 0} &- \frac{1}{N} \int_{\mathcal{C}} dt \zeta(t) \sum_{(i_1 \cdots i_p)} J_{i_1 \cdots i_p} m_{i_1}(t) \cdots m_{i_p}(t) \\
&+ \frac{iJ^2}{2p} \int_{\mathcal{C}} dt dt' \big( \zeta(t) + \zeta(t') \big) \bigg[ \mathcal{G}(t, t')^p + (p-1) Q(t, t')^p - p \mathcal{G}(t, t') Q(t, t')^{p-1} \bigg] + O(\zeta^2).
\end{aligned}
\end{equation}
Thus
\begin{equation} \label{eq:TAP_potential_energy_actual_expression}
\begin{aligned}
\frac{1}{N} \sum_{(i_1 \cdots i_p)} J_{i_1 \cdots i_p} \big< \sigma_{i_1}(t) \cdots \sigma_{i_p}(t) \big> &= \frac{1}{N} \sum_{(i_1 \cdots i_p)} J_{i_1 \cdots i_p} m_{i_1}(t) \cdots m_{i_p}(t) \\
&\quad - \frac{iJ^2}{p} \int_{\mathcal{C}} dt' \bigg[ \mathcal{G}(t, t')^p + (p-1) Q(t, t')^p - p \mathcal{G}(t, t') Q(t, t')^{p-1} \bigg],
\end{aligned}
\end{equation}
and the total energy per spin is given by
\begin{equation} \label{eq:TAP_full_energy_expression}
\begin{aligned}
\epsilon = -\frac{\mu}{2} \partial_t^2 \mathcal{G}(t^+, t) &+ \frac{1}{N} \sum_{(i_1 \cdots i_p)} J_{i_1 \cdots i_p} m_{i_1}(t) \cdots m_{i_p}(t) \\
&- \frac{iJ^2}{p} \int_{\mathcal{C}} dt' \bigg[ \mathcal{G}(t, t')^p + (p-1) Q(t, t')^p - p \mathcal{G}(t, t') Q(t, t')^{p-1} \bigg].
\end{aligned}
\end{equation}

Writing out the various components of $\mathcal{G}$ explicitly and using time translation invariance (as well as taking $t$ to be far from the thermal branch), we have that
\begin{equation} \label{eq:TAP_full_energy_Keldysh_expression}
\epsilon = -\frac{\mu}{2} \partial_t^2 \mathcal{G}_{\alpha \alpha}(0^+) + Jq_{\textrm{EA}}^{p/2} \mathcal{E} - \frac{iJ^2}{p} \int dt' \sum_{\alpha'} (-1)^{\alpha'} \bigg[ \mathcal{G}_{\alpha \alpha'}(t - t')^p - p q_{\textrm{EA}}^{p-1} \mathcal{G}_{\alpha \alpha'}(t - t') + (p-1) q_{\textrm{EA}}^p \bigg],
\end{equation}
where $\mathcal{E}$ is defined as in the main text (Eq.~\eqref{eq:normalized_energy_definition}).
Despite the appearance, this expression is independent of both $t$ and $\alpha$.
Compare it to the argument of the filter function from Sec.~\ref{sec:nonergodic_ramp} (Eq.~\eqref{eq:TAP_resolved_energy_density_definition}), reproduced here:
\begin{equation} \label{eq:TAP_resolved_energy_density_reproduction}
\epsilon_{\alpha}[\lambda, G] = -\frac{\mu}{2} \partial_t^2 G_{\alpha \alpha}(0^+) - \frac{iJ^2}{p} \int_0^T dt' \sum_{\alpha'} (-1)^{\alpha'} G_{\alpha \alpha'}(t')^p - \frac{iJ \lambda}{pq[G]^{p/2}} \left( \frac{1}{T} \int_0^T dt G_{\alpha u}(t, 0) \right)^p.
\end{equation}
Evaluated at the saddle point given in Sec.~\ref{subsec:TAP_resolved_connected_solutions}, Eq.~\eqref{eq:TAP_resolved_energy_density_reproduction} comes out to be (up to corrections small in $T^{-1}$)
\begin{equation} \label{eq:TAP_resolved_energy_density_equivalence}
\epsilon_{\alpha}[\lambda, G] = -\frac{\mu}{2} \partial_t^2 \mathcal{G}_{\alpha \alpha}(0^+) + Jq_{\textrm{EA}}^{p/2} \mathcal{E} - 2J^2 q_{\textrm{EA}}^{p-1} \Lambda - \frac{iJ^2}{p} \int dt' \sum_{\alpha'} (-1)^{\alpha'} \Big( \mathcal{G}_{\alpha \alpha'}(t')^p - q_{\textrm{EA}}^p \Big),
\end{equation}
which, since $2i \Lambda \equiv \int dt' \sum_{\alpha'} (-1)^{\alpha'} \mathcal{G}_{\alpha \alpha'}(t)$, is precisely the energy density of the TAP state, Eq.~\eqref{eq:TAP_full_energy_Keldysh_expression}.

\section{Accounting for filter functions}
\label{sec:filter_functions}

The purpose of this section is to derive Eqs.~\eqref{eq:SFF_formal_path_integral} through~\eqref{eq:energy_density_definition} of the main text.
In the original definition of the SFF,
\begin{equation} \label{eq:generic_SFF_definition_restated}
\textrm{SFF}(T, f) = \overline{\textrm{Tr} f(H) e^{-iHT} \, \textrm{Tr} f(H) e^{iHT}},
\end{equation}
we need to be careful in how we treat the filter functions, since $f(H)$ does depend on the specific disorder realization.
It is convenient to assume that $f(H)$ is an analytic function of energy \textit{density}, i.e., it can be expanded as
\begin{equation} \label{eq:filter_function_expansion_definition}
f(H) = \sum_{n=0}^{\infty} \frac{c_n}{n!} \left( \frac{H}{N} \right)^n,
\end{equation}
where the coefficients $c_n$ do not depend on $N$.
This is not a significant restriction, since we can still choose $f(H)$ to be as tightly concentrated around any given energy density $\epsilon_0$ as we wish.

Inserting Eq.~\eqref{eq:filter_function_expansion_definition} into Eq.~\eqref{eq:generic_SFF_definition_restated} gives
\begin{equation} \label{eq:SFF_expanded_expression}
\textrm{SFF}(T, f) = \sum_{n_u n_l} \frac{c_{n_u} c_{n_l}}{n_u! n_l!} \frac{1}{N^{n_u + n_l}} \overline{\textrm{Tr} H^{n_u} e^{-iHT} \, \textrm{Tr} H^{n_l} e^{iHT}}.
\end{equation}
We construct the SFF path integral as usual: write $H = H_{\textrm{q}} + H_{\textrm{cl}}$ with $H_{\textrm{q}} \equiv \sum_i \pi_i^2 / 2\mu$ the kinetic energy and $H_{\textrm{cl}}$ the potential energy (including a Lagrange multiplier for the spherical constraint), approximate
\begin{equation} \label{eq:generic_Trotterization}
e^{i H T} \sim (e^{i H_{\textrm{q}} \Delta t} e^{i H_{\textrm{cl}} \Delta t})^\frac{T}{\Delta t},
\end{equation}
insert resolutions of the identity alternating between the $|\sigma \rangle$ and the $|\pi \rangle$ basis, perform the Gaussian integrals over momenta, and finally take $\Delta t \rightarrow 0$.
The only subtlety is in where and how we insert the factors of $H$ during this process.
While any arrangement has to ultimately give the same answer, we can simplify things by making judicious choices.

First write each $H$ as $\Delta t \sum_t H/T$ and insert one term at every time slice, i.e.,
\begin{equation} \label{eq:H_insertion_distributing}
\begin{aligned}
\textrm{Tr} H^{n_u} e^{-iHT} \textrm{Tr} H^{n_l} e^{iHT} = \left( \frac{\Delta t}{T} \right)^{n_u + n_l} &\sum_{t_1 \cdots t_{n_u}} \textrm{Tr} e^{-iH(T - t_{n_u})} H e^{-iH(t_{n_u} - t_{n_u-1})} \cdots e^{-iH(t_2 - t_1)} H e^{-iHt_1} \\
\cdot &\sum_{t_1 \cdots t_{n_l}} \textrm{Tr} e^{iH(T - t_{n_l})} H e^{iH(t_{n_l} - t_{n_l-1})} \cdots e^{iH(t_2 - t_1)} H e^{iHt_1}.
\end{aligned}
\end{equation}
Note that, to leading order in $\Delta t$, no two factors of $H$ coincide at the same time.
Furthermore, each factor of $H$ is itself two terms, whose operators we arrange within the Trotterized exponential (Eq.~\eqref{eq:generic_Trotterization}) as indicated:
\begin{equation} \label{eq:exponential_insertion_placement}
\cdots e^{i H_{\textrm{q}} \Delta t} e^{i H_{\textrm{cl}} \Delta t} \left( \frac{1}{2\mu} \sum_i \underbracket{\pi_i} e^{i H_{\textrm{q}} \Delta t} e^{i H_{\textrm{cl}} \Delta t} \underbracket{\pi_i} \, + \, e^{i H_{\textrm{q}} \Delta t} e^{i H_{\textrm{cl}} \Delta t} \underbracket{H_{\textrm{cl}}} \right) e^{i H_{\textrm{q}} \Delta t} e^{i H_{\textrm{cl}} \Delta t} \cdots.
\end{equation}
With this arrangement, when we carry out the rest of the path integral procedure, we end up with the integrand
\begin{equation} \label{eq:H_insertion_resulting_integrand}
\frac{\mu}{2 \Delta t^2} \sum_i \big[ \sigma_i(t + 2 \Delta t) - \sigma_i(t + \Delta t) \big] \big[ \sigma_i(t + \Delta t) - \sigma_i(t) \big] + \sum_{(i_1 \cdots i_p)} J_{i_1 \cdots i_p} \sigma_{i_1}(t) \cdots \sigma_{i_p}(t).
\end{equation}
While these expressions are for the lower contour, those for the upper contour are exactly analogous.
Thus the $n_u n_l$ term of the SFF becomes
\begin{equation} \label{eq:SFF_single_expansion_term_path_integral}
\textrm{Tr} H^{n_u} e^{-iHT} \textrm{Tr} H^{n_l} e^{iHT} = \int \mathcal{D}\sigma^N e^{iS} \left( \frac{\Delta t}{T} \right)^{n_u + n_l} \sum_{t_1 \cdots t_{n_u}} H_u(t_1) \cdots H_u(t_{n_u}) \sum_{t_1 \cdots t_{n_l}} H_l(t_1) \cdots H_l(t_{n_l}),
\end{equation}
where the action $S$ is, in continuum notation,
\begin{equation} \label{eq:SFF_original_spin_action}
S = \int_0^T dt \sum_{\alpha} (-1)^{\alpha} \left[ \sum_i \left( \frac{\mu}{2} \big( \partial_t \sigma_i(t) \big)^2 - \frac{z(t)}{2} \big( \sigma_{i \alpha}(t)^2 - 1 \big) \right) - \sum_{(i_1 \cdots i_p)} J_{i_1 \cdots i_p} \sigma_{i_1 \alpha}(t) \cdots \sigma_{i_p \alpha}(t) \right],
\end{equation}
and we've defined (in a slight abuse of notation) the random function
\begin{equation} \label{eq:path_integral_energy_function}
H_{\alpha}(t) \equiv \frac{\mu}{2 \Delta t^2} \sum_i \big[ \sigma_{i \alpha}(t + 2 \Delta t) - \sigma_{i \alpha}(t + \Delta t) \big] \big[ \sigma_{i \alpha}(t + \Delta t) - \sigma_{i \alpha}(t) \big] + \sum_{(i_1 \cdots i_p)} J_{i_1 \cdots i_p} \sigma_{i_1 \alpha}(t) \cdots \sigma_{i_p \alpha}(t).
\end{equation}

Eq.~\eqref{eq:SFF_single_expansion_term_path_integral} must now be averaged over the random couplings.
Due to the term of the action linear in $J_{i_1 \cdots i_p}$, each coupling acquires an expectation value:
\begin{equation} \label{eq:annealed_coupling_expectation_value}
\overline{J_{i_1 \cdots i_p} e^{iS}} = -\frac{iJ^2 (p-1)!}{C_{i_1 \cdots i_p} N^{p-1}} \int_0^T dt' \sum_{\alpha'} (-1)^{\alpha'} \sigma_{i_1 \alpha'}(t') \cdots \sigma_{i_p \alpha'}(t').
\end{equation}
Thus in the average of Eq.~\eqref{eq:SFF_single_expansion_term_path_integral}, one contribution is the fully disconnected term in which each insertion of $H_{\alpha}(t)$ gets replaced by its mean value, denoted $N \epsilon_{\alpha}(t)$:
\begin{equation} \label{eq:SFF_energy_density_expression}
N \epsilon_{\alpha}(t) \equiv \frac{\mu}{2} \sum_i \big[ \partial_t \sigma_{i \alpha}(t^+) \big] \big[ \partial_t \sigma_{i \alpha}(t) \big] - \frac{iNJ^2}{p} \int_0^T dt' \sum_{\alpha'} (-1)^{\alpha'} \left( \frac{1}{N} \sum_i \sigma_{i \alpha}(t) \sigma_{i \alpha'}(t') \right)^p.
\end{equation}
In fact, all other contributions, which come from contractions of pairs of couplings about their means, are necessarily subleading in $1/N$: whereas the disconnected term contributes one factor of $N$ for every insertion of $H$, each contraction accounts for two insertions but contributes only one factor of $N$ between the two.
Thus, to leading order,
\begin{equation} \label{eq:SFF_single_expansion_term_average}
\overline{\textrm{Tr} H^{n_u} e^{-iHT} \textrm{Tr} H^{n_l} e^{iHT}} \sim \int \mathcal{D}\sigma^N \big( N \epsilon'_u \big)^{n_u} \big( N \epsilon'_l \big)^{n_l} \overline{e^{iS}},
\end{equation}
where we've defined $\epsilon'_{\alpha} \equiv T^{-1} \int dt \epsilon_{\alpha}(t)$.

Returning to the SFF, we see that we can simply resum the series for the filter functions, but now with the \textit{deterministic} function $\epsilon'$ in place of the random operator $H$:
\begin{equation} \label{eq:SFF_expression_filter_function_treated}
\textrm{SFF}(T, f) = \int \mathcal{D}\sigma^N f \big( \epsilon'_u \big) f \big( \epsilon'_l \big) \overline{e^{iS}}.
\end{equation}
The average of $e^{iS}$ is treated exactly as in Subsec.~\ref{subsec:Schwinger_Keldysh_path_integral}, albeit with the extra index $\alpha$.
It results in the effective action given by Eq.~\eqref{eq:SFF_formal_action} in the main text.
Note that, when we introduce the order parameter $G_{\alpha \alpha'}(t, t')$ (see Eq.~\eqref{eq:partition_function_fat_unity}), we can also substitute it into the expression for $\epsilon'$, giving the function
\begin{equation} \label{eq:energy_density_first_definition}
\epsilon_{\alpha}[G] \equiv -\frac{\mu}{2T} \int dt \partial_t^2 G_{\alpha \alpha}(t^+, t) - \frac{iJ^2}{pT} \int_0^T dt dt' \sum_{\alpha'} (-1)^{\alpha'} G_{\alpha \alpha'}(t, t')^p.
\end{equation}
The only difference from Eq.~\eqref{eq:energy_density_definition} is that the latter assumes $G_{\alpha \alpha'}(t, t')$ is time-translation invariant.
This is true of the saddle-point value, and the replacement is justified here because $f(\epsilon[G])$ is independent of $N$ and is thus evaluated solely at the saddle point in the large-$N$ limit.
We have therefore arrived at Eqs.~\eqref{eq:SFF_formal_path_integral} through~\eqref{eq:energy_density_definition} of the main text.

\end{document}